%% file: main.tex
\documentclass[twocolumn]{aastex631}

\shorttitle{Seasonal variation of Saturn’s Lyman-$\alpha$ brightness}
\shortauthors{Stephenson et al.}

\include{preamble}
\begin{document}

\title{Seasonal variation of Saturn’s Lyman-$\alpha$ brightness}

\author{P. Stephenson}
\affiliation{Lunar and Planetary Laboratory, University of Arizona, Tucson, AZ, USA}

\author{T. T. Koskinen}
\affiliation{Lunar and Planetary Laboratory, University of Arizona, Tucson, AZ, USA}

\author{Z. Brown}
\affiliation{Lunar and Planetary Laboratory, University of Arizona, Tucson, AZ, USA}

\author{E. Qu\'emerais}
\affiliation{LATMOS/IPSL, Université de Versailles Saint-Quentin, Guyancourt,France}

\author{P. Lavvas}
\affiliation{GSMA, Université de Reims, Reims, France}

\author{J. I. Moses}
\affiliation{Space Science Institute, Boulder, CO, USA}

\author{B. Sandel}
\affiliation{Lunar and Planetary Laboratory, University of Arizona, Tucson, AZ, USA}
\author{R. Yelle}
\affiliation{Lunar and Planetary Laboratory, University of Arizona, Tucson, AZ, USA}

\begin{abstract}
We examine Saturn's non-auroral (dayglow) emissions at Lyman-$\alpha$ observed by the {Cassini/UVIS} instrument from 2004 until 2016, to constrain meridional and seasonal trends in the upper atmosphere. We separate viewing geometry effects from trends driven by atmospheric properties, by applying a multi-variate regression to the observed emissions. The Lyman-$\alpha$ dayglow brightnesses depend on the incident solar flux, solar incidence angle, emission angle, and observed latitude. The emissions across latitudes and seasons show a strong dependence with solar incidence angle, typical of resonantly scattered solar flux and consistent with no significant internal source. We observe a bulge in Ly-$\alpha$ brightness that shifts with the summer season from the southern to the northern hemisphere. We estimate atomic hydrogen optical depths above the methane homopause level for dayside disk observations (2004-2016) by comparing observed Lyman-$\alpha$ emissions to a radiative transfer model. We model emissions from resonantly scattered solar flux and a smaller but significant contribution by scattered photons from the interplanetary hydrogen (IPH) background. During northern summer, inferred hydrogen optical depths steeply decrease with latitude towards the winter hemisphere from a northern hemisphere bulge, as predicted by a 2D seasonal photochemical model. The southern hemisphere mirrors this trend during its summer. However, inferred optical depths show substantially more temporal variation between 2004 and 2016 than predicted by the photochemical model.
\end{abstract}



\section{Introduction} \label{sec:intro}

\noindent
Lyman-$\alpha$ is the brightest ultraviolet emission line in the solar system and has been observed at Saturn since 1976, initially with sounding rockets and the Copernicus satellite \citep{Weiser1977Copernicus, Barker1980Copernicus}. Subsequent observations of Lyman-$\alpha$ emissions from Saturn were made by the Interplanetary Ultraviolet Explorer \citep[IUE,][]{Clarke1981IUESaturn, McGrath1992H1980-1990} and during the Voyager flybys with the Ultraviolet Spectrometer \citep[UVS,][]{Broadfoot1981ExtremeSaturn, Sandel1982ExtremeSaturn, Ben-Jaffel1995NewSaturn}. Over its 13-year mission, Cassini/Huygens \citep{Matson2002TheSystem} orbited Saturn, compiling an extensive dataset of Lyman-$\alpha$ emissions with the Ultraviolet Imaging Spectrograph \citep[UVIS,][]{Esposito2005TheInvestigation}. 

Recently, \cite{Ben-Jaffel2023TheAtmosphere} identified a bulge in Lyman-$\alpha$ emissions from Saturn's thermosphere in the northern hemisphere between latitudes of 5 and 35$^\circ$ N, with observations from the Hubble Space Telescope's Space Telescope Imaging Spectrograph (HST/STIS) and Cassini/UVIS. They also identified the same bulge in Voyager/UVS observations that probed Saturn's atmosphere close to the northern spring equinox 35 years earlier \citep{Yelle1986AltitudeAtmosphere}. They concluded that the observed northern hemisphere bulge is a permanent feature of the thermosphere. We note that this bulge shows no longitudinal variation and therefore differs significantly from the Lyman-$\alpha$ bulge that has been observed on Jupiter, which is fixed in system III longitude \citep{Sandel1980DiscoveryJupiter, Clarke1980SpatialJupiter,Dessler1981TheConvection, Skinner1988Temporal19791986}. \cite{Ben-Jaffel2023TheAtmosphere} proposed two primary mechanisms to drive the emission bulge: variation of the temperature profile in the lower thermosphere and upper stratosphere or a previously unidentified suprathermal atomic hydrogen population at high altitudes, both of which could vary seasonally. They proposed that the suprathermal population could be created, for example, by a significant influx of material from the rings or Enceladus into the upper atmosphere. In this study, we examine Lyman-$\alpha$ emissions over the duration of the Cassini mission, in order to examine the cause of the bulge, whether it is a permanent feature of Saturn's thermosphere, and the source of the increased emissions. 

At Lyman-$\alpha$, {Cassini/UVIS} consistently observed much lower disk brightnesses compared to the observations of Voyager/UVS, with peak brightnesses of about 1 kR outside the auroral oval compared to 3-4 kR during the Voyager flybys \citep{Ben-Jaffel1995NewSaturn, Gustin2010CharacteristicsCassini-UVIS, Shemansky2009ThePlume, Koskinen2020SaturnObservations}. \cite{Gustin2010CharacteristicsCassini-UVIS} suggested that the disparity was a result of ring-reflected light during the Voyager observations, while \cite{Shemansky2009ThePlume} suggested strong electroglow emissions could reconcile the differences. The Voyager brightnesses have since been questioned and revised downward by \citet{Quemerais2013CrosscalibrationHeliosphere}, who concluded that the sensitivity of the Voyager/UVS instruments were underestimated by a factor of 1.5-2.5. The revised Voyager Lyman-$\alpha$ brightnesses are roughly consistent with those observed by {Cassini/UVIS} \citep{Koskinen2020SaturnObservations}. 
Using HST observations from Earth orbit as a calibration standard, \cite{Ben-Jaffel2023TheAtmosphere} challenged the downward revision of the Voyager brightnesses and proposed instead a recalibration of the {Cassini/UVIS} instrument at Lyman-$\alpha$ that would increase the observed brightnesses by 70\%. After scaling with the solar Lyman-$\alpha$ flux at different times, they compared several observations by {Cassini/UVIS} (in 2007, 2013 and 2014), HST/STIS (in 2017) and HST's Goddard High Resolution Spectrograph (in 1996) to arrive at this conclusion.

Cross-calibration of UV instruments between missions at Lyman-$\alpha$ remains difficult. Observations of Lyman-$\alpha$ emissions from interplanetary background hydrogen provide one method to facilitate it. Observations of the IPH Lyman-$\alpha$ by the Voyager/UVS instruments \citep{Katushkina2016Remote1993-2003, Katushkina2017Voyager1/UVSEmission}, New Horizons Alice \citep{Gladstone2018TheHorizons, Gladstone2021NewBackground} and future observations by PHEBUS on Bepi/Colombo \citep{Quemerais2020PHEBUSPerformance}, have and will continue to constrain models of the IPH background \citep[e.g.][]{Quemerais2002EffectsSun, Quemerais2013Cross-calibrationHeliospherec, Izmodenov2013DistributionLyman-, Pryor2022SupportingData}.
Models of the interaction between the local interstellar medium and the solar wind \citep{Quemerais2006InterplanetaryMaps, Izmodenov2001InterstellarHeliosphere, Izmodenov2013DistributionLyman-} are dependent on the hydrogen density of the local interstellar medium (LISM) and at the terminator shock near 90 au, with estimates of the LISM Hydrogen density varying from 0.12 to 0.195\,cm$^{-3}$ \citep{Dialynas2019PlasmaCrossing, Swaczyna2020DensityNeighborhood}. 
In addition to scattered solar flux, a galactic contribution to the background of 40\,R has been identified at large heliocentric distances \citep{Gladstone2021NewBackground, Pryor2022SupportingData}. While there remains uncertainty on the density of the LISM, we compare Cassini/UVIS Lyman  $\alpha$ observations  to the model of \citep{Quemerais2013CrosscalibrationHeliosphere} and find good agreement between Cassini/UVIS observed and modelled brightnesses, without the proposed recalibration by a factor 1.7 \citep{Ben-Jaffel2023TheAtmosphere, Pryor2024Modeling2017}. 
\comments{The situation with regard to calibration, however, is clearly confusing and based on this, we cannot entirely discount the possibility of a significant uncertainty in Saturn’s Lyman-$\alpha$ brightness.} 

Limb observations and solar occultations, in conjunction with photochemical models, can provide further constraints on the atomic hydrogen columns and Lyman-$\alpha$ emissions.


In addition to the Lyman-$\alpha$ bulge, Saturn's thermosphere also exhibits latitudinal variation in temperature \citep{Brown2020AData} and exobase altitude that are likely to be seasonally variable \citep{Koskinen2021empirical}.
Stellar occultations by {Cassini/UVIS} have constrained the temperature of Saturn's thermosphere, allowing the retrieval of density profiles of upper atmospheric constituents, including \ce{H2, He and CH4} \citep{Koskinen2013TheOccultations, Koskinen2015SaturnsOccultations, Koskinen2016TheAtmosphere, Koskinen2018uviscirs, Shemansky2012SaturnOccultations, Brown2020AData, Brown2022EvidenceCirculation}. Stellar occultations, when stable, are essentially self-calibrating and independent of instrument calibration. We note, however, that stellar occultations cannot be used to retrieve the density of H.

Latitudinal and seasonal trends in the upper atmosphere have been predicted by photochemical models \citep{Moses2000PhotochemistryObservations, Moses2000TheIonosphere, Moses2005LatitudinalIRTF/TEXES, Hue20152DTransport, Hue20162DTemperature}. This is because, for example, hydrocarbons in the stratosphere are influenced by meridionally-varying insolation, including changes due to the motion of the ring shadow across Saturn's disk \citep{Moses2005LatitudinalIRTF/TEXES}. The variation of methane in the stratosphere also has substantial impact on atomic hydrogen through both the production of H through photolysis \citep{Moses2000PhotochemistryObservations} and the location of the homopause. Comparison of photochemical models with Lyman-$\alpha$ observations from {Cassini/UVIS} can constrain the hydrogen column above the methane homopause, with methane a strong absorber at Lyman-$\alpha$. This can subsequently constrain eddy mixing and circulation near the homopause level \citep{Atreya1982EddySaturn, Sandel1982ExtremeSaturn, Atreya1984TheorySaturn, Emerich1993OnUranus, Moses2000PhotochemistryObservations, Moses2005LatitudinalIRTF/TEXES}. 

The emissions of Lyman-$\alpha$ from Saturn's disk, observed by Cassini/UVIS, provide an extensive dataset over 13 years, with coverage across all latitudes. \cite{Koskinen2020SaturnObservations} examined one observational sequence of Lyman-$\alpha$ from the Saturn disk in 2007. They found the brightnesses were consistent with resonance scattering of solar flux by a hydrogen column of $3\times10^{16}$\,cm$^{-3}$, in agreement with columns calculated with a photochemical model \citep{Moses2000TheIonosphere, Moses2000PhotochemistryOxygen, Moses2005LatitudinalIRTF/TEXES}. 
Several other case studies have been examined \citep{Mitchell2009IonSaturn, Shemansky2009ThePlume, Gustin2010CharacteristicsCassini-UVIS, Koskinen2020SaturnObservations, Ben-Jaffel2023TheAtmosphere}, but the full Lyman-$\alpha$ emission dataset has not yet been fully explored.

By comparing results from a radiative transfer model with Cassini UVIS observations, we can estimate the effective optical depth of the atomic hydrogen layer. We follow the approach of \citet{Yelle1989ResonanceTechniques} who modelled resonance scattering of a deep atmosphere by the iterative doubling and adding of thin layers, using angle-averaged partial frequency redistribution. We examine the variation of the effective hydrogen optical depths with latitude and season, under the assumption that emissions are dominated by resonant scattering by the ambient, thermal hydrogen population. In addition, we directly compare the inferred effective optical depths to the results of a seasonal photochemical model, to identify processes not included in the photochemical model, and to identify discrepancies that might indicate the presence of suprathermal atoms or internal emissions generated by photoelectron or energetic particle impact.

In this study, we examine the extensive dataset of Lyman-$\alpha$ emissions from Saturn's dayside disk collected by Cassini UVIS from 2004 to the end of 2016. We consider the Lyman-$\alpha$ observations through three approaches, with the methods used outlined in Section \ref{sec: methods}. In section \ref{sec: UVIS IPH comparison}, we compare observation of the IPH Lyman-$\alpha$ background to the model of \cite{Quemerais2013Cross-calibrationHeliospherec}. In Section \ref{sec: MVR results}, we employ a multi-variate analysis of the Lyman-$\alpha$ observations to confirm that resonance scattering of solar flux is a dominant source of the emissions from Saturn's non-auroral, dayside disk. Finally, in Section \ref{sec: rt model results}, we compare the radiative transfer model, based on doubling and adding of thin layers, to the Lyman-$\alpha$ observations, and we retrieve the H optical depth above the methane homopause across the mission, to determine seasonal variation of Saturn's thermosphere. We discuss the results in Section \ref{sec: discussion}, in particular with relation to the nature of the Lyman-$\alpha$ bulge. We also compare a seasonal photochemical model to the effective H optical depths retrieved in Section \ref{sec: rt model results}.

\section{Methods}\label{sec: methods}

\subsection{Cassini UVIS data}
\begin{figure*}
    \centering
    \includegraphics[width =0.8\textwidth]{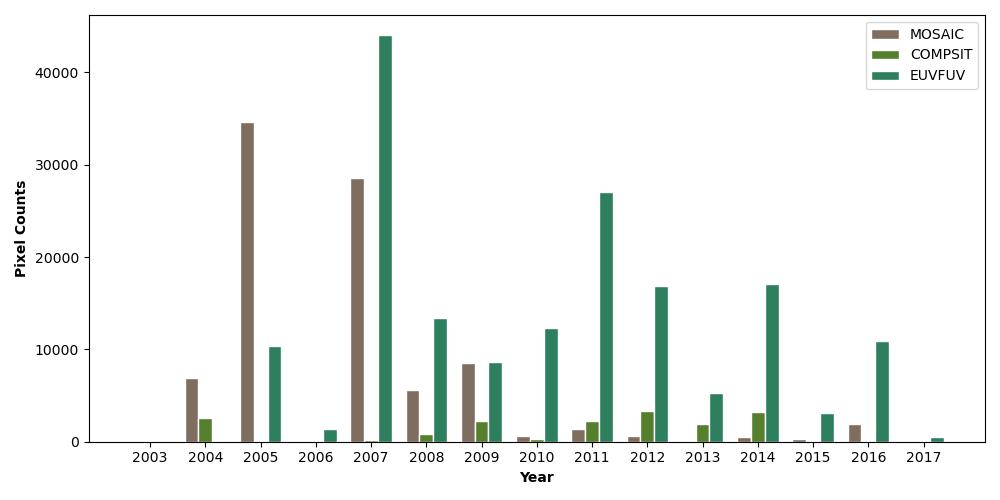}
    \caption{Number of observation pixels on Saturn's dayside disk, outside the auroral regions ($\lvert \phi_{lat}\rvert<65^\circ$) by year and observation type (see text for an explanation of the types).}
    \label{fig: UVIS pixel counts}
\end{figure*}

\noindent
We consider {Cassini/UVIS} observations from 2003 until 2016 during Cassini's orbits of Saturn, including MOSAIC, EUVFUV and COMPSIT/CIRS scans identified in the Planetary Data System. This first survey of the emission data focuses on nadir observations. Limb scans will not be considered in this work and are instead the subject of a future paper. The {Cassini/UVIS} instrument comprised 64 spatial bins along the slit and 1024 spectral bins with a resolution of 0.78\,{\AA}. We focus on Lyman-$\alpha$ emissions from Saturn's disk using the FUV channel of UVIS (1115 to 1912\,{\AA}), integrating the emission brightness between 1205 and 1225\,{\AA}. We use the time-dependent sensitivity and flat field updates that indicate a degradation of the signal by 30\% from launch to the end of the mission but in our default models, we do not use the proposed recalibration of the Cassini/UVIS instrument by a factor 1.7 at Lyman-$\alpha$ \citep[][p24]{Ben-Jaffel2023TheAtmosphere}. While additional degradation at Lyman-alpha would not be surprising, cross-calibrating instruments with different viewing geometries and the use of IPH models for calibration also includes uncertainties. For example, our IPH model produces Lyman-$\alpha$ brightnesses consistent with UVIS observations without the additional calibration factor (see Section \ref{sec: UVIS IPH comparison}). The effective optical depths predicted by photochemical models of Saturn’s atmosphere are also more consistent with UVIS observations with the previous calibration (see Section \ref{sec: rt model results}). In addition, we note that a calibration uncertainty of a factor 2 does not affect our results with respect to both seasonal and meridional trends observed in the Lyman-$\alpha$ emission.


In order to reduce the downlink data volume and increase spatial coverage, the wavelength range of observations was sometimes reduced and neighbouring spectral pixels were combined (typically with a bin width of 4 pixels). In cases with no spectral binning, we use the Cube Generator from the {Cassini/UVIS} team and the pipeline flatfield to process the data. When spectral binning was applied, the pipeline flatfield does not appropriately address the presence of `evil' pixels on the detector, which returned much lower signals than those surrounding them. In the pipeline calibration the intensity is interpolated to the `evil' pixels, which are assigned the value NaN in the flatfield. With binned spectra, one evil pixel in the bin results in a NaN value for the whole bin, losing information from the adjacent pixels. In these cases, we process the data with the Cube Generator with no flatfield, applying the derived 2007 flatfield correction of \citep{Koskinen2020SaturnObservations} for binned data. The only observations that use spectral binning occur between 2004 and 2008, and we do not expect a substantial change in the flatfield correction over this period.

\begin{figure}
\centering
\includegraphics[width=0.5\textwidth]{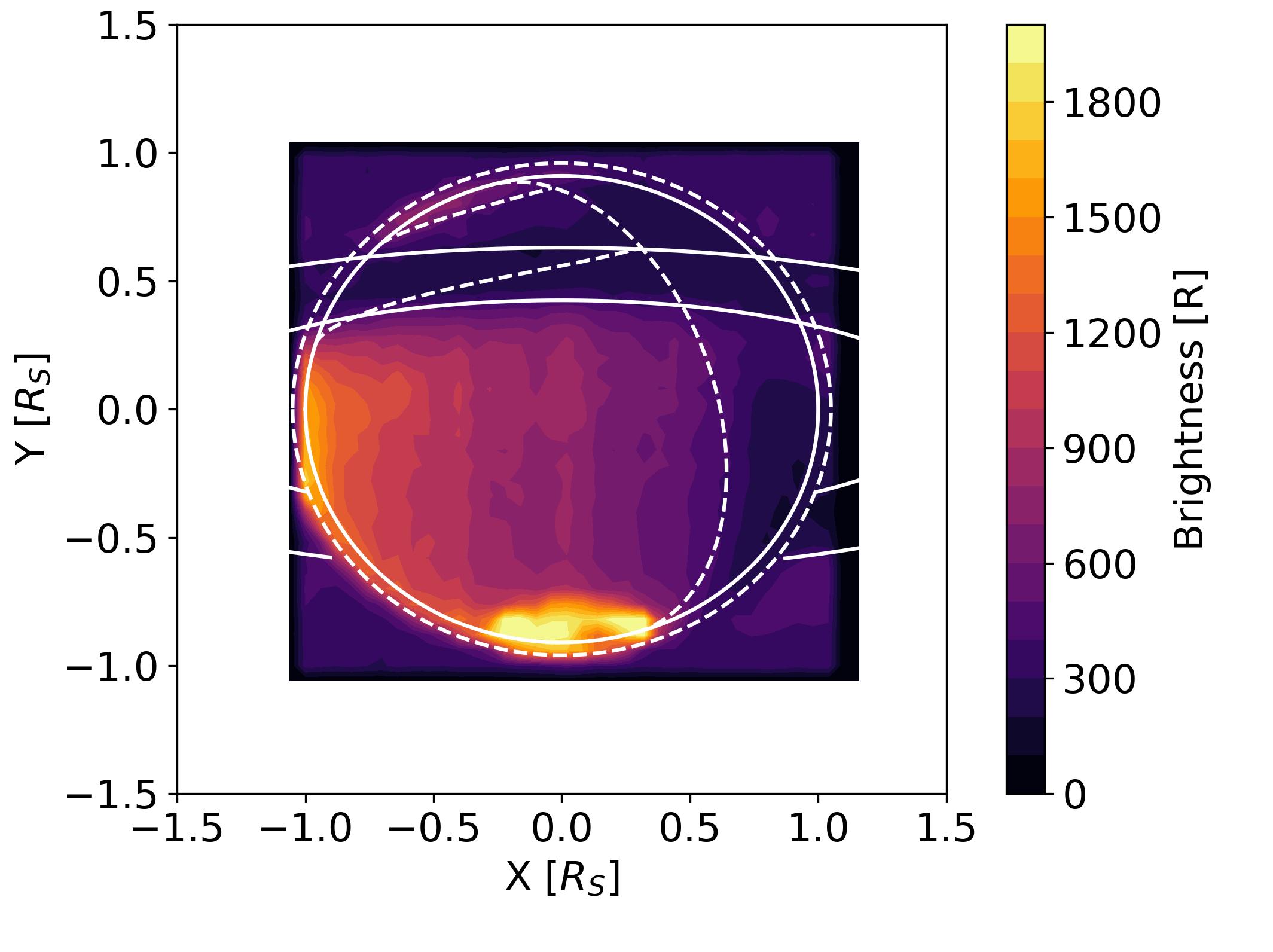}
\caption{Observed Lyman-$\alpha$ emission brightness from Saturn on 21 Jun 2005. The solid white lines outline Saturn's disk and its rings. Dashed lines show the ring shadow region (at this time in the northern hemisphere), the terminator, and the position of the exobase.}
\label{fig: observation example}
\end{figure}

The three observation classes (MOSAIC, EUVFUV and COMPSIT/CIRS) had different aims and characteristics. MOSAIC observations were designed to maximise spatial coverage, observing much or all of the Saturn system (including the rings). Prior to 2008, spectral binning of width 4 was common allowing integration times of 25 to 95s. After 2008, the MOSAIC observations did not use any spectral binning and required longer integration times from 120 to 900\,s. During EUVFUV observations, spectra from both the EUV and FUV channels were retrieved, with no spectral binning. These often capture a substantial part of the Saturn system, such as in Figure \ref{fig: observation example}, using an integration time between 180 and 260s. Finally, the COMPSIT/CIRS observations were paired with observations by the  Cassini/Composite Infrared Spectrometer \citep[CIRS,][]{Flasar2005ExploringSpectrometer}. These do not use any spectral binning and have much smaller spatial coverage, often only observing along a single line across Saturn's disk and limb. However, the long integration times (1200 or 2400s) provide excellent signal-to-noise ratios for the spectra.

The UVIS dataset comprises 636 observations over 14 years comprising 140,925 scans and 8,032,725 pixels, of which 3,000,000 are on the dayside of Saturn's disk. In this study, we focus on airglow emissions and therefore exclude the auroral regions with latitudes poleward of 60$^\circ$. Figure \ref{fig: UVIS pixel counts} shows the number of observation pixels by observation type and year. Additionally, over large periods much of Saturn's disc was shadowed by the Saturnian rings. We do not have a good constraint on the solar flux entering the atmosphere after absorption in the ring atmosphere, so these points are removed from the dataset. For each observation, the ring shadow region and rings are mapped onto the surface of Saturn (e.g. see Figure \ref{fig: observation example}).

\subsection{Multivariate quadratic regression of UVIS data} \label{sec: MVR outline}

\noindent
In this section, we outline a data based approach to the analysis of the Lyman-$\alpha$ emissions from Saturn, which is independent of radiative transfer modelling. This helps to identify the key emission source of Lyman-$\alpha$ at Saturn and to test the assumptions used in the RT model. 
For this purpose, we consider emission observations from 2014 until the end of mission, from MOSAIC, EUVFUV, and COMPSIT/CIRS observations. We primarily focus on part of the northern hemisphere summer (2014-2016), such that the meridional trends in the atmosphere and emissions do not change substantially with time. 

We use a multi-variate regression with independent variables of emission angle, incidence angle and latitude, which are used to predict the observed Lyman-$\alpha$ brightness. For simplicity, we have removed the dependence on the solar flux variation by scaling all the brightnesses to the solar flux applicable on 1 Jan 2016 (see Section~\ref{subsc:scatsol}). We choose these variables because resonance scattering is strongly dependent on the solar incidence angle and the emission angle of the observation, and because previous studies have identified meridional trends in the Lyman-$\alpha$ emissions.
The model includes a quadratic in emission and incidence angles, in addition to a cubic in latitude. The full expression used for the regression is:

\begin{multline}\label{eq: MVR fit equation}
 B = p_0 + p_1 \cdot \theta_{\text{{em}}} + p_2 \cdot \theta_{\text{{in}}} + p_3 \cdot \phi_{\text{{lat}}} + p_4 \cdot \theta_{\text{{em}}}^2 + \\ p_5 \cdot \theta_{\text{{em}}} \cdot \theta_{\text{{in}}} + p_6 \cdot \theta_{\text{{in}}}^2 + p_7 \cdot \phi_{\text{{lat}}}^2 + p_8 \cdot \phi_{\text{{lat}}}^3, 
\end{multline}
with $p_i$ the coefficients, $\theta_{\text{{em}}}$ is the emission angle, $\theta_{\text{{in}}}$ is the solar incidence angle, and $\phi_{\text{{lat}}}$ is the planetocentric latitude.
We prepare the data by standardizing the independent variables ($\theta_{\text{{em}}}, \theta_{\text{{in}}}, \phi_{\text{{lat}}}$), and then transforming them into the required polynomial expressions in Eq. \ref{eq: MVR fit equation}. The 86,913 pixels in the NH summer were randomly split by 80\% to 20\% into training and testing sets, respectively. The model was then trained using a least-squares regression. Monte-Carlo analysis was used to retrieve confidence intervals for the coefficients, resampling the dataset 1000 times.

\subsection{Radiative transfer modeling}\label{sec: rt model methods}

\noindent
In order to constrain the properties of the atmosphere, we model the brightness of scattered solar and IPH Lyman-$\alpha$ using a radiative transfer model based on doubling and adding, which includes an angular dependent frequency redistribution function \citep{Yelle1988AAtmospheres, Yelle1989Function, Wallace1989ResonanceFunction}. The model is based on assumptions of a plane-parallel and isothermal atmosphere. The plane-parallel assumption breaks down for observations close to either the limb or the terminator, as the incidence or emission angles near 90$^\circ$. Consequently, we only apply the model to cases of $\theta_{em}<60^\circ$ and $\theta_{in}<65^\circ$. Thin layer approximations of scattering and transmission functions \citep[outlined in Appendix \ref{sec: thin layer approx}][]{Yelle1988AAtmospheres} are iteratively doubled in thickness (see Appendix \ref{sec: RT model appendix}), computing new scattering, transmission and extinction functions until the required optical depth is reached. 

The scattering and transmission functions are dependent on the atmospheric temperature (via the Lyman-$\alpha$ lineshape in the thermosphere) and the optical depth. We use a thermospheric temperature that varies with latitude, based on the pole-to-pole map of stellar occultations of {Cassini/UVIS} throughout 2017 \citep[see Section \ref{sec: optical depth methods};][]{Brown2020AData}.
\begin{figure}
    \centering
    \includegraphics[width=0.45\textwidth]{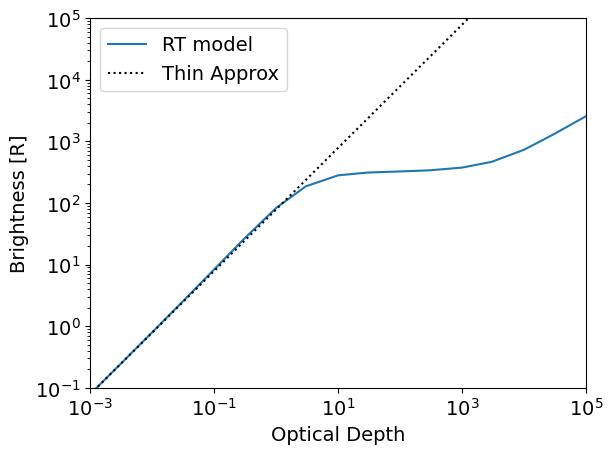}
    \caption{(blue) Brightness vs. optical depth for resonance scattering of Lyman-$\alpha$ by atomic hydrogen for $(\theta_{in}, \theta_{em})=(0,0)$. The optically thin approximation is given by the black dotted line.}
    \label{fig: function of gain}
\end{figure}
Figure \ref{fig: function of gain} shows the variation in brightness as the atomic H optical depth increases, for the nadir case ($\theta_{in}, \theta_{em})=(0,0)$. The approximation for an optically thin atmosphere is given by the dotted line. The RT model is identical to the optically thin approximation at small optical depths, before deviating as it approaches $\tau=1$. Beyond this, the atmosphere is optically thick and the output brightness varies little as the column grows. At $\tau=10^3$, the brightness then begins to increase once again, as frequency redistribution becomes more effective and photons in the Lorentzian wings begin to be multiply scattered.
For deep atmospheres, like Saturn's, the frequency redistribution is critical to computing the scattered brightness, due to the scattering of photons initially far from the line centre.

\subsubsection{Scattered solar flux}
\label{subsc:scatsol}

\begin{figure}
    \centering
    \includegraphics[width =0.45\textwidth]{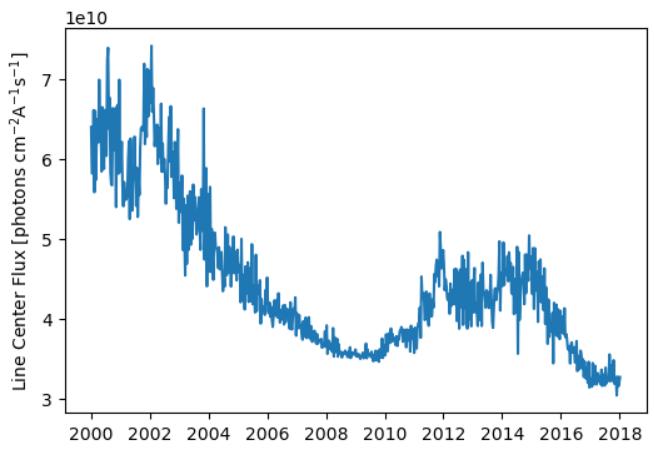}
    \caption{Lyman-$\alpha$ flux at Saturn throughout the Cassini mission, extrapolated from LISIRD fluxes at 1 AU. This does not include absorption of the solar flux between 1 AU and Saturn, which is accounted for separately.}
    \label{fig: solar LC flux}
\end{figure}

\begin{figure}
    \centering
    \includegraphics[width=0.5\textwidth]{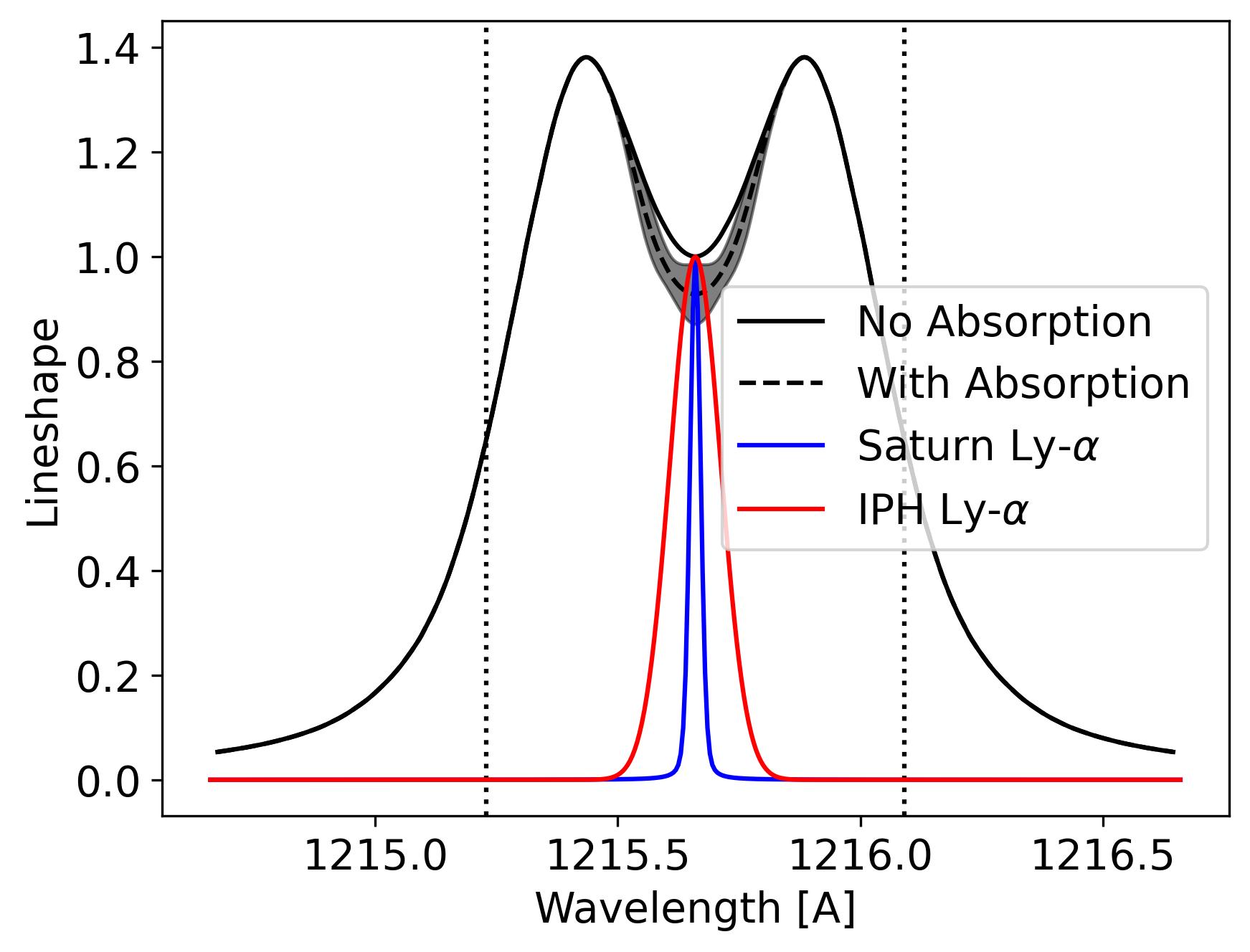}
    \caption{(black solid line) Lineshape of the solar Lyman $\alpha$ flux (black) at 1au normalised to 1 at the linecentre. (black dashed line and shaded region) Lineshape of solar Lyman-$\alpha$ after absorption of solar flux by the IPH background. (blue) The Voigt profile of Lyman-$\alpha$ in Saturn's thermosphere for a temperature of 350K. (red) The lineshape of the IPH Lyman-$\alpha$ entering the upper atmosphere at $T_{IPH}=10,000$\,K. The dotted vertical lines show the integration limits of $\pm44\Delta\nu_{D, Sat}$, used in the radiative transfer model.}
    \label{fig: lya lineshape at 1au}
\end{figure}

\noindent
For scattered solar flux, the brightness of Lyman-$\alpha$ is dependent on the magnitude and shape of the solar flux entering the top of the atmosphere. The RT code is normalised to the solar flux at line center. For the magnitude of the flux, we use the LISIRD composite Lyman-$\alpha$ database \citep{Machol2019AnComposite}, which is based on fluxes measured at 1 AU by various instruments \citep[e.g.][]{Hinteregger1981ObservationalAEE, Barth1983SolarResults, Woods2000ImprovedObservations, McClintock2005Solar-StellarCalibrations}. First, we correct the LISIRD date for the flux to match the solar longitude at Saturn during the UVIS observation, which requires a shift of up to 15 days depending on the relative positions of Earth and Saturn. This is converted to a line center flux using the line shape from \cite{Lemaire2005Variation23}, imposed at 1 AU (see Figure \ref{fig: lya lineshape at 1au}).
After this, we extrapolate the flux to Saturn using an inverse square law with heliocentric distance (see Figure \ref{fig: solar LC flux}). We include absorption of the solar flux between 1\,AU and Saturn by the IPH background, using the IPH model described in Section \ref{sec: IPH model description}. Once normalised, the modelled brightness based on scattered solar flux from Saturn's atmosphere is interpolated to the emission and solar incidence angle of a given observation, giving brightness as a function of temperature, effective H column optical depth, and viewing geometry. The temperature is constrained by a fit with latitude to the results retrieved from stellar occultations of the thermosphere (see Section \ref{sec: optical depth methods}).

\subsubsection{Lyman-$\alpha$ from the interplanetary background}\label{sec: IPH model description}

\noindent
Lyman-$\alpha$ flux emitted by the IPH background is also scattered by Saturn's upper atmosphere. Unlike the solar flux, the IPH Lyman-$\alpha$ radiation covers the sky, so we integrate the IPH flux that enters from all directions (see Appendix \ref{sec: iph model brightness} for further details). The flux entering the atmosphere from each direction is calculated with an IPH background model of \cite{Quemerais2013CrosscalibrationHeliosphere}. This model incorporates angular dependent partial frequency redistribution to treat resonant scattering of solar Lyman-$\alpha$ radiation by the interplanetary hydrogen distribution \citep{Quemerais2000AngleAlpha, Quemerais2003VoyagerHeliosphere}. The distribution is calculated from the interaction of the local interstellar medium with the solar wind \citep{Izmodenov2001InterstellarHeliosphere, Izmodenov2013DistributionLyman-}, using a hydrogen density of $n_{H, TS}=0.09$\,cm$^{-3}$ and $n_H=0.14$\,cm$^{-3}$ in the local interstellar medium \citep{Bzowski2009NeutralResults, Izmodenov2020MagnitudeModel}. 
We do not include an additional contribution of a 40\,R galactic contribution \citep{Gladstone2021NewBackground, Pryor2022SupportingData} that is isotropic. However, at distances of 9-10\,au the galactic contribution is small in comparison to IPH-scattered solar flux. The brightness, line width and temperature of the IPH Lyman-$\alpha$ is calculated by integrating the emissions along the line-of-sight. 

\begin{figure*}
    \centering
 
    \includegraphics[width=\textwidth]{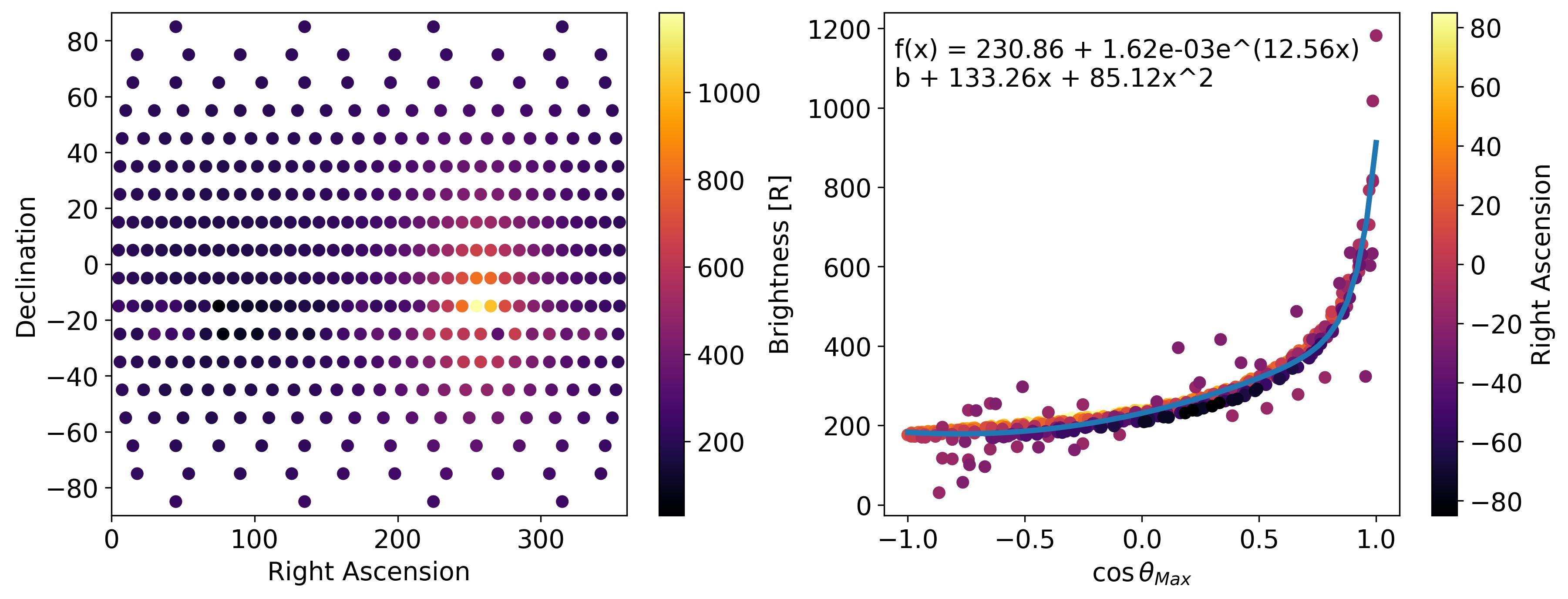}
    \caption{(left) Full sky map of IPH brightness at Saturn in 2016, modelled as described in Section \ref{sec: IPH model description}. (right) Modelled IPH Lyman-$\alpha$ brightness against angle to direction of maximum brightness. The fit to the IPH brightness is shown in blue.}
    \label{fig: IPH map 2016}
\end{figure*}

The brightness of the IPH emission line varies strongly with direction of observation, being largest close to the sunward direction when at large heliocentric distances. The position of the observer relative to the flow of the local interstellar medium also impacts the variation of Lyman-$\alpha$ with line-of-sight, although to a lesser extent. Here, we consider two positions in the IPH: the position of Saturn in 2006 and 2016. In both cases, we construct a full sky map of the Lyman-$\alpha$ brightness at the position of Saturn (e.g. Figure \ref{fig: IPH map 2016}a). The IPH Lyman-$\alpha$ brightness can be parameterised as a function of the angle to the direction of maximum brightness ($\theta_{Max}$, Fig. \ref{fig: IPH map 2016}b). We fit this as a sum of a quadratic and an exponential with respect to $\cos\theta_{max}$. At heliocentric distances of 10~AU, the maximum in the IPH density and Lyman-$\alpha$ brightness are closely aligned with the sunward direction.

\subsubsection{Hydrogen effective optical depths}\label{sec: optical depth methods}

\noindent
The radiative transfer model, combining scattering of solar and IPH Lyman-$\alpha$ photons is used to calculate brightness as a function of the optical depth of the effective H column in the atmosphere, emission and incidence angle, and atmospheric temperature $B_{RT}(\tau_H, \theta_{em}, \theta_{in}, T_{Sat})$. By effective optical depth, we refer to the vertical Lyman-$\alpha$ line center optical depth of the H column above the methane homopause. It is important to note that we assume all light at Lyman-$\alpha$ to be absorbed below the homopause. We do not self-consistently simulate absorption by CH$_4$ because seasonal changes to the depth of the homopause are not known a priori. This effort represents the first, zeroth order effort to constrain variations in the effective H scattering column in the thermosphere. We have therefore chosen to follow this simplified, retrieval-type approach in our modeling.

We use a variable temperature profile with latitude, based on stellar occultations by UVIS \citep{Brown2020AData} combined with CIRS limb scans \citep{Brown2024AMesosphere, Guerlet2018EquatorialEpoch, Koskinen2018uviscirs}. Thermal structure is also expected to be seasonally variable but the general characteristics of the temperature distribution, including the gradient between the auroral regions and the equator, are relatively stable, even on multi-decadal timescales \citep{Koskinen2021empirical}. We calculate the pressure averaged temperatures ($\bar{T}_P = 1/[\ln(p_1/p_0)]\int_{P_0}^{P_1} T (d\ln P)$ ) above the methane homopause, which  are then fit with a 6th-order polynomial (red, Figure \ref{fig: temp variation}). Here, the CH$_4$ homopause is defined as the location where $\tau_{CH_4} =$~ 1 at Lyman-$\alpha$ in the atmospheric structure models fitted to the occultations and CIRS observations \citep[e.g.,][]{Brown2024AMesosphere, Koskinen2018uviscirs}. As noted above, the homopause location is obviously expected to change over time but the pressure-averaged temperatures are nevertheless assumed to remain relatively stable. The polynomial fit is used to convert the latitude of observation into an atmospheric temperature in the radiative transfer model.

\begin{figure}
    \centering
    \includegraphics[width = 0.5\textwidth]{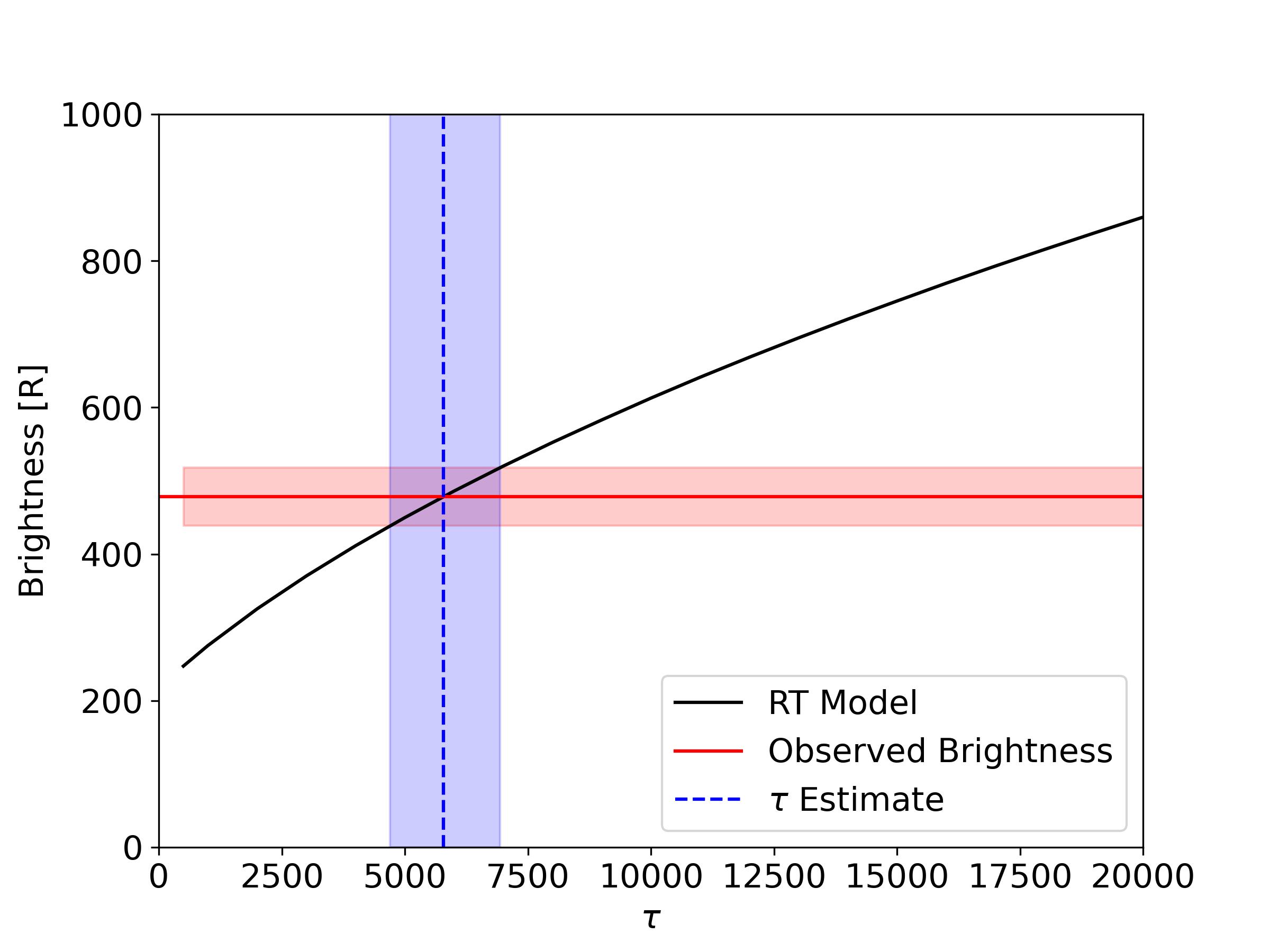}
    \caption{Retrieval of optical depth estimates (blue) by comparison of {Cassini/UVIS} observations (red) with the radiative transfer model (black) for a single pixel. }
    \label{fig: optical depth retrieval}
\end{figure}

\begin{figure}
    \centering
    \includegraphics[width =0.5\textwidth]{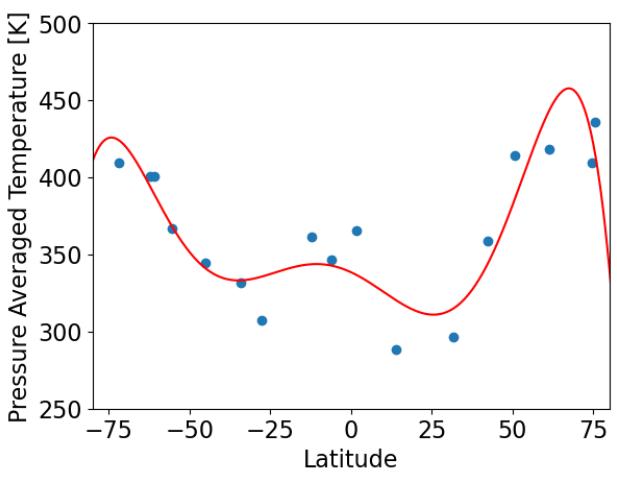}
    \caption{(blue) Pressure averaged temperatures above the methane homopause, from a 1D photochemical model (see Section \ref{sec: photochemical model}), constrained by stellar UVIS occultations throughout 2017 \citep{Brown2020AData}. (red) The temperatures are fitted with a 6th order polynomial. }
    \label{fig: temp variation}
\end{figure}

For each pixel, characterised by incidence angle, emission angle and latitude (temperature), the brightness against optical depth is interpolated to the viewing geometry. Using this monotonic relationship, the observed Lyman-$\alpha$ brightness for each pixel is converted to an effective hydrogen optical depth for each pixel (Figure \ref{fig: optical depth retrieval}). 
We use Monte Carlo error estimation, allowing for uncertainty in the Lyman-$\alpha$ brightness from the UVIS instrument, as well as uncertainty in the solar flux normalisation and absorption of the solar flux between the Sun and Saturn.

\subsubsection{Other sources of Lyman-$\alpha$}

\noindent
In comparing the radiative transfer model with the observed brightnesses, we ignore the possibility of internal sources and scattering by suprathermal atomic H. However, we considered the following potential emission sources that are likely to be insignificant:
\begin{itemize}
    \item photoelectron impact excitation
    \item \ce{H+} recombination
    \item scattering by hot hydrogen in the thermosphere, produced via \ce{H_3+} recombination.
\end{itemize}

\cite{Waite1983ElectronIonosphere} used a two-stream model to examine Lyman-$\alpha$ emissions driven by photoelectrons at Jupiter, finding a total contribution of 26\,R. Extrapolating this to Saturn, we expect a factor 3-4 reduction in the photoelectrons based on the reduced solar flux at Saturn and the emitting column of H at Saturn is approximately a factor 3 smaller than at Jupiter. Therefore, we expect photoelectron-induced Lyman $\alpha$ emissions to be negligible compared to scattered solar flux. 

Recombination is a well known astrophysical source of Lyman-$\alpha$ emissions, and a possible source in hot exoplanet atmospheres. In cold planetary atmospheres, however, it is negligible. For \ce{H+} recombination, we calculate the case-B recombination rate using \ce{H+} and \ce{e-} profiles from the 1D photochemical model \citep{Moses2023SaturnsMeasurements}, to estimate an upper limit for the production of excited H atoms ($n\geqslant 2$). Integrating this over the atmospheric column gives a column emission of only $\sim10^{-3}$\,R. 

\comments{
Observations of Lyman-$\alpha$ with high spectral resolution should be able to identify, or provide an upper limit on, a hot population of hydrogen in the atmosphere, such as HST/STIS echelle observations \citep[see Fig. 3 in ][]{Ben-Jaffel2023TheAtmosphere}. However, the uncertainties in the high-spectral resolution HST observations are significant, partly due to the subtraction of the strong geocoronal Lyman-$\alpha$ emission. Consequently, retrieving a useful upper limit on a hot atomic hydrogen population does not seem feasible. \cite{Ben-Jaffel2023TheAtmosphere} compare a reference and boosted (with 2-3 times the atomic hydrogen) photochemical model  to the HST/STIS observations. The additional 280\,R of emissions in the boosted model could similarly be attained by including a hot atomic hydrogen population ($\tau_{H,Hot}=1$, $T_{Hot}=25,000$\,K), with minimal impact on the resulting lineshape. However, the mechanism for producing a significant suprathermal hydrogen population remains unclear. }

One potential source of hot hydrogen in the upper atmosphere is recombination of \ce{H3+}, producing \ce{H + H +H } and \ce{H + H_2}. We constrain this source by using the photochemical model profiles \citep{Moses2023SaturnsMeasurements}. At zeroth order, we assume local production of hot H \citep[rates listed in ][]{Larsson2008TheEnd} balanced by local cooling through elastic collisions of the hot H with thermal H and \ce{H_2}, with collision cross sections from \cite{Krstic1999ConsistentMolecules}. Under this assumption, the hot H population generated by recombination of \ce{H3+} in the upper atmosphere is negligible, with column densities a factor $10^7$ smaller than the ambient atomic hydrogen column density. This hot H population is too small to generate Lyman-$\alpha$ emissions comparable to scattering by the thermal H population (a column of hot atoms a factor of $10^4$ smaller than thermal H could generate similar emissions). 


$H_3^+$ rain from the rings has been observed at Saturn \citep{Odonoghue2019rain}. However, the ring rain latitudes show no obvious correlation with the observed Lyman-$\alpha$ bulge and it is difficult to anticipate a related mechanism that could generate enough hot atoms to explain the bulge. While we cannot rule out a hot population generated by another means, the observations and modelling at present appear most consistent with emissions driven by resonance scattering of solar flux by thermal atomic hydrogen.

\subsection{Photochemical models of Saturn's upper atmosphere}\label{sec: photochemical model}

\noindent
The upper atmospheric hydrogen is a probe of photochemistry deeper in the atmosphere, as it is primarily produced through photolysis of methane \citep{Moses2000PhotochemistryObservations}, with only a small contribution from thermospheric chemistry. We compare the inferred H optical depths (outlined in Section \ref{sec: rt model methods}) to two photochemical models: (a) 1D models tuned to the results of stellar occultations observed by {Cassini/UVIS} in 2016-2017 \citep{Brown2020AData, Brown2024AMesosphere} and (b) a 2D model identical to \cite{Moses2005LatitudinalIRTF/TEXES}, except we adopt solar-cycle average incident ultraviolet flux (including at Lyman-$alpha$), rather than tracking the $\sim$11 year solar cycle variation from the recent era.

\paragraph{1D models}
Throughout 2016 and 2017, stellar occultations provided a pole-to-pole map of the thermosphere \citep{Brown2020AData} and were used to retrieve temperature profiles as well as the densities of many atmospheric constituents \citep[e.g. \ce{H2}, \ce{CH4}][]{Brown2024AMesosphere}. However, as we noted before, it is not possible to retrieve the H density profile from stellar occultations, due to absorption of starlight by the interstellar medium. The H density profile is calculated with the 1D photochemical model, constrained by the profiles of other molecules. 

For this study, we are interested in the hydrogen and methane profiles, particularly the H column above the methane homopause. We consider optical depths of $\tau_{CH4}=1$ and $\tau_{CH4}=10$ at Lyman-$\alpha$, using a cross-section of $1.79\times10^{-17}$\,cm$^{2}$  \citep{Chen2004Temperature-dependentEthane}. For the purposes of the comparison between the inferred optical depths and the model, the H optical depth is integrated above the homopause altitude for each occultation location, in line with the pressure averaged temperatures above the homopause (see Figure \ref{fig: temp variation}).

\begin{figure}
    \centering
    \includegraphics[width =0.5\textwidth]{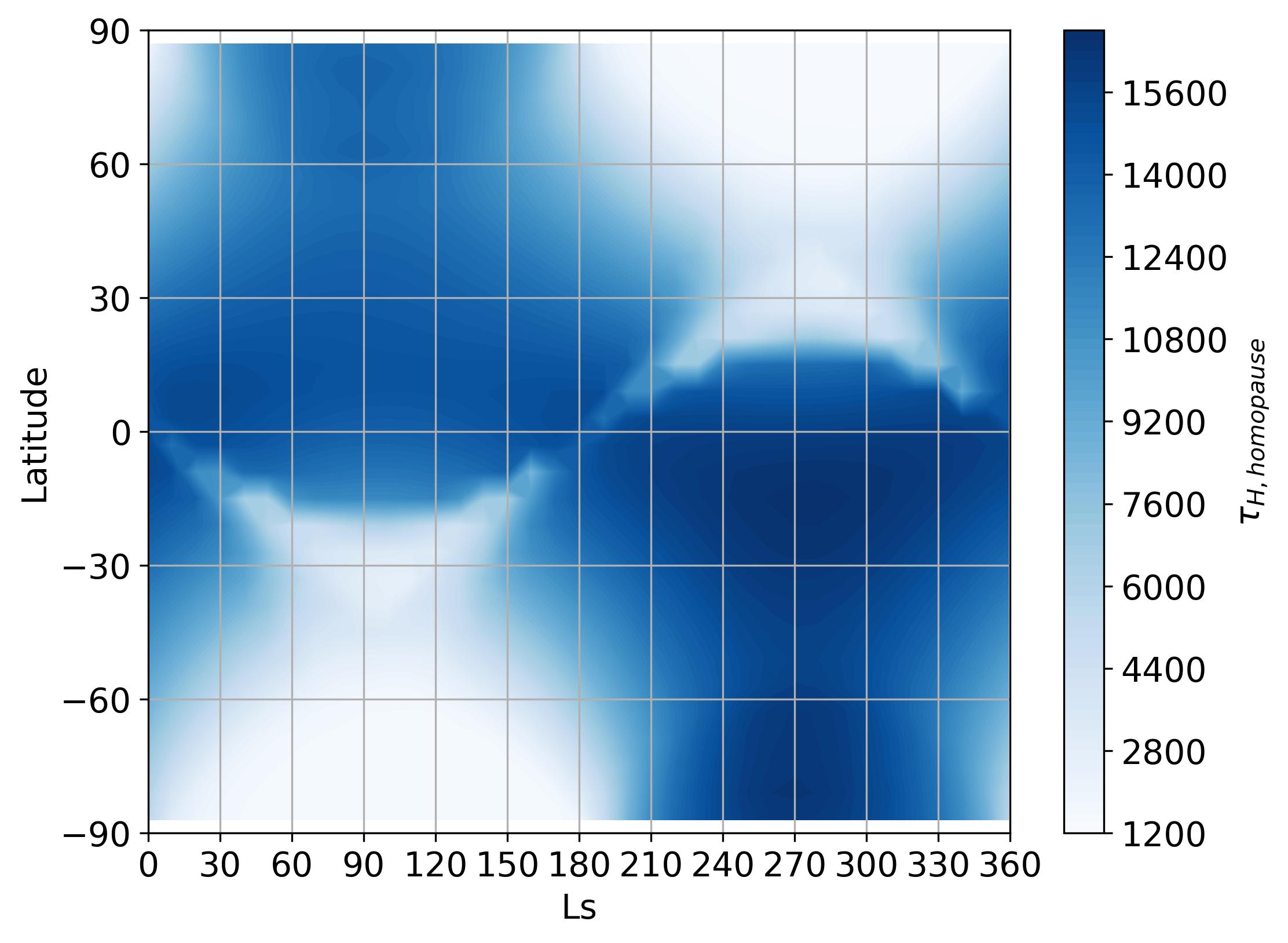}
    \caption{Optical depth of H above the methane homopause ($\tau_{CH4}=1$) from the 2D photochemical model, similar to \cite{Moses2005LatitudinalIRTF/TEXES}. The $x$-axis is solar longitude with the northern solstice occurring at $L_S=90^\circ$ and southern solstice at $L_S=270^\circ$.}
    \label{fig: 2d photochem model}
\end{figure}

\paragraph{2D model}
We use the seasonally and meridionally varying 2D model of \cite{Moses2005LatitudinalIRTF/TEXES}, which incorporates the time varying insolation to represent seasonal changes with solar-cycle averaged ultraviolet fluxes. In this model, the temperature-pressure profile is constant across latitudes and with season. It is important to note that this model parameterizes vertical transport by using an eddy diffusion profile that is constant with latitude and does not include any meridional transport. We integrate the H optical depth above the methane homopause ($\tau_{CH4}=1$) as a function of latitude and time (with the time mapped to the solar longitude; see Figure \ref{fig: 2d photochem model}). At present, photochemical models that account for seasonal changes to the temperature structure and mixing rates in the middle and upper atmosphere do not exist.
The differences between the inferred H column from the Lyman-$\alpha$ observations and the photochemical model predictions identified in this study represent the first step to guide the development of such models in the future.

\section{Results}\label{sec: results}

\subsection{A comparison of UVIS observations with the IPH model}\label{sec: UVIS IPH comparison}

\begin{figure*}
    \includegraphics[width=\textwidth]{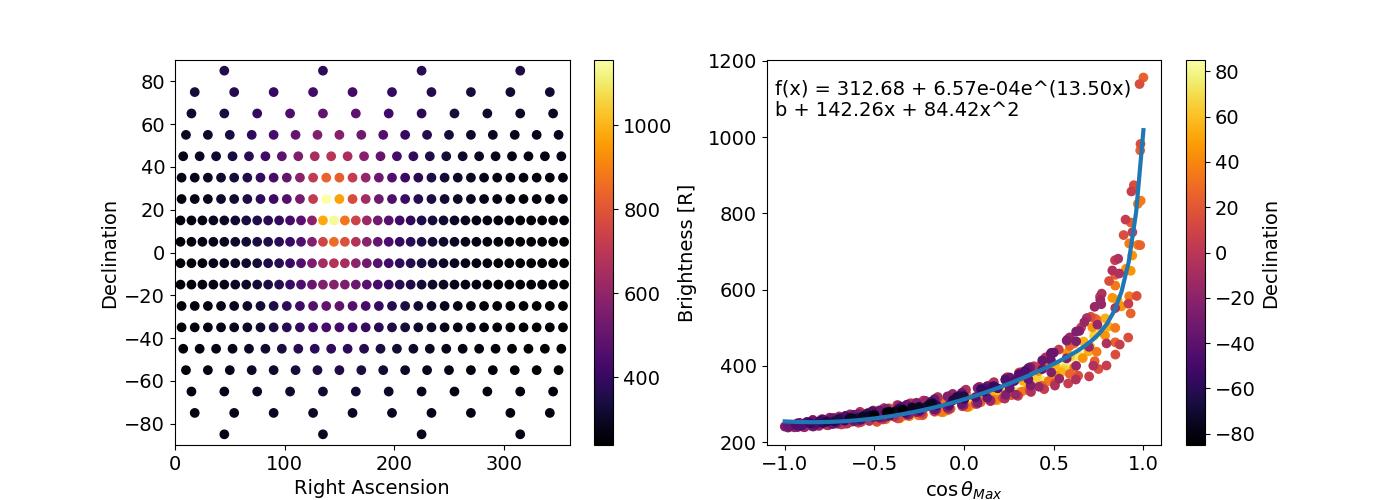}
    \caption{(left) Full sky map of the modelled IPH brightness at Saturn in 2006 (see Section \ref{sec: IPH model description}). (right) Modelled IPH Lyman-$\alpha$ brightness against angle to direction of maximum brightness. The fit to the IPH brightness is shown in blue.}
    \label{fig: IPH model 2006}
\end{figure*}

\begin{figure}
    \centering
    \includegraphics[width=0.5\textwidth]{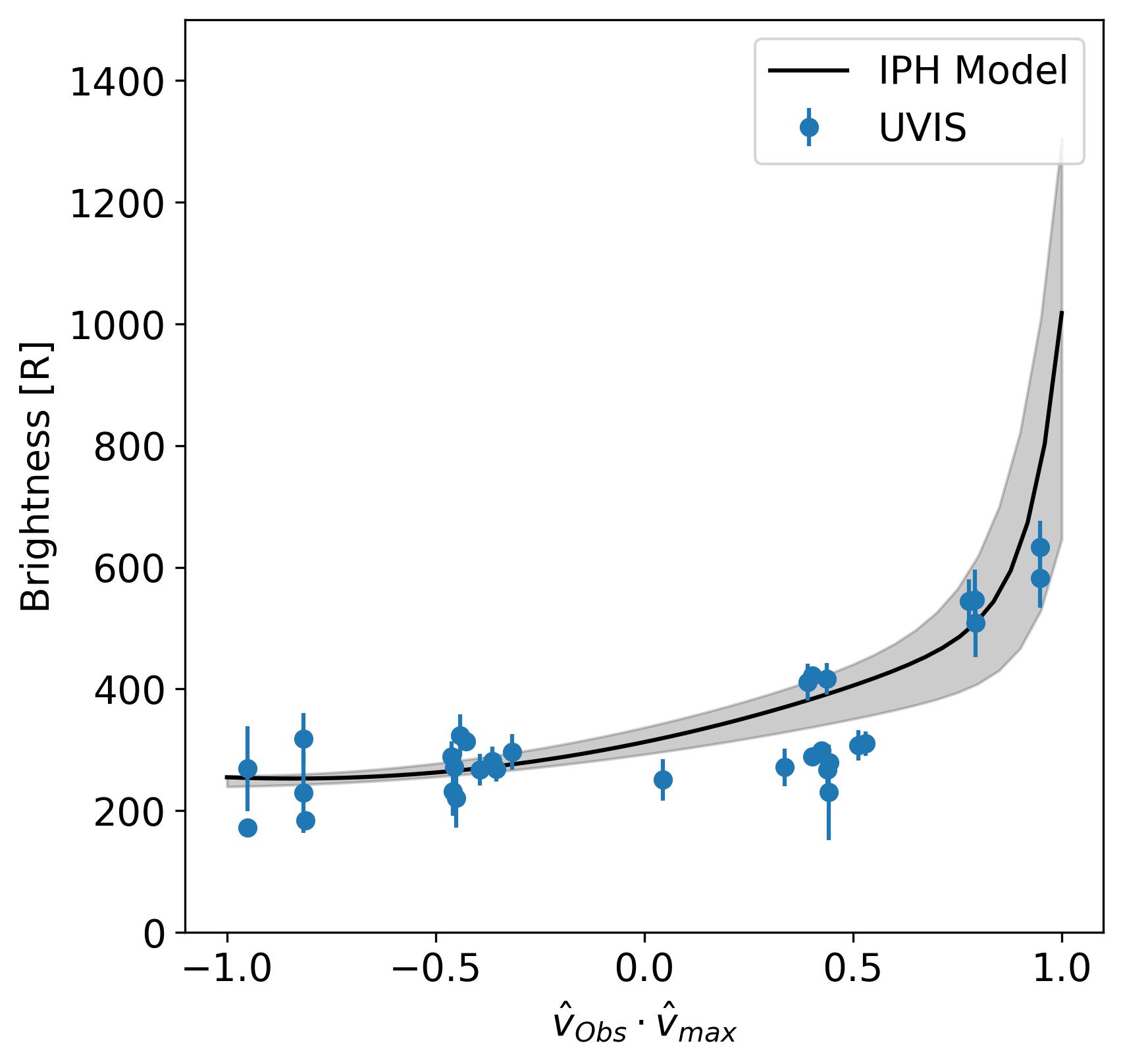}
    \caption{Comparison of IPH Lyman-$\alpha$ brightness from {Cassini/UVIS} observations throughout 2006 (blue) with the modelled brightness from the IPH background model (black, see Section \ref{sec: IPH model description} and Figure \ref{fig: IPH model 2006}). The $x$ axis is the cosine of the angle between the observation direction and direction of maximum brightness The black line gives the best fit, with the shaded region depicting minimum and maximum modelled values.}
    \label{fig: iph comparison plot}
\end{figure}

\noindent
In this section, we compare observations of the IPH Lyman-$\alpha$ background throughout 2006 with modelled IPH brightnesses (see Figures \ref{fig: IPH model 2006} and \ref{fig: iph comparison plot}). The set of observations is listed in Table \ref{tab: IPH observations}. The brightnesses of the IPH observations have been scaled to the solar flux on Jan 1st 2009, when the solar activity was at a minimum. This corrects for variation of the IPH that results from the Sun's rotation. The scaled brightnesses are also given in Table \ref{tab: IPH observations}.
The modelled IPH brightnesses (black) agree well with the observations from {Cassini/UVIS} (blue), with an $R^2=0.64$. The brightness increases from 250 R in the anti-sunward direction ($\cos\theta_{Max}=-1$) to 650 R near the sunward direction ($\cos\theta_{Max}=0.8$). 

A cluster of 8 points at $\cos\theta_{Max}=0.4$ with $B=250\pm40$~R is 100~R smaller than predicted by the model. However, there are several observations at similar angles with $B=450$~R, suggesting the IPH model is not overestimating the brightnesses at this angle.
 Overall, the IPH observations by Cassini are consistent with the predictions of the IPH model at a heliocentric distance of 9.1\,au. 

\subsection{Emission trends of Saturnian Lyman-$\alpha$ from multi-variate regression analysis}\label{sec: MVR results}

\begin{figure*}
    \centering
    \begin{tikzpicture}
    \node (Figure) at (0,0) {\includegraphics[width=0.8\textwidth]{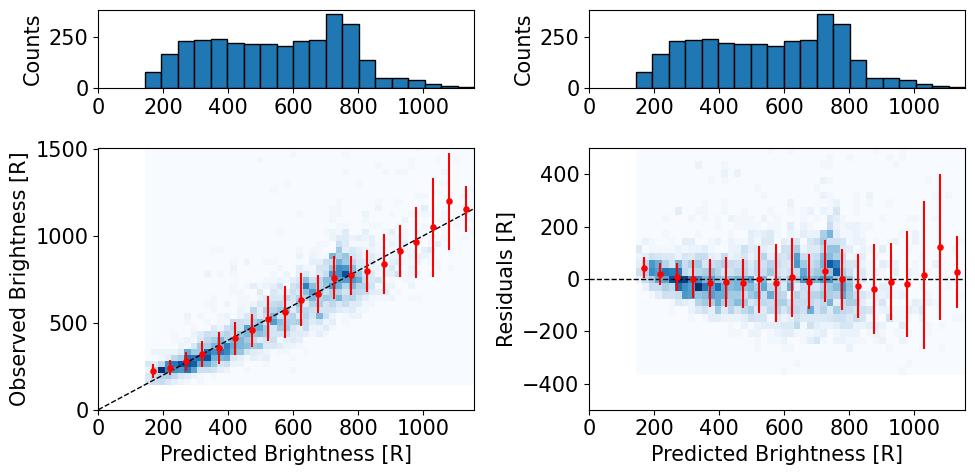}};
     \node [above left = 6 and 4.8 of Figure.base] (a) {(a)};
     \node [above right = 6 and 1.5 of Figure.base] (b) {(b)};
     \node [above left = 4 and 4.8 of Figure.base] (c) {(c)};
     \node [above right = 4 and 1.5 of Figure.base] (d) {(d)};
    \end{tikzpicture}
    
    \caption{(a,b) Testing data counts vs predicted brightness from the MVR. (c) Observed brightnesses and (d) residuals vs predicted brightnesses for the testing dataset (blue) with the averages and standard deviations  shown in red. }
    \label{fig: MVR prediction vs actual}
\end{figure*}
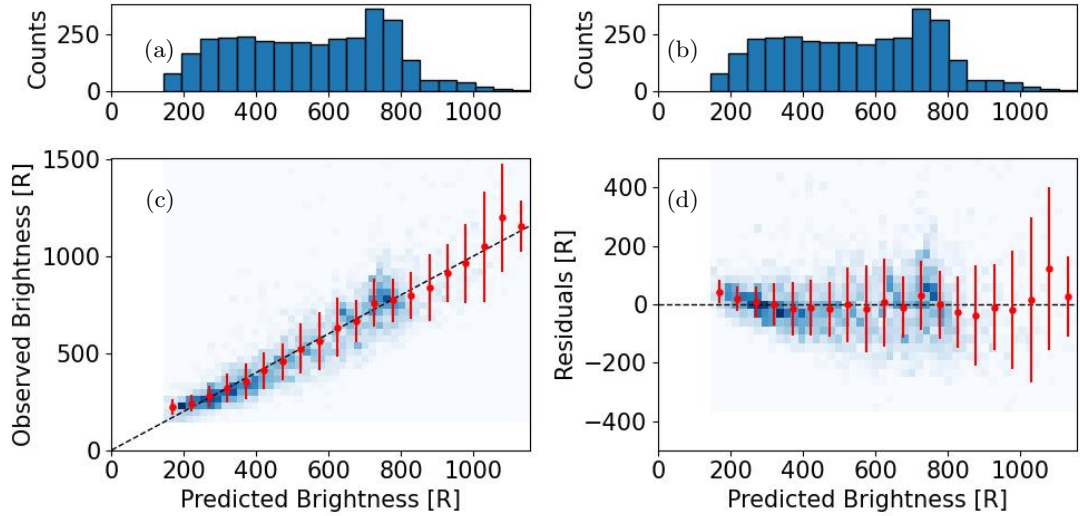

\noindent
 Figure \ref{fig: MVR prediction vs actual} shows the predicted brightnesses vs the actual observed brightnesses of Lyman-$\alpha$ for the multi-variate regression model outlined in Section \ref{sec: MVR outline} for the testing dataset (20\% of points). The coefficients of the fit to Eq. \ref{eq: MVR fit equation} are given in Table \ref{tab: MVR coefficients}. The fit has an $R^2= 0.791$ demonstrating that the observed brightness can be well parameterised by the latitude, incidence and emission angle once the observed brightness is scaled with the time-dependent solar flux. The 3-variable regression does not capture the brightnesses at small values, but above 250~R the model accurately predicts the observed brightnesses. 

 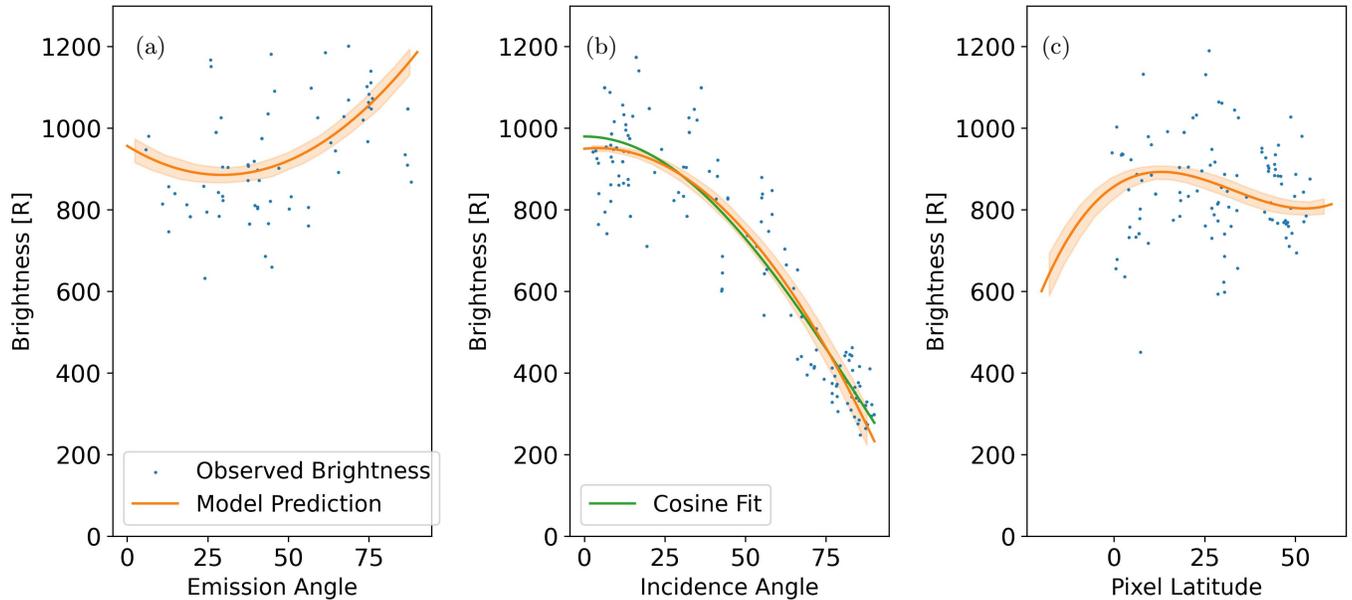
\begin{figure*}
    \centering
    \begin{tikzpicture}
        \node (Figure) at (0,0)  {\includegraphics[width=\textwidth]{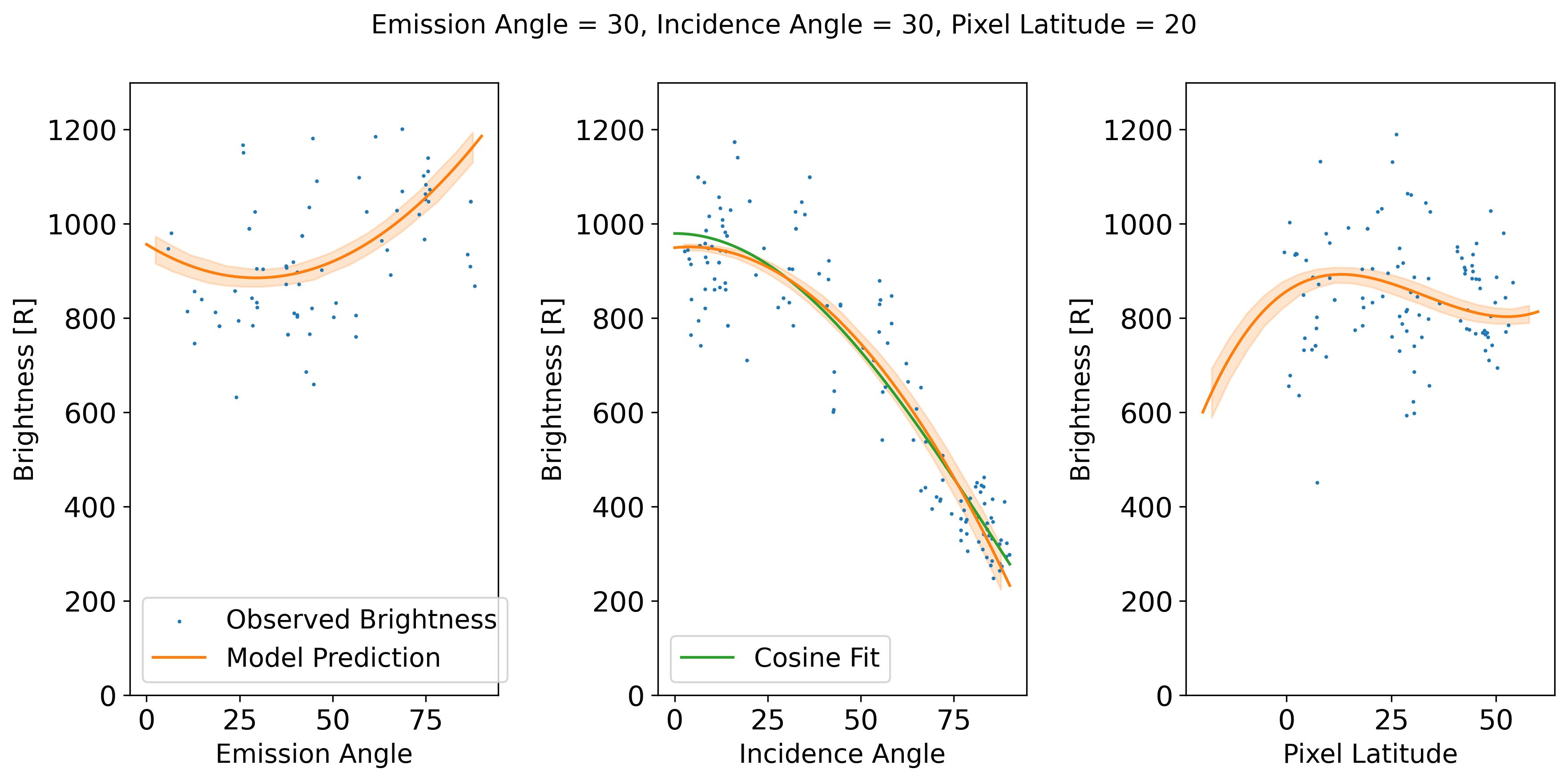}};
            \node [above left = 7.2 and 6.7 of Figure.base] (a) {(a)};
            \node [above left = 7.2 and 0.7 of Figure.base] (b) {(b)};
            \node [above right = 7.2 and 4.7 of Figure.base] (c) {(c)};
    \end{tikzpicture}
    \caption{Comparison of the observed brightnesses from 2014-2016 in the testing dataset (blue) to the multi-variate regression predictions (orange) as a function of (a) emission angle, (b) solar incidence angle and (c) latitude. The observed data are within the intervals $\theta_{em}=[27, 33]$, $\theta_{in}=[27, 33]$ and $\phi_{Lat}=[18,22]$ for the other two independent variables. In panel (b), the best fit of a cosine to the observed brightnesses is given (green).}
    \label{fig: MVR prediction vs vars.}
\end{figure*}

Figure \ref{fig: MVR prediction vs vars.} shows the dependence of the brightness on each of the independent variables around the constant values of $[\theta_{em}, \theta_{in}, \phi_{lat}] = [30, 30, 20]^\circ$ with the observed brightnesses shown for points within 10\% of the constant values. The Lyman-$\alpha$ brightness is most strongly dependent on the incidence angle of light arriving from the Sun, with brightness decreasing strongly from the subsolar point ($\theta_{in}=0$) to the terminator plane at $\theta_{in} = 90$. Conversely, the brightness increases with emission angle, with the brightest regions near the limb of Saturn's disk. The dependence on the incidence angle closely fits a $\cos\theta_{in}$ dependence.

\begin{figure*}
    \centering
    \begin{tikzpicture}
        \node (Figure) at (0,0) {\includegraphics[width=0.8\textwidth]{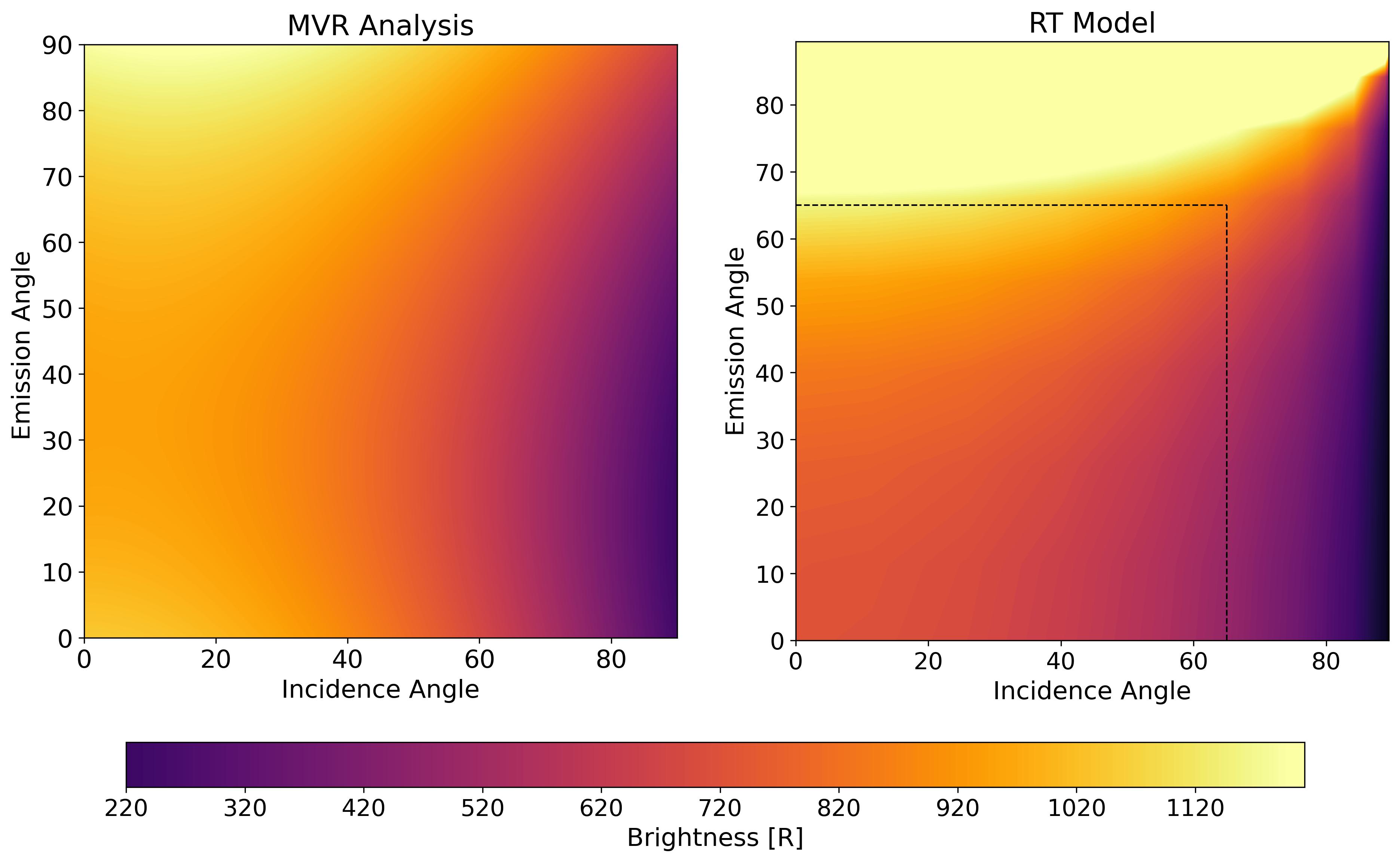}};
        \node [above left = 8 and 6.9 of Figure.base] (a) {(a)};
        \node [above right = 8 and 0.1 of Figure.base] (b) {(b)};
    \end{tikzpicture}
    
    \caption{(a) Predicted brightness as a function of solar incidence and emission angles from the MVR analysis at a latitude of 20$^\circ$N for the 2014-2016 dataset. (b) Modelled brightness as a function of solar incidence and emission angles from the radiative transfer model (see Section \ref{sec: rt model methods}) for an optical depth of $\tau_{H}=10000$ and $T_{Sat}=350$\,K. The dashed lines illustrate the region in which we compare observations to the radiative transfer model in Section \ref{sec: rt model results}.}
    \label{fig: MVR radiation field}
\end{figure*}

Figure \ref{fig: MVR radiation field}a shows the radiation field predicted by the trained model at a latitude of $20^\circ$N, with the combined dependence of the brightness on the incidence and emission angles. It again demonstrates substantial darkening towards the terminator and brightening of the disk near the limb. The dependences of the observed brightness on latitude, emission and incidence angles are consistent when using different multi-variate regression models, such as support vector regression using radial basis fields.

In Figure \ref{fig: MVR prediction vs actual}b, there is non-linearity to the residuals, with a slight parabolic shape between 200 and 1000R. Beyond 1000 R, the MVR model overestimates the observed brightnesses, but the statistics at these values are much lower ($<150$) compared to the intermediate brightnesses (200-1000\,R). The structure in the residuals is not replicated when considering the behaviour with respect to the independent variables (Figure \ref{fig: MVR prediction vs vars.}). Each variable shows no discernable behaviour in the residuals, suggesting higher order terms for each variable are not required. 

\begin{table*}[htbp]
\centering
\caption{Coefficients for the fit used in the multi-variate regression (see Eq. \ref{eq: MVR fit equation}), for the 2014-2016 observations. }
\label{tab: MVR coefficients}
\begin{tabular}{cccc}
\hline
Variable & Coefficient & Mean Value & Confidence Interval \\
\hline
Constant & $p_0$ & $6.51 \times 10^{2}$ & ($6.49 \times 10^{2}$, $6.53 \times 10^{2}$) \\
$\theta_{em}$ & $p_1$ & $3.73 \times 10^{0}$ & ($3.68 \times 10^{0}$, $3.77 \times 10^{0}$) \\
$\theta_{in}$ & $p_2$ & $-1.01 \times 10^{1}$ & ($-1.02 \times 10^{1}$, $-1.01 \times 10^{1}$) \\
$\phi_{lat}$ & $p_3$ & $-3.54 \times 10^{0}$ & ($-3.64 \times 10^{0}$, $-3.44 \times 10^{0}$) \\
$\theta_{em}^2$ & $p_4$ & $7.95 \times 10^{-2}$ & ($7.74 \times 10^{-2}$, $8.17 \times 10^{-2}$) \\
$\theta_{em} \cdot \theta_{in}$ & $p_5$ & $3.05 \times 10^{-2}$ & ($2.85 \times 10^{-2}$, $3.25 \times 10^{-2}$) \\
$\theta_{in}^2$ & $p_6$ & $-9.36 \times 10^{-2}$ & ($-9.54 \times 10^{-2}$, $-9.17 \times 10^{-2}$) \\
$\phi_{lat}^2$ & $p_7$ & $2.52 \times 10^{-2}$ & ($2.25 \times 10^{-2}$, $2.80 \times 10^{-2}$) \\
$\phi_{lat}^3$ & $p_8$ & $2.55 \times 10^{-3}$ & ($2.43 \times 10^{-3}$, $2.66 \times 10^{-3}$) \\
\hline
\end{tabular}
\end{table*}

\begin{figure}
    \centering
    \includegraphics[width=0.5\textwidth]{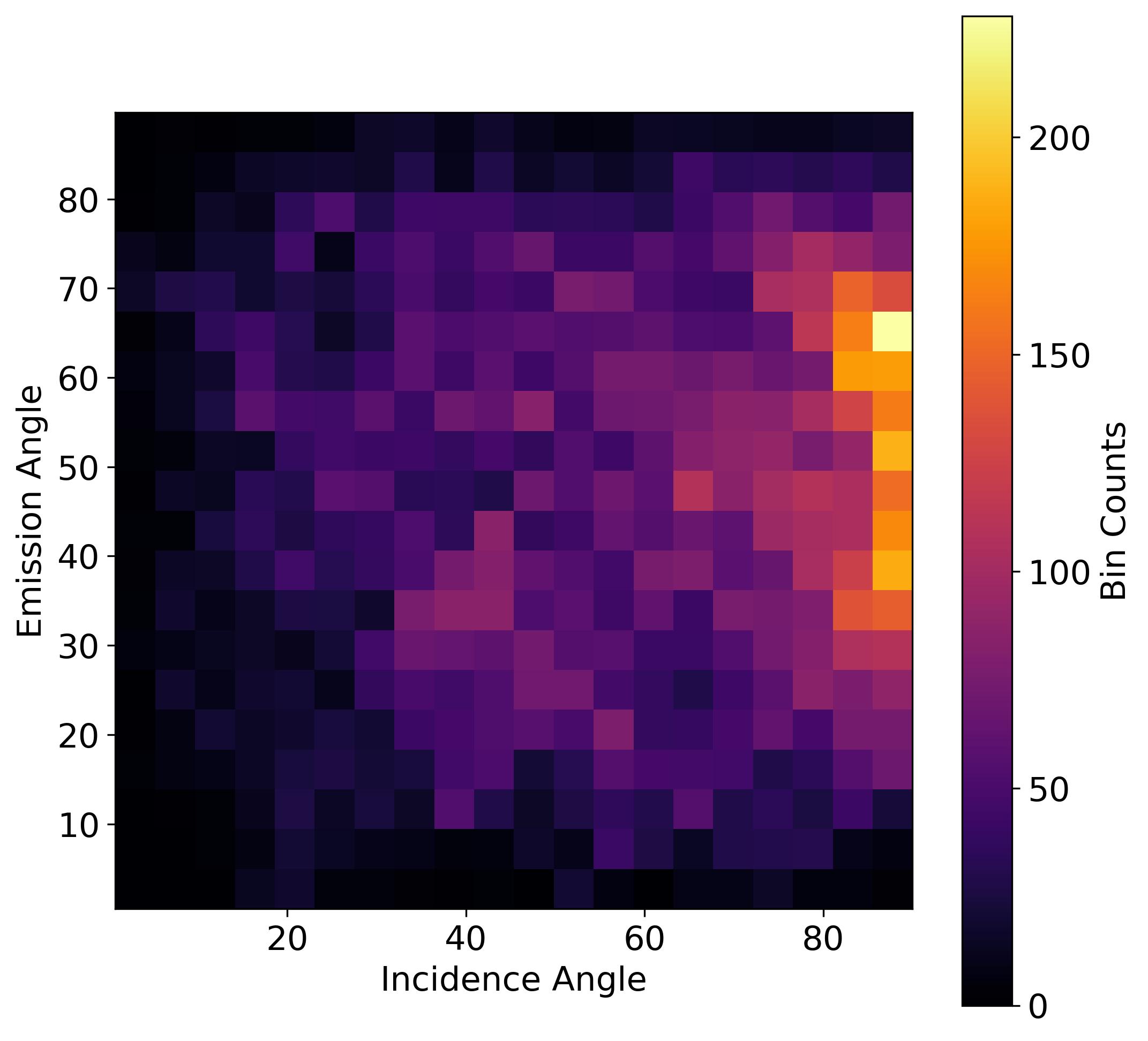}
    \caption{Histogram of observation geometries (solar incidence angles and emission angles) in the training dataset for the multi-variate regression from 2014-2016 (see Figures \ref{fig: MVR prediction vs vars.} and \ref{fig: MVR radiation field}). Regions with fewer observations are less well constrained in the MVR model.}

    \label{fig: MVR radiation field counts}
\end{figure}

Figure \ref{fig: MVR radiation field}b similarly shows the modelled brightness of Lyman-$\alpha$ from the radiative transfer model (Section \ref{sec: rt model methods}) at a temperature of 350\,K and an atomic Hydrogen optical depth of $\tau_H = 10,000$. The trained MVR model shows very similar behaviour to the brightness from the RT model, with a bright limb and dark terminator region, indicating that the observed brightnesses can be mostly explained by resonance scattering of Lyman-$\alpha$ by upper atmospheric atomic hydrogen. The MVR and RT model differ in two main areas: the bright limb of the RT model and the increase in brightness at low emission angles observed in the MVR model. The very bright limb observed in the RT model is a result of the plane-parallel assumption, which breaks down near the terminator and near the limb of the disk. With a plane parallel-model, the gas column the photons travel through is greatly overestimated, inflating the modelled brightness. The increase in brightness at low emission angles for the MVR may be a result of the limited data available at small emission angles (see Figure \ref{fig: MVR radiation field counts}), which do not strongly constrain the observations at low emission and incidence angles. 

\begin{figure*}
    \centering
    \begin{tikzpicture}
        \node (Figure) at (0,0)  {\includegraphics[width=\textwidth]{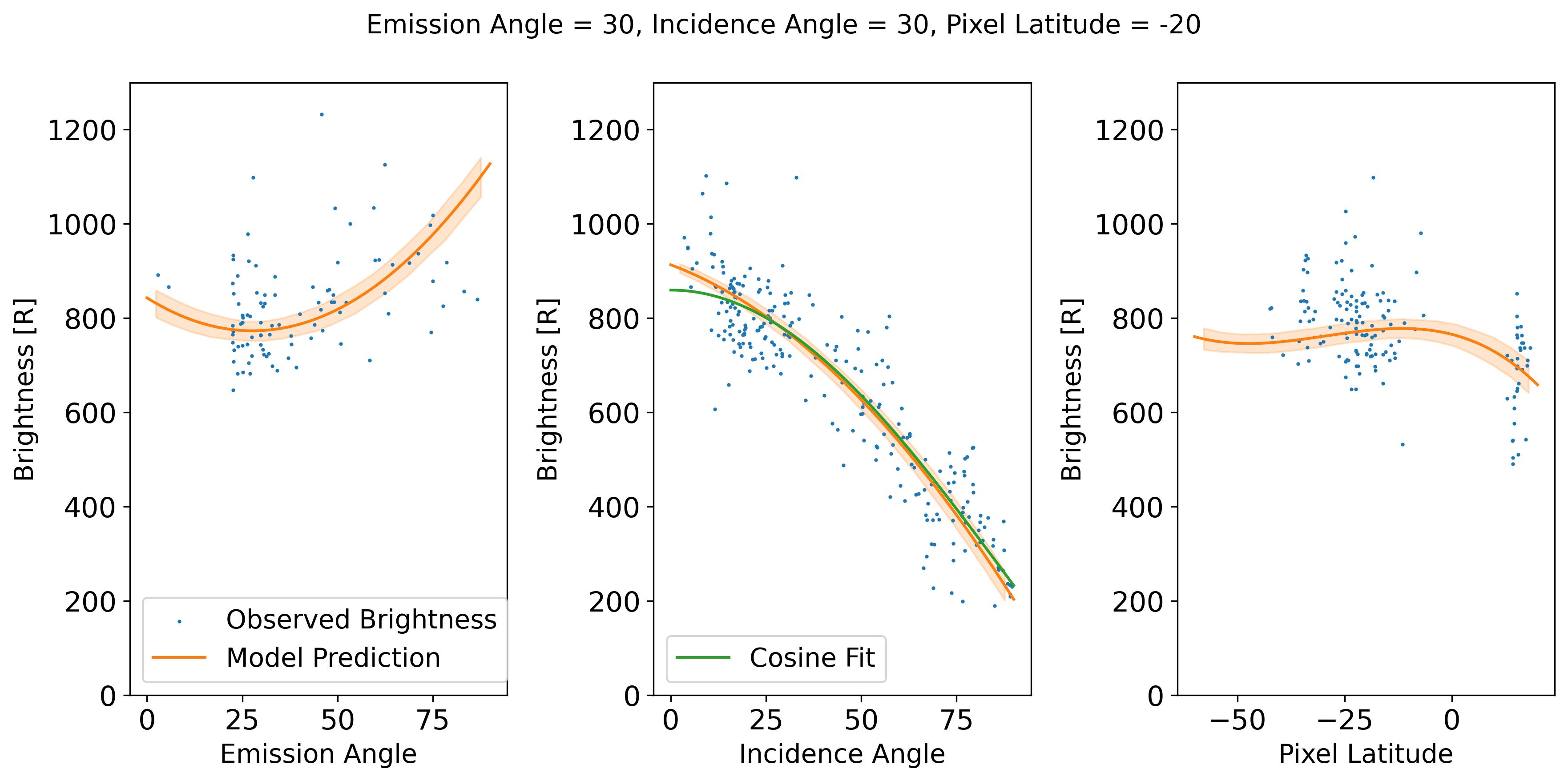}};
        \node [above left = 7 and 6.5 of Figure.base] (a) {(a)};
        \node [above left = 7 and 0.7 of Figure.base] (b) {(b)};
        \node [above right = 7 and 4.4 of Figure.base] (c) {(c)};
    \end{tikzpicture}
    
    \caption{Comparison of the observed brightnesses in the 2004-2006 testing dataset (blue) to the multi-variate regression predictions (orange) as a function of (a) emission angle, (b) solar incidence angle and (c) latitude. The observed data are within the intervals $\theta_{em}=[27, 33]$, $\theta_{in}=[27, 33]$ and $\phi_{Lat}=[-22,-18]$ for the other two independent variables. In panel (b), the best fit of a cosine to the observed brightnesses is given (green).}
    \label{fig: MVR prediction vs vars early.}
    
\end{figure*}
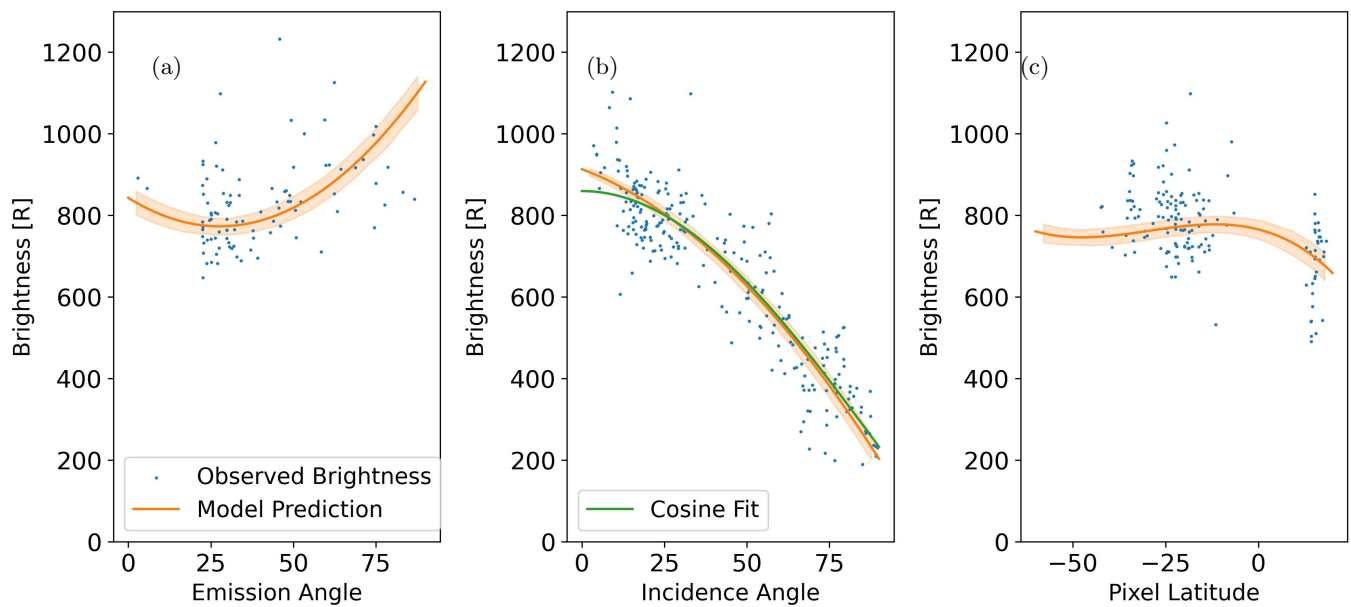

The latitudinal variation seen in Figure \ref{fig: MVR prediction vs vars.}c, has been disentangled from variation with the geometry of the observations without reliance on a radiative transfer model. The northern hemisphere bulge at low latitudes is a result of a meridional structure in the upper atmosphere. Using the same analysis for the 2004-2006 period, the latitudinal profile shows the reverse trend, with the bulge appearing near 20$^\circ$S (see Figure~\ref{fig: MVR prediction vs vars early.}). The brightness also decreases into the winter hemisphere and towards the summer pole. This shows a clear reversal with season in the latitudinal behaviour, and shows that the bulge is not a permanent feature of the northern hemisphere. Instead, it is a seasonally changing feature that appears in the summer or spring hemisphere.

The source of the increased brightness at 20$^\circ$ in the summer hemisphere could be driven by either enhanced H column in the region, or additional sources of emission within the bulge region. We split the dataset into latitude bins of 10$^\circ$ and applied the same multi-variate regression (see Figures \ref{fig: NH summer B vs inc em lat} and \ref{fig: SH summer B vs inc em lat}). At all latitudes, the same dependence of the brightness on solar incidence and emission angles was observed, with $R^2>0.7$ and $R^2>0.6$ throughout northern and southern summers respectively. Some latitude bins were limited by the phase space coverage of the data (Figures \ref{fig: MVR radiation field counts}, \ref{fig: SH summer counts vs inc em lat} and \ref{fig: NH summer counts vs inc em lat}), with small emission angles not probed in the winter hemisphere or close to the poles. The consistency of the relation between the incidence and emission angles with latitude implies that there is not a substantial additional internal emission source in the northern hemisphere bulge. Therefore, the variation in Lyman-$\alpha$ brightness can likely be attributed to variation of the H column above the methane homopause. The alternative would be to attribute the variation to a seasonal change in the hot atomic H population, although no source of hot atoms that would have the required behaviour has been identified to date.

\subsection{Seasonal Variation of Saturn's atmosphere}\label{sec: rt model results}

\subsubsection{A case study of northern hemisphere summer}\label{sec: NH summer}
\begin{figure*}
    \centering
    \begin{tikzpicture}
        \node (Figure) at (0,0)  {\includegraphics[width=\textwidth]{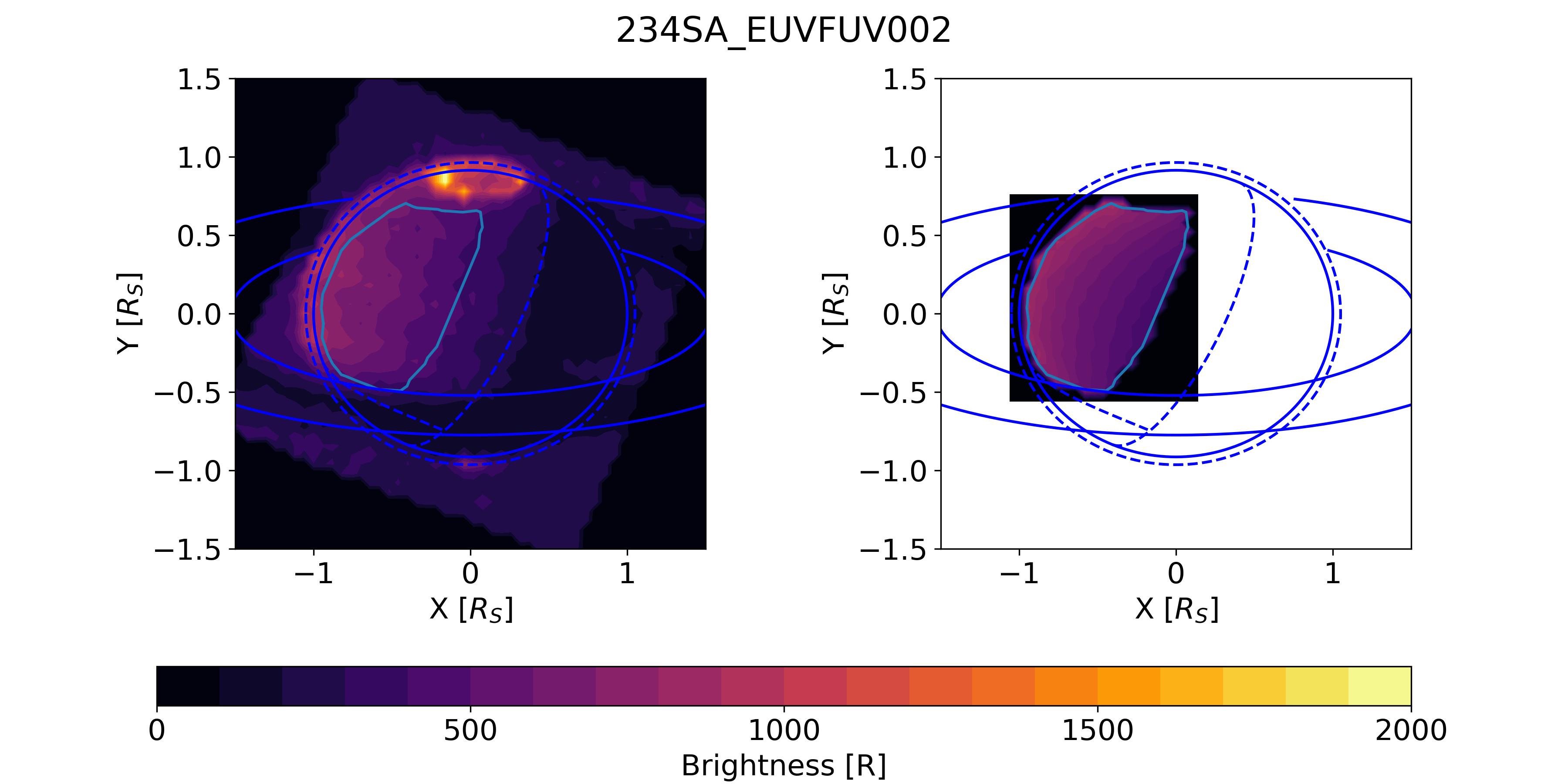}};
        \node [above left = 7 and 7 of Figure.base] (a) {(a)};
        \node [above right = 7 and 0.1 of Figure.base] (b) {(b)};
    \end{tikzpicture}
    
    \caption{(a) Lyman-$\alpha$ brightness in observation 234SA\_EUVFUV002, with the ring geometry, limb and terminator of Saturn highlighted in blue. The considered region for comparison to the radiative transfer model is bounded by the light blue line. (b) Modelled Lyman-$\alpha$ brightness using $\tau_H=10,000$ and the temperature dependence in Fig. \ref{fig: temp variation} in the RT model (see Section \ref{sec: rt model methods}). }
    \label{fig: case late xy}
\end{figure*}

\begin{figure*}
    \centering
     \begin{tikzpicture}
        \node (Figure) at (0,0)  {\includegraphics[width =0.95\textwidth]{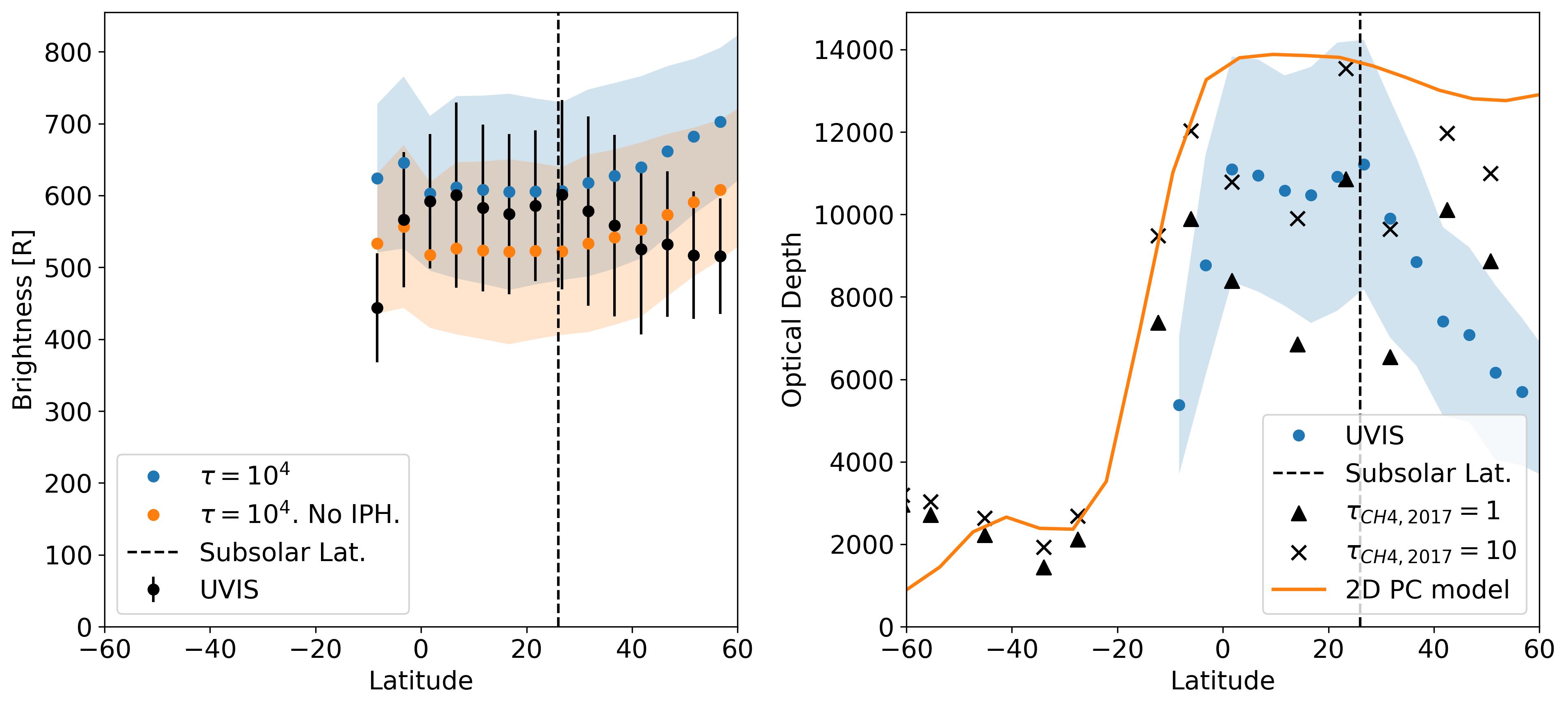}};
        \node [above left =  7.5 and 7.5 of Figure.base] (a) {(a)};
        \node [above right = 7.5 and 0.1 of Figure.base] (b) {(b)};
    \end{tikzpicture}
    \caption{(a) Observed Lyman-$\alpha$ brightness (black) for observation 234SA\_EUVFUV002 against latitude. The modelled brightness for the observation geometry using $\tau=10000$ and the temperature variation in Fig.\ref{fig: temp variation} is shown with (blue) and without (orange) the contribution of scattered IPH Lyman-$\alpha$. (b) Optical depth (blue) vs. latitude retrieved from comparison of {Cassini/UVIS} observations. The optical depths of the methane homopause (using $\tau_{CH4}=1$, triangles; and $\tau_{CH4}=10$, crosses) predicted by 1D photochemical models tuned to 2017 stellar occultations are shown. (orange) H optical depth from the 2D photochemical model on 26 Mar 2016 at subsolar latitude of 26.0$^\circ$ and $L_S=81.2^\circ$ (see Figure \ref{fig: 2d photochem model}).}
    \label{fig: case late}
\end{figure*}

\noindent
We consider first an observation of Saturn on 25th March 2016, when the subsolar latitude was 26.0$^\circ$, with incidence and emission angles from 0 to 65$^\circ$. We exclude latitudes poleward of 60$^\circ$, and in the ring shadow region which extends from  14$^\circ$S to the south pole. This includes 1617 pixels across the dayside region highlighted in Figure \ref{fig: case late xy}, accumulated over 4.5 hours. The observed brightness is constant between the latitudes of 0 and 20$^\circ$N at $596 \pm 15$ R, before decreasing to $518\pm12$\,R 26$^\circ$N to  60$^\circ$N. The observed brightness also drops sharply to 421\,R at $9.4^\circ$S towards the ring-shadowed region (see Figure~\ref{fig: case late}). 

The modelled brightness for a constant H Lyman-$\alpha$ line center optical depth (hereafter, the H optical depth) of 10,000 also shown in Figure~\ref{fig: case late} agrees well with the brightness between 0 and 25$^\circ$, but outside this range the modelled brightness increases in contrast to the observation. The increased brightness is a result of larger emission angles for the southern and higher latitudes. The scattered Lyman-$\alpha$ from the IPH background has a fairly constant contribution of $78.5\pm4.4$\,R across latitudes, which is 12\% of the brightness of the scattered solar flux.

Figure \ref{fig: case late}b shows the H optical depths retrieved from comparison of the observations with the RT model. The close agreement between the modelled and observed brightness between 0 and 25$^\circ$ is reflected by the constant optical depth of $10,800 \pm 300$\, R over this range. Northward of 25$^\circ$N the optical depth decreases continuously towards $\tau=5000$ at 60$^\circ$N, as the observed brightness declines relative to the expected values from the RT model based on constant H optical depth. This decrease in the H optical depth is not predicted by the 1D photochemical model, which predicts that the H optical depth is approximately constant at 10,000 up to 60$^\circ$N. However, the 2D photochemical model (orange) does show a slight decrease in the H optical depth from the peak at 10$^\circ$N, but only by 10\%, compared to the 50\% reduction retrieved from the UVIS data. Southward of the equator, the H optical depth sharply decreases between 5.6$^\circ$N and $9.4^\circ$S latitude towards the ring shadow, in close agreement with the photochemical model. 

\subsubsection{A case study of southern hemisphere summer}\label{sec: SH summer case}
\begin{figure}
    \centering
    \includegraphics[width=0.5\textwidth]{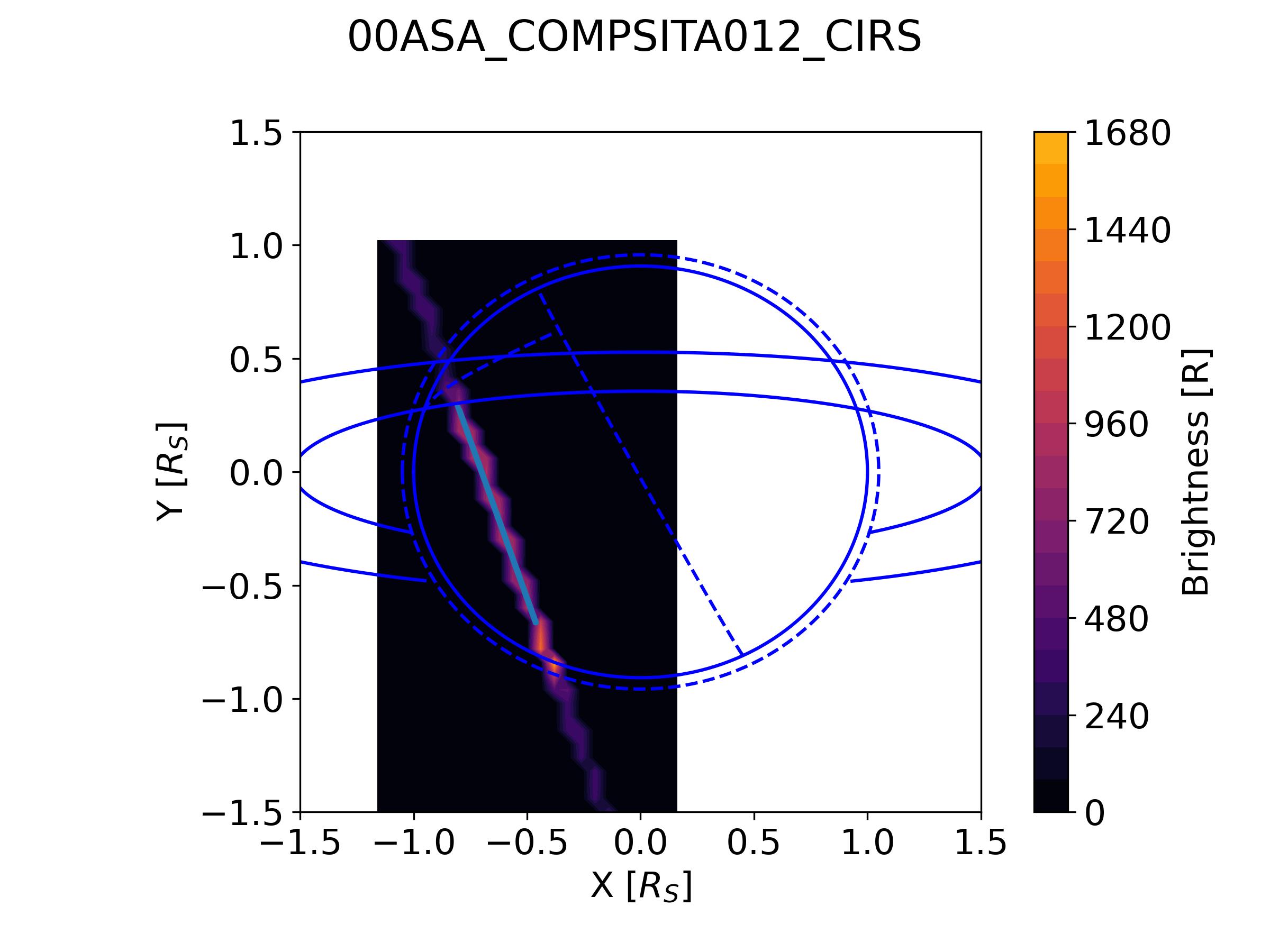}
    \caption{Observed Lyman-$\alpha$ brightness for the 00ASA\_COMPSITA012\_CIRS observation on 8 Nov 2004 with a subsolar latitude of -23.5$^\circ$. The region modelled by the radiative transfer code is highlighted by the light blue line, with Saturn, the rings and ring shadow regions superimposed in dark blue.}
    \label{fig: xy COMPSIT CIRS}
\end{figure}
\begin{figure*}
    \centering
    \begin{tikzpicture}
        \node (Figure) at (0,0) {\includegraphics[width = \textwidth]{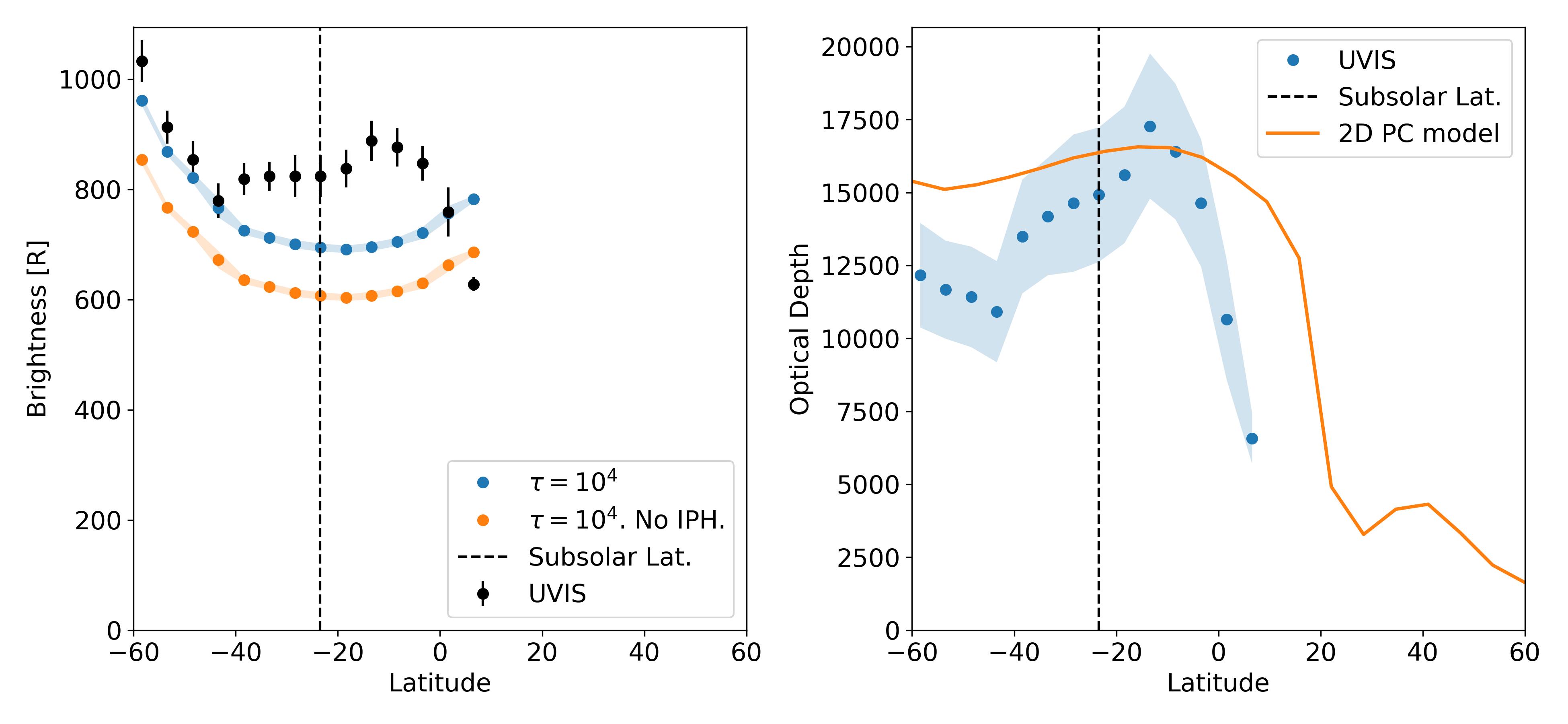}};
        \node [above left = 7.3 and 8.3 of Figure.base] (a) {(a)};
        \node [above = 7.3 of Figure.base] (b) {(b)};
    \end{tikzpicture}
    
    \caption{(a) Observed Lyman-$\alpha$ brightness (black) for observation 00ASA\_COMPSITA012\_CIRS against latitude. The modelled brightness for the observation geometry using $\tau=10000$ and the temperature variation in Fig.\ref{fig: temp variation} is shown with (blue) and without (orange) the contribution of scattered IPH Lyman-$\alpha$. (b) Optical depth (blue) vs. latitude retrieved from comparison of {Cassini/UVIS} observations with the radiative transfer model. (orange) H optical depth from the 2D photochemical model on 8 Nov 2004 at subsolar latitude of -23.5$^\circ$ and $L_S=302^\circ$.} 
    \label{fig: case early}
\end{figure*}

\noindent
Figures \ref{fig: xy COMPSIT CIRS} and \ref{fig: case early} show results for an observation from 8 Nov 2004, during southern hemisphere summer, with a subsolar latitude of 23.5$^\circ$S. The spatial coverage of Saturn's disk was much narrower for this case (see Figure \ref{fig: xy COMPSIT CIRS}), so the variation in the modelled brightnesses is much smaller at each latitude. 
The observed brightness decreases sharply from $1030\pm38$\,R at $60^\circ$S to $780\pm31$\,R at $40^\circ$S as a result of the decreasing emission angle. The lower brightness is closely matched by the modelled brightness based on constant H optical depth, and can be attributed wholly to changes in the viewing geometry. 

The  observations and constant H optical depth model, however, diverge between $40^\circ$S and the equator, with the modelled brightness continuing to decrease to $690\pm7$\,R while the observed brightness increases to $888\pm36$\,R in this region. The retrieved H optical depths (blue, Figure \ref{fig: case early}b) therefore increase substantially over this latitude range, peaking at $\tau_H=17,500\pm2500$ at $13^\circ$S. Equatorward of the peak, the H optical depths decrease sharply towards the ring shadow region, closely mirroring the behaviour observed in the northern hemisphere summer hemisphere (see Figure \ref{fig: case late}), and predicted by the 2D seasonal photochemical model (orange). The magnitude of the H optical depths in the summer hemisphere are also broadly consistent with the model, which predicts peak H optical depths of $\tau_H=16,600$ at $3^\circ$S.

\subsubsection{Implications for upper atmospheric hydrogen}\label{sec: seasonal overview results}

\begin{figure}
    \centering
    \includegraphics[width = 0.5\textwidth]{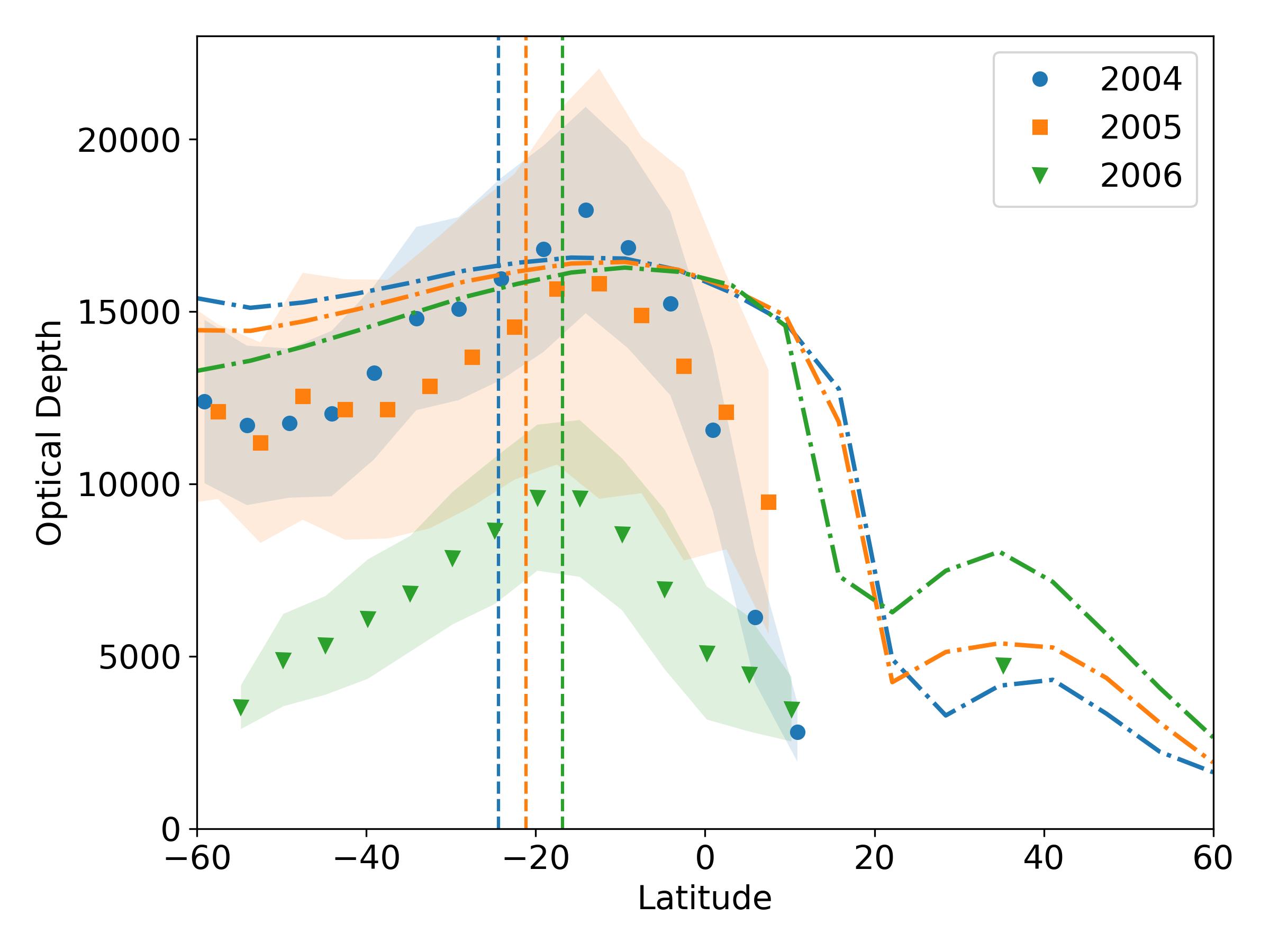}
    \caption{Yearly averages of the optical depths during the southern hemisphere summer (2004-2006) (colored circles), with shaded regions showing the uncertainty (see Section \ref{sec: rt model methods}). The subsolar latitude for each year is given by the dashed vertical lines. The dash-dotted lines show $\tau_H$ predictions from the 2D photochemical model in each year.}
    \label{fig: early mission tau}
\end{figure}

\begin{figure}
    \centering
    \includegraphics[width=0.5\textwidth]{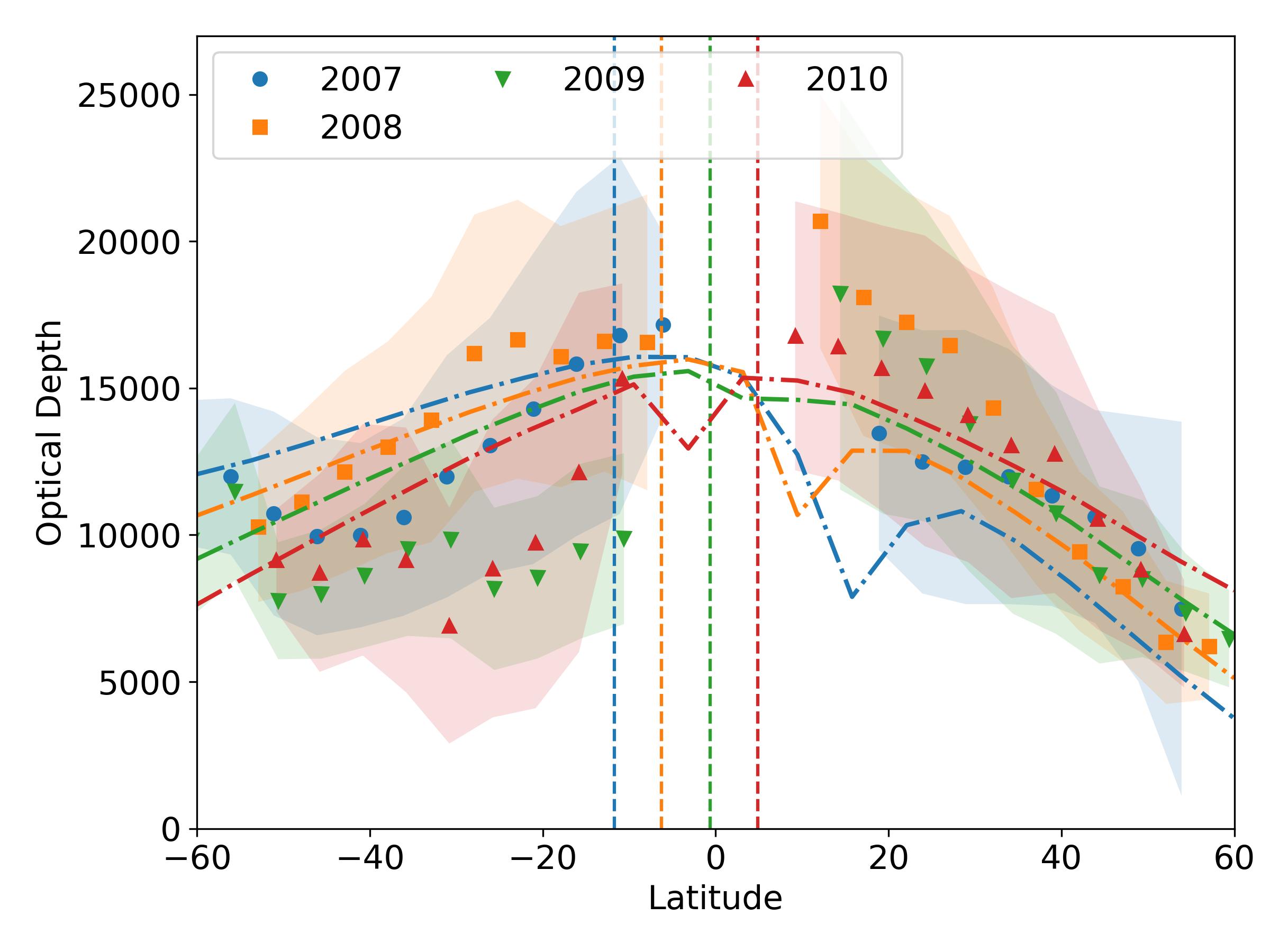}
    \caption{Yearly averages of the optical depths around equinox (2007-2010) (colored points), with shaded regions showing the uncertainty (see Section \ref{sec: rt model methods}). The subsolar latitude for each year in given by the dashed vertical lines.The dash-dotted lines show $\tau_H$ predictions from the 2D photochemical model in each year. }
    \label{fig: mid period tau}
\end{figure}

\begin{figure}
    \centering
    \includegraphics[width = 0.5\textwidth]{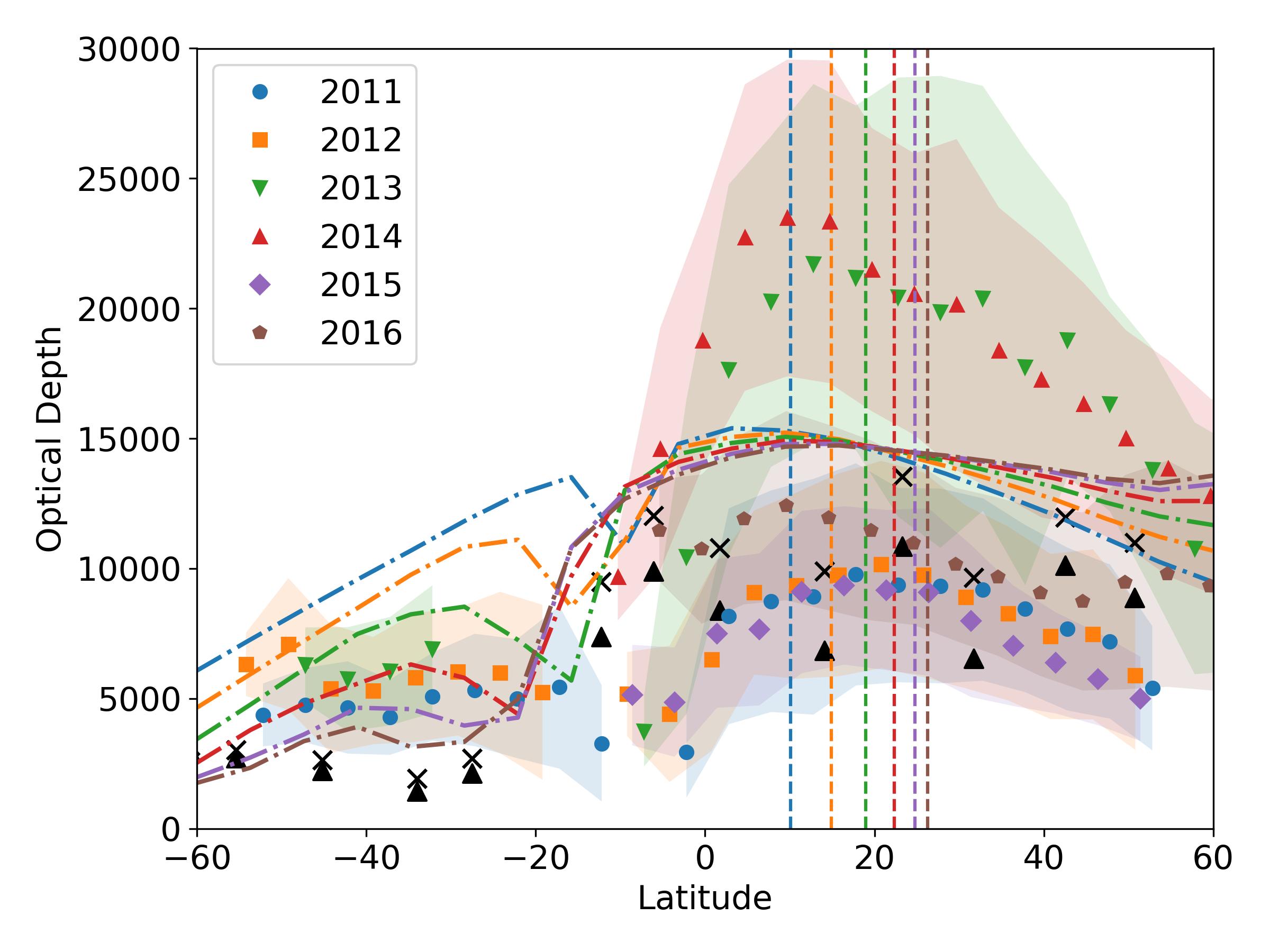}
    \caption{Yearly averages of the H optical depths during the northern hemisphere summer (2011-2016, colored points), with shaded regions showing the uncertainty (see Section \ref{sec: rt model methods}). The subsolar latitude for each year in given by the dashed vertical lines, with the corresponding color. (black points) H optical depths above the methane homopause (at $\tau_{CH4}=1$, triangles; and $\tau_{CH4}=1$, crosses) from 1D photochemical models based on 2017 UVIS occultations (see Section \ref{sec: photochemical model}). The dash-dotted lines show $\tau_H$ predictions from the 2D photochemical model in each year.}
    \label{fig: end of mission tau}
\end{figure}

\noindent
Having examined two observations from the NH and SH summers, we now apply the same analysis to all the dayside disk observations from {Cassini/UVIS} from 2004 through to 2016. We exclude 2017 as there is little coverage of Lyman-$\alpha$ emissions from the dayside Saturn disk. We have calculated H optical depths for each pixel (as outlined in Section \ref{sec: rt model methods}) and taken annual averages and standard deviations. We have separated the SH summer (2004-2006, Figure \ref{fig: early mission tau}), equinox period (2007-2010, Figure \ref{fig: mid period tau}) and NH summer periods (2011-2016, Figure\,\ref{fig: end of mission tau}).

Much like in the case study (Figure \ref{fig: case early}), the annual averages exhibit a substantial latitudinal variation during the southern summer (Figure~\ref{fig: early mission tau}), with H optical depths increasing from 60$^\circ$S towards the hydrogen bulge in the SH between 10$^\circ$S and 25$^\circ$S. Northward of the bulge, the H optical depth declines sharply with latitude. There are few available observations north of 20$^\circ$N throughout 2004-2006, as these latitudes are typically obscured by either the rings or the ring shadow region. The sharp decrease in H optical depth around the equator is also observed in the photochemical model. The decrease in the H optical depth at the equator is shifted southward relative to the photochemical model, and the gradient shifts southward from 2004 to 2006. This may result from the movement of the southern boundary of the ring shadow, which begins at 6.6-17.3$^\circ$N in 2004 and reaches 4.4-12.3$^\circ$N in 2006. The peak H optical depth, seen in the southern hemisphere bulge, decreases from 2004 to 2006. There is substantially more variation than expected based on the photochemical model \citep{Moses2005LatitudinalIRTF/TEXES} and is discussed in Section \ref{sec: discussion seasonal var}. 

During the equinox period (Figure \ref{fig: mid period tau}), the H optical depths increase continuously from the polar regions at $\pm60^\circ$ towards the equator. They agree well with the modelled depths from the 2D photochemical model (dash-dotted lines) in both magnitude and meridional structure. The equatorial latitudes have been excluded as they are obscured by the ring shadow and ring atmosphere during these years. The observed brightness decreases around the equator, but we cannot determine whether the H column is smaller due to the uncertainty in the illumination conditions. The H optical depths throughout this period show less variation than seen in the summer hemisphere (Figures \ref{fig: early mission tau} and \ref{fig: end of mission tau}), particularly in the northern hemisphere where the maximum H optical depth is $\tau_H=17,200\pm2300$. 

In the southern hemisphere during the equinox period, the peak H optical depths are similar, except in 2009 and 2010, which both exhibit significantly reduced H optical depths at mid-latitudes. The lower brightnesses compared to 2007 and 2008 are unexpected, much like the substantial variation during the southern hemisphere summer. The retrieved H optical depths are also 40\% smaller than the photochemical model predictions at 20$^\circ$S. We note that the highest optical depths are not consistently located in the more illuminated hemisphere. While this is the case in 2007 and 2010, the optical depths in 2008 are largest in the northern hemisphere despite a subsolar latitude of 6$^\circ$S. 

During the northern summer (Figure \ref{fig: end of mission tau}), the hydrogen bulge has shifted from the southern hemisphere (Figure \ref{fig: early mission tau}) to the northern hemisphere, along with the subsolar latitude. The H optical depth decreases continuously towards 60$^\circ$N, as seen in the case study (Fig. \ref{fig: case early}). From 2011-2013, the southern hemisphere is not completely shadowed by Saturn's rings. The retrieved H optical depths are roughly constant at $\tau_H=5000$ in the winter hemisphere and much smaller than the northern hemisphere peaks. This is also in good agreement with the 1D models based on 2017 stellar occultations (see Section \ref{sec: photochemical model}), which predicted H optical depths of $\tau_H=3000$ in the shadowed winter hemisphere. As seen during the earlier periods, there is substantial variability in the peak H optical depths over time. Large H optical depths of $\tau_H=22,100\pm6990$ and $\tau_H=24,000\pm6,200$  are observed in 2013 and 2014, while in 2012 and 2015 the peak H optical depths are $\tau_H=10,000\pm4100$. The H optical retrieved from the observations in 2013 and 2014 are twice as large as those predicted by the 1D photochemical models and 50\% larger than the predictions of the 2D model. 


\section{Discussion}\label{sec: discussion}

\subsection{Source of Lyman-$\alpha$ emissions}

\noindent
The multivariate regression analysis of the Lyman-$\alpha$ brightness observations strongly supports the dominance of resonance scattering as the source of non-auroral Lyman-$\alpha$ emissions, with scattered solar flux comprising the majority of the observed emissions. The strong dependence of the Lyman-$\alpha$ brightness on the solar incidence angle (see Figure \ref{fig: MVR prediction vs vars.}b) demonstrates the importance of solar photons in generating the observed emissions. Scattered photons from the IPH background would have a much weaker dependence on solar incidence angle, as they enter the atmosphere from all directions. Potential internal sources of Lyman-$\alpha$ photons would also not exhibit the variation with solar incidence angle, such as the electroglow proposed by \citep{Shemansky2009ThePlume} and others. While solar flux dominates, the contribution of scattered IPH Lyman-$\alpha$ photons is important for estimating of the H optical depth.

The radiation field retrieved through the multivariate regression analysis remains consistent across latitudes and mission periods. At all latitudes, the observed brightness can be fit with a quadratic function of incidence and emission angles, with brightness decreasing strongly towards larger incidence angles. For internal sources of emissions, we would not expect a strong dependence on solar incidence angle. The peak in the brightness and optical depth at latitudes of 10-20$^\circ$ in the summer hemisphere is therefore unlikely to be driven by an internal emission source. Our results, however, do not discount the possibility of hydrogen or water entering the upper atmosphere from the rings in the equatorial region, such as that detected by INMS \citep{Waite2018ChemicalRings, Yelle2018ThermalMeasurements, Serigano2020CompositionalINMS, Serigano2022CompositionalOrbits}, and leading to local increases in the H density, although {Cassini/UVIS} occultations and photochemical modelling during the Grand Finale do not suggest that a substantial fraction of the influx materials vaporise in the thermosphere \citep{Moses2023SaturnsMeasurements}. A smaller suprathermal (25,000 K) hydrogen population (0.1\% of the ambient H) could drive substantial emissions. The existence of such a population is currently hypothetical but can be better constrained with detailed modelling of emission profiles from Saturn's limb. Assuming no hot hydrogen population, the $\tau_H$ variation is similar to that predicted by the 2D photochemical model, which predicts a sharp decrease in the H column in the winter hemisphere (see Fig \ref{fig: 2d photochem model}). This strongly suggests that the low-latitude brightness peak is seasonal in nature and origin.

\subsection{The IPH model and {Cassini/UVIS} calibration}

\noindent

The brightnesses predicted by our IPH model (Section \ref{sec: IPH model description}) agree well in both viewing direction and magnitude with observations of the  IPH Lyman-$\alpha$ emissions by {Cassini/UVIS} (see Figure \ref{fig: iph comparison plot}). 

A proposed recalibration of {Cassini/UVIS} at Lyman-$\alpha$ by a factor 1.7 \citep{Ben-Jaffel2023TheAtmosphere, Pryor2024Modeling2017} would result in much larger observed brightnesses for the IPH (1255\,R at $\cos\theta_{Max}=0.94$) compared to those predicted by the IPH model (762\,R). This would require a significant increase in the hydrogen density at the termination shock, or galactic emissions beyond the scale of current estimates \citep{Gladstone2021NewBackground}. The angular dependence of the disparity could not be rectified by additional galactic Lyman-$\alpha$ flux. We note that \citet{Ben-Jaffel2023TheAtmosphere} obtained the calibration factor by comparing {Cassini/UVIS} observations with Saturn disk brightnesses observed by HST/STIS in 2017 from Earth orbit. There were, however, no simultaneous dayside observations of Saturn's disk obtained by {Cassini/UVIS} in 2017 and the comparison also depends on viewing geometry as indicated here. The required corrections for IPH absorption and geocorona also make it difficult to extract the line shape from HST/STIS observations.

Our IPH model, which is constrained by observations of New Horizons Alice and SWAN/SOHO \citep{Quemerais2013CrosscalibrationHeliosphere, Izmodenov2013DistributionLyman-}, predicts the brightness of the IPH Lyman-$\alpha$ to reach 250\,R in the anti-sunward direction during the 2006 observations and rise to 900~R in the sunward direction. \comments{SWAN/SOHO was cross-calibrated with HST/STIS in 2001, and later observations in March 2023 show the instruments agree within 15\%. \citep{Quemerais2013CrosscalibrationHeliosphere}.} The good agreement between the model and UVIS observations suggests that the calibrations of {Cassini/UVIS} and the other instruments are consistent, \comments{although we cannot discount the possibility of recalibration by up to 20\%.}

\subsection{Seasonal variability in thermospheric atomic hydrogen} \label{sec: discussion seasonal var}

\begin{figure}
    \centering
    \begin{tikzpicture}
        \node (left) at (0,0) {\includegraphics[width=0.45\textwidth]{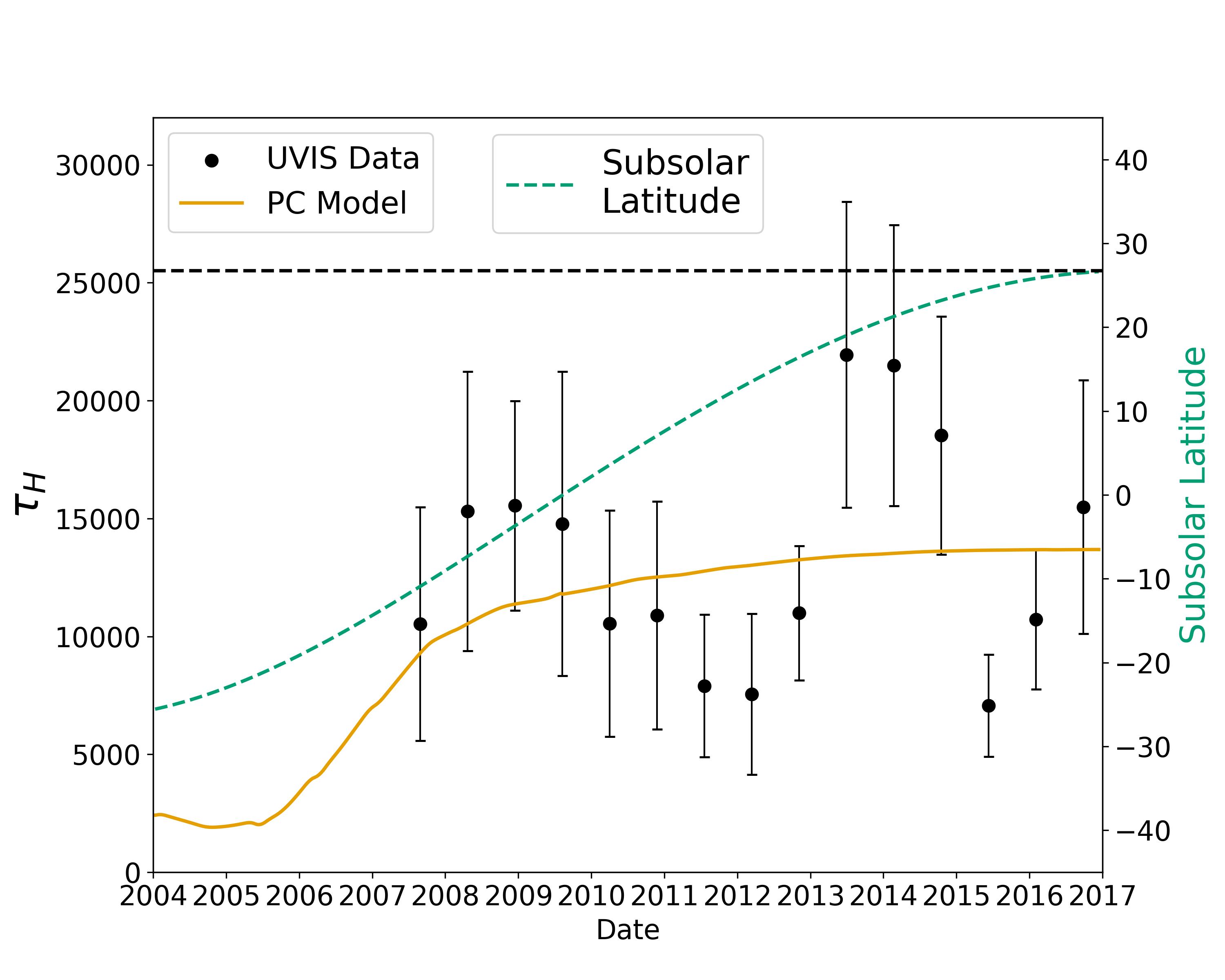}};
        \node [below = 2.8 of left.center] (right) {\includegraphics[width=0.45\textwidth]{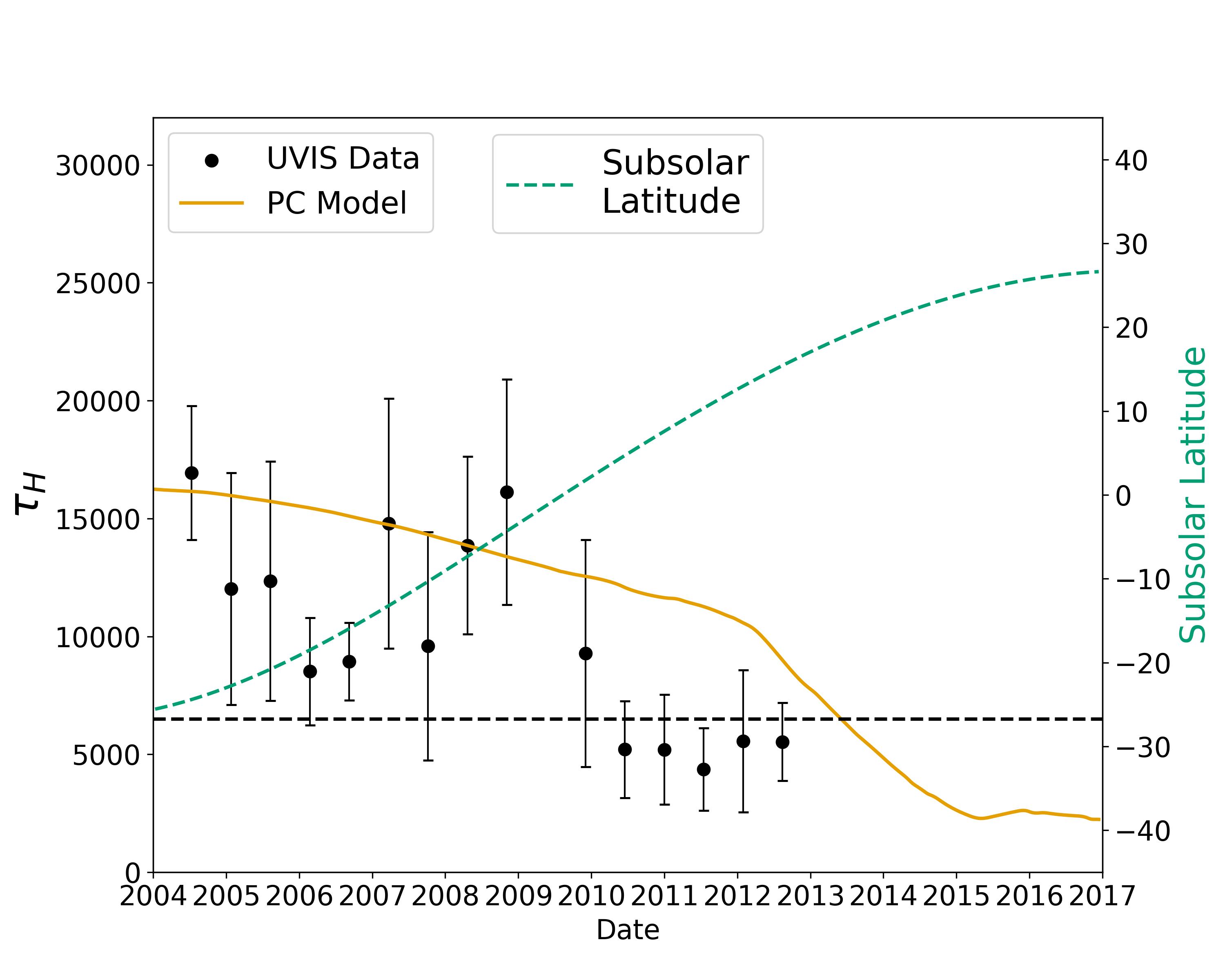}};
        \node [above left = 5 and 3.7 of left.base] (a) {(a)};
        \node [above left = 5 and 3.7 of right.base] (b) {(b)};
    \end{tikzpicture}
   
    \caption{(a) Optical depth at 26.7$^\circ$N$\pm5$ (northern solstice latitude, black dashed line), throughout the Cassini mission from the (black) UVIS observations and RT model comparison (see Section \ref{sec: rt model methods}) and (orange) the 2D photochemical model (see Section \ref{sec: photochemical model}). The subsolar latitude is shown in green. (b) The same as above but for $-26.7^\circ$ (southern solstice latitude ). \comments{The large error bars from the UVIS retrieval reflect shorter term variability in the Lyman-$\alpha$ brightness and atomic hydrogen atmosphere, within each temporal bin. The trends in the optical depths shown here demonstrate the long timescale variation of the atmosphere.}}
    \label{fig: seasonal tau variation}
\end{figure}

\noindent
Assuming that a population of hot atoms is not significant, the H optical depths retrieved from the UVIS observations with the RT model (see Section \ref{sec: rt model results}) agree well with the optical depths predicted by the photochemical model. The peak in brightness, identified in the northern hemisphere by \cite{Ben-Jaffel2023TheAtmosphere}, translates to an increase in the atomic hydrogen optical depth above the methane homopause. However, the bulge at 20$^\circ$ N is not a permanent feature of the northern hemisphere, but shifts with season and is observed in the southern hemisphere during the southern summer (see Figure \ref{fig: early mission tau}). A similar peak in the H optical depth is also predicted by the 2D photochemical model, which clearly shows a reversal in season (see Figure \ref{fig: 2d photochem model}). 

The poleward decrease in H optical depth from 20 to 60$^\circ$N in northern summer is also predicted by the 2D photochemical model (Figure \ref{fig: 2d photochem model}), although not to the same extent as estimated by the UVIS/RT model comparison. The poleward decrease disagrees with the 1D photochemical models, which were tuned to stellar occultations in 2017 \citep[black points, Figure \ref{fig: end of mission tau}; ][]{Brown2024AMesosphere}. These models do not show a significant decrease in the H optical depth towards higher latitudes, and also do not exhibit the same bulge around 20$^\circ$N.
The magnitude of the H optical depths is consistent with the results of the photochemical models (both 1D and 2D) with typical optical depths between $\tau_H=5,000-15,000$ (see Figures \ref{fig: 2d photochem model}, \ref{fig: case early}, and \ref{fig: end of mission tau}). The proposed revision of the {Cassini/UVIS} sensitivity by \cite{Ben-Jaffel2023TheAtmosphere}, increasing observed Lyman-$\alpha$ brightnesses by a factor 1.7, would result in more significant enhancements in the retrieved optical depths. While the latitudinal and seasonal trends would be unaffected, the non-linear relationship between optical depth and brightness (see Figure \ref{fig: function of gain}) would see an increase of 2.5 to $3\times$ in the H optical depth compared to the current estimates. As a result, the retrieved hydrogen column densities would be inconsistent with either the 1D or 2D photochemical models, albeit at the same time roughly following the seasonal behaviour predicted by the models.

The proposed calibration factor means that an additional emission source or an alternative source of H would be required. The strong dependence of brightness on solar incidence angle, identified in the multi-variate regression (see Section \ref{sec: MVR results}), excludes an emission source of the required magnitude other than resonance scattering of solar flux. While an additional source of hydrogen is not excluded by the current analysis, the magnitude of atomic H required would be beyond the estimates of what is possible from equatorial ring inflow \cite[up to $1.75\times$,][]{Ben-Jaffel2023TheAtmosphere}, and would be required planet-wide. Photochemical modelling suggests that ring rain entering the atmosphere as solid particles is unlikely to vaporize and contribute to the background gaseous atmosphere, and therefore should not significantly enhance the H column. A smaller population of hot H in the thermosphere is not excluded by our analysis but remains unidentified. 

The year-to-year variability of the optical depths retrieved, however, is highly unexpected, with variation by a factor 2 observed from one year to the next (see Figure \ref{fig: seasonal tau variation}). The 2D photochemical model (orange), which incorporates the variation of the subsolar latitude but neglects EUV solar-cycle variation, predicts minimal short-term variation of the H column above the homopause. Between solar minimum and maximum, a variation by about a factor of 2 is expected in the H column \citep{Ben-Jaffel2023TheAtmosphere}, but on the shorter timescales of the observed behaviour, the variability of the H optical depth is far greater than expected. 

To illustrate the variability, Figure \ref{fig: seasonal tau variation}a shows the optical depth at $26.7\pm5^\circ$N and Figure \ref{fig: seasonal tau variation}b for $26.7\pm5^\circ$S, the subsolar latitudes of the northern and southern hemisphere solstices. 
At $26.7\pm5^\circ$S, the 2D photochemical model predicts a continuous and monotonic reduction in the H optical depth above the methane homopause from 2004 to the end of mission in 2017, while the reverse is true for the northern hemisphere.  

The retrieved H optical depths from {Cassini/UVIS} show the same overall trend with time for $\phi_{lat}=26.7^\circ$S, initially large at values of $\tau_H=18,000$ and subsequently decreasing into the NH summer, when the ring shadows much of the southern hemisphere. However, the optical depth appears to oscillate around a mean value similar to the photochemical model. The trend is repeated in the northern hemisphere, with a sinusoidal time dependence around the photochemical model average. This indicates that there may be additional time-dependent behaviour that is modifying the upper atmospheric hydrogen content with a timescale of several years. The source of this time-variability is not incorporated into the photochemical model and requires further investigation.

The solar cycle is a natural candidate for this temporal dependence, which was at a minimum in 2009 and peaked in 2013 and 2014 (see Figure \ref{fig: solar LC flux}). The variation of the UV flux could modulate chemistry (e.g. methane photolysis) and temperature profiles in the thermosphere leading, which in turn would impact the upper atmospheric hydrogen.  While some years could be explained by this process (e.g. large optical depths in 2014, Figure \ref{fig: end of mission tau}), the variation of the thermospheric hydrogen often bucks trends with the solar cycle (e.g. 2015 vs 2016). As such, there is likely an alternative or additional driver of changes in the thermosphere.

One possible source of the variation in the inferred H optical depths could be changes in the thermospheric temperature. The 2D seasonal photochemical model does not include temperature changes with either season or latitude. Occultations and circulation models have shown that thermospheric temperatures vary both seasonally and meridionally \citep{Muller-Wodarg2019AtmosphericThermosphere, Brown2020AData, Koskinen2021empirical}. Increases in thermospheric temperatures lead to increases in the scale height and the brightness from resonant scattering of solar flux.
The factor of two variation in the inferred optical depth (see  Figure \ref{fig: seasonal tau variation}) could be reconciled by a 40\% temperature change (i.e., 350~K to 500~K), for the same density of H above the methane homopause. For example, a temperature variation of the order of 150 K between 2006 and 2017, with a peak temperature around 2010-2012, has been observed in the middle thermosphere around the equator in the UV occultation data \citep{Koskinen2021empirical}. In addition, the homopause depth likely changes over time, due to the effect of the changing seasons on dynamics in the middle atmosphere. For example, downwelling near the homopause would drive methane deeper in the atmosphere and increase optical depths. Disentangling the possible drivers of atomic hydrogen variability will require a seasonal model in which the temperature structure and homopause depth also change with time, combined with an analysis of the occultations, H$_2$ emission data, and limb scans from the duration of the Cassini orbital mission. We note that the analysis of solar occultations and limb scans can be used to retrieve vertical profiles of H and therefore further constrain additional degradation of the Cassini/UVIS instrument and calibration at Lyman-$\alpha$.

\section{Conclusion}

\noindent
We have examined the extensive dataset of Lyman-$\alpha$ emissions from the Saturn disk throughout the Cassini mission, as well as observations of Lyman-$\alpha$ emitted from the interplanetary  hydrogen background. We compared the IPH Lyman-$\alpha$ observations with the model of \cite{Quemerais2013CrosscalibrationHeliosphere}, which is calibrated with observations from SWAN, New Horizons Alice, and other platforms \citep{Quemerais2013CrosscalibrationHeliosphere, Izmodenov2013DistributionLyman-}. The {Cassini/UVIS} observations and the IPH model agree well with each other, suggesting that the calibration of {Cassini/UVIS} at Lyman-$\alpha$ is consistent with the instruments underlying the IPH model. 

We applied a multi-variate regression analysis to the observed brightness throughout the mission, to disentangle variations as a result of observation  geometry and those from meridional changes in the atmosphere. The emission brightness from Saturn's disk is dependent on four key variables: solar flux at the top of the atmosphere, latitude, and the emission and incidence angles applicable to the observations. The dependence of the observed Lyman-$\alpha$ brightness on the emission and incidence angles agrees closely with a model of resonant scattering of solar photons, exhibiting a strong decrease of the observed brightness with increasing solar incidence angle. We therefore exclude the possibility of a substantial internal source of Lyman-$\alpha$ emissions outside auroral regions.

We observe a bulge in the Lyman-$\alpha$ emissions in both the northern and southern hemispheres, during their respective summer seasons. Therefore, we conclude that the bulge previously reported by \citet{Ben-Jaffel2023TheAtmosphere} in the northern hemisphere during spring and summer is a seasonally modulated feature. Around the equinox, the Lyman-$\alpha$ brightness and effective H optical depths increase towards the equator (once the ring and ring shadow are excluded) in both hemispheres, giving the equatorial region an appearance of a bulge.

We compared observations of Lyman-$\alpha$ emissions taken during the Cassini orbital mission to a radiative transfer model, using the model to retrieve estimates of the H optical depth above the strongly-absorbing methane homopause level. The magnitude and latitudinal variation of the H column agree well with the predictions of the photochemical model (2D in particular), albeit with much more substantial temporal variability in the observations. A comparison of the temporal changes in the southern and northern hemispheres again show further evidence of seasonal change in the upper atmospheric hydrogen, with a peak in the effective H optical depths at a latitude of 20$^\circ$ in the summer hemisphere. The atomic H column decreases sharply into the winter hemisphere towards the ring shadow and more gradually toward higher latitudes ($60^\circ$). While the seasonal photochemical model of \cite{Moses2005LatitudinalIRTF/TEXES} predicts roughly constant peak optical depths of $\tau_H=15,000$ with season, our peak optical depth estimates varied by up to a factor 2 year-to-year. At latitudes of $\pm26.7^\circ$, the optical depth appears to show a sinusoidal variation in time around the mean depth of the photochemical model, especially in the northern hemisphere. The source of this variation requires further examination.

Observations of emissions from the limb and terminator of the planet have been avoided in this study due to the plane-parallel assumption used in the radiative transfer model. Further work is required to analyse the limb and near-limb observations, which would provide a constraint on the vertical profile of the atomic hydrogen in the thermosphere. This would provide much greater constraints on the possibility of a suprathermal hydrogen layer in the upper atmosphere than nadir scans that probe scattering by the bulk of the H column. Additionally, the Lyman and Werner bands of H$_2$ and He\,584{\AA} line emission should also be addressed to infer constraints on energy deposition as well as the \ce{H2} and He densities in the upper atmosphere.

\section*{Acknowledgments}
PS, TTK and JM acknowledge support by the NASA Cassini Data Analysis Program grant 80NSSC22K0306. ZB acknowledges support by the NASA/CDAP grant 80NSSC19K0902. PL acknowledges support from the project ATMOHAZE within the framework of the CNRS-UArizona collaboration initiative Searching for Habitable Worlds, in the Solar System and Beyond.
\appendix
\section{Radiative transfer model}\label{sec: RT model appendix}
The radiative transfer model is a plane-parallel model based on iterative doubling and adding of layers. Two layers of the atmosphere, $A$ and $B$ with $A$ above $B$ (see Figure \ref{fig: layer diagram}), are combined with
\begin{figure}
  \centering
  \begin{tikzpicture}
    \fill[blue!20] (0,0) rectangle (6,1);
    \fill[red!20] (0, -1) rectangle (6, 0);
    \node at (3, 0.5) {A};
    \draw[->] (3.3, 0) -- (3.3, -0.5) node[right] {$D_A$};
    \draw[->] (0.7, 0.5) -- +(40:0.5) node[right] {$S_A$};
    \draw[->] (0.7, 0.5) -- +(230:0.5) node[left] {$T_A$};
    \draw[-, dashed] (0.7, 0.5) -- +(-90:0.5) node {};
    \node at (3, -0.5) {B};
    \draw[->] (2.7, 0) --  (2.7,  0.5) node[left] {$U_A$};
    \draw[-latex] (0.7,0.2) arc
    [
        start angle=-90,
        end angle=40,
        x radius=0.3cm,
        y radius =0.3cm
    ] ;
    \draw[-latex] (0.7,0.2) arc
    [
        start angle=-90,
        end angle=-130,
        x radius=0.3cm,
        y radius =0.3cm
    ] ;
    \draw[->] (5.0, 0.5) -- +(-40:0.5) node[right] {$S^*_A$};
    \draw[->] (5.0, 0.5) -- +(110:0.5) node[left] {$T^*_A$};
    \draw[-, dashed] (5.0, 0.5) -- +(90:0.5) node {};
    \draw[-latex] (5.0,0.8) arc
    [
        start angle=90,
        end angle=-40,
        x radius=0.3cm,
        y radius =0.3cm
    ] ;
    \draw[->] (4.5, -0.5) -- +(40:0.5) node[right] {$S_B$};
    \draw[->] (4.5, -0.5) -- +(230:0.5) node[left] {$T_B$};
    \draw[-, dashed] (4.5, -0.5) -- +(-90:0.5) node {};
    \draw[-latex] (4.5,-0.8) arc
    [
        start angle=-90,
        end angle=40,
        x radius=0.3cm,
        y radius =0.3cm
    ] ;
    \draw[-latex] (4.5,-0.8) arc
    [
        start angle=-90,
        end angle=-130,
        x radius=0.3cm,
        y radius =0.3cm
    ] ;
  \end{tikzpicture}
  \caption{Diagram of layers A and B, with optical depths $\tau_A$ and $\tau_B$. Upward ($U_{A,B}$) and downward fluxes for each layer are shown, in addition to the scattering ($S_{A,B}$) and transmission functions ($T_{A,B}$). Starred variables are oriented opposite to the typical direction and are important for multiple scattering.}
  \label{fig: layer diagram}
\end{figure}
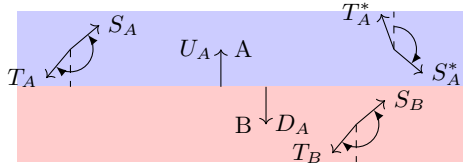
\begin{equation}\label{eq: upward flux}
U_A = S_B\,\exp\bigg[-\tau_A\frac{\psi(x_{in})+\bar\omega_a}{\mu_{in}}\bigg]+S_B\cdot D_A 
\end{equation}
\begin{equation}\label{eq: downward flux}
D_A = T_A + S^*_A\cdot U_A
\end{equation}
\begin{equation}\label{eq: adding scattering}
    S_{A+B} =S_{A}+\exp\bigg[-\tau_A\frac{\psi(x_{em}) +\bar\omega_a}{\mu_{em}}\bigg] U_A + T^*_A\cdot U_A
\end{equation}

\begin{multline}\label{eq: adding transmission}
    T_{A+B} = \exp\bigg[-\tau_A\frac{\psi(x_{in}) +\bar\omega_a}{\mu_{in}}\bigg]T_B +\\ D_A\exp\bigg[-\tau_B\frac{\psi(x_{em}) +\bar\omega_a}{\mu_{em}}\bigg] + T_B\cdot D_A
\end{multline}
where
\begin{itemize}
    \item $U, D = U, D(x_{em}, \theta_{em}, \phi_{em}; x_{in}, \theta_{in}, \phi_{in})$ are the upward and downward fluxes at the layer boundaries.
    \item $x_{em}, x_{in}$ are the emitted and incident frequency
    \item $\theta, \theta_0$ are the angles of emission and incidence, with $\mu=\cos\theta$. 
    \item $\tau$ is the optical depth of a layer
    \item $S$ and $T = S, T(\tau; x_{em}, \theta_{em}, \phi_{em}; x_{in}, \theta_{in}, \phi_{in})$ are the scattering and transmission functions for each layer with the starred values indicating they are defined opposite to the typical direction (scattering is typically upwards, transmission downwards).
    \item $\psi(x)$ is the Lyman-$\alpha$ lineshape, characterised by a Voigt profile at the upper atmospheric temperature
    \item $\bar{\omega}_a$ is the absorption albedo, which has been set to 0 throughout

\end{itemize}
We note that Equations \ref{eq: downward flux} and \ref{eq: adding scattering} differ slightly from Equations 8 and 9 in \cite{Yelle1989ResonanceTechniques}, as we more explicitly define the directions of the scattering function. The final term in Eq. 10 of \cite{Yelle1989ResonanceTechniques} is a typo and is corrected in Equation \ref{eq: adding transmission}.

The dot product is defined as
\begin{multline}\label{eq: dot product}
S_A\cdot U_A = \frac{1}{4\pi\mu}\int\limits_{0}^{2\pi}d\phi'\int\limits_{0}^{1}d\mu'\int\limits_{-\infty}^{\infty}dx'\bigg\{ \\S_A(\tau_A; \bm{x}_{em};\bm{x}')U_A(\tau_A; \bm{x}';\bm{x}_{in}).\bigg\}
\end{multline}
where the vectors $\bm{x}=(x, \mu, \phi)$ for each set.

The brightness of scattered solar flux is given by


\begin{multline}\label{eq: RT solar brightness}
    4\pi I_{Sol}(\mu_{em}, \mu_{in}) [R] = \Delta\nu_{D, Sat} \times10^{-6} \times\\ \int_{-\infty}^{\infty} dx_{in}\int_{-\infty}^{\infty}dx_{em} S(\tau;\bm{x}_{em};\bm{x}_{in})  F_{Sol}(x_{in})
\end{multline}
where $F_{Sol}$ is the solar Lyman-$\alpha$ flux incident on the atmosphere, with an incidence angle of $\theta_{in}$.
The Doppler width in Saturn's thermosphere is
\begin{equation} \label{eq: Doppler width}
    \Delta \nu_{D,Sat} = \frac{\nu_0v_{th}}{c}=\frac{\nu_0}{c}\sqrt{\frac{2k_BT_{Sat}}{m}}.
\end{equation}. 

\section{Thin Layer approximations}\label{sec: thin layer approx}
 From \cite{Yelle1988AAtmospheres}, we have the equation for the scattering function, $S$:
\begin{multline}
 \Big[\frac{\psi(x_{em})}{\mu_{em}} +\frac{\psi(x_{in})}{\mu_{in}}\Big]S(\tau; \bm{x}_{em}; \bm{x}_{in})+ \frac{\partial S(\tau; \bm{x}_{em}; \bm{x}_{in})}{\partial \tau} = \\ 
 R(\tau; \bm{x}_{em}; \bm{x}_{in}) + \frac{1}{4\pi}\int\limits_0^{2\pi}\mathrm{d}\phi'\int\limits_0^1\frac{\mathrm{d}\mu'}{\mu'}\int\limits_{-\infty}^{\infty} \mathrm{d}x'\bigg\{\\ R(\tau; \bm{x}_{em}; x', \mu',\phi')S(\tau; x', \mu', \phi'; \bm{x}_{in})\bigg\}\\
 + \frac{1}{4\pi}\int\limits_0^{2\pi}\mathrm{d}\phi'\int\limits_0^1\frac{\mathrm{d}\mu'}{\mu'}\int\limits_{-\infty}^{\infty}\mathrm{d}x'\bigg\{ \\S(\tau; \bm{x}_{em}; x', \mu',\phi')R(\tau; x', \mu', \phi'; \bm{x}_{in})\bigg\}\\
 + \frac{1}{16\pi^2}\int\limits_0^{2\pi}\mathrm{d}\phi'\int\limits_0^1\frac{\mathrm{d}\mu'}{\mu'}\int\limits_{-\infty}^{\infty}\mathrm{d}x' \int\limits_0^{2\pi}\mathrm{d}\phi''\int\limits_0^1\frac{\mathrm{d}\mu''}{\mu''}\int\limits_{-\infty}^{\infty}\mathrm{d}x''\bigg\{ \\
 S(\tau; \bm{x}_{em}; x', \mu',\phi')\\
 \times R(\tau; x', -\mu', \phi'; x'', \mu'', \phi'')S(\tau; x'', \mu'',\phi'';  \bm{x}_{in})\bigg\}
\end{multline}
 with 
 \begin{outline}
    \1 $\bm{x}_{in}  = (x_{in}, \mu_{in}, \phi_{in})$ and $\bm{x}_{em}$ is the corresponding vector for the emitted flux
    \1 $x_{in, em}$ are the incident and emitted frequencies
     \1 $\mu_{in, em} = \cos\theta_{in, em}$ with incidence and emission angles $\theta_{in, em}$
     \1  $\phi_{in, em}$ are the azimuthal incidence and emission angles
     \1 $\tau$ - optical depth of the layer at the line centre
     \1 $\psi(x)$ - line shape as a function of frequency
     \1 $S$ is the scattering function of the incident flux from $(x_{in}, \mu_{in}, \phi_{in})$ to $(x_{em}, \mu_{em}, \phi_{em})$ by the layer of thickness $\tau$
     \1 $R$ is the angular dependent partial frequency redistribution function $R_{II}$ from \cite{Hummer1962Non-CoherentBroadening}.
 \end{outline}
 Neglect the multiple scattering terms as we consider a thin layer and set
 \begin{equation}
     \eta = \frac{\psi(x_{em})}{\mu_{em}} +\frac{\psi(x_{in})}{\mu_{in}}
 \end{equation}
 An integration factor of $\exp(\eta\tau)$ gives
 \begin{multline}
     \frac{\partial \big(S(\tau, x_{em}, \mu_{em}, \phi_{em}, x_{in}, \mu_{in}, \phi_{in})\exp(\eta\tau)\big)}{\partial \tau} =\\ R(\tau, x_{em}, \mu_{em}, x_{in}, \mu_{in})\exp(\eta\tau)
 \end{multline}

Integrating from zero to optical depth $\tau$ and simplifying we get:
\begin{equation}
S(\tau; \bm{x}_{em}; \bm{x}_{in}) = \frac{R(\tau; \bm{x}_{em}; \bm{x}_{in})}{\eta}[1-\exp(-\eta\tau)]
\end{equation}
This is then expanded using a MacLaurin series and using $a = \eta \tau$ to get 
\begin{equation}
S(\tau; \bm{x}_{em}; \bm{x}_{in}) = \frac{R(\tau; \bm{x}_{em}; \bm{x}_{in})\tau}{a} \sum\limits_{n=1}^\infty\frac{(-a)^n}{n!}
\end{equation}
Expanding this to 8th order gives the thin layer approximation:

\begin{multline}
S(\tau; \bm{x}_{em}; \bm{x}_{in}) =R(\tau; \bm{x}_{em}; \bm{x}_{in})\tau(1-a(1- \\ \frac{a}{2}(1-\frac{a}{3}(1-\frac{a}{4}(1-\frac{a}{5}(1-\frac{a}{6}(1-\frac{a}{7}(1-\frac{a}{8}))))))))
\end{multline}

For the thin layer transmission function, $T$, the equation of transfer is:
\begin{multline}
    \frac{\psi(x_{in})}{\mu_{in}}T(\tau; \bm{x}_{em}; \bm{x}_{in}) + \frac{\partial T(\tau; \bm{x}_{em}; \bm{x}_{in})}{\partial \tau} = \\\exp(-\tau\frac{\psi(x_{em})}{\mu})R(\bm{x}_{em}; \bm{x}_{in})+...
\end{multline}
  
Setting $\alpha=\frac{\psi(x_{in})}{\mu_{in}}$ and $\beta=\frac{\psi(x_{em})}{\mu_{em}}$ and using an integrating factor of $\exp(\alpha\tau)$ gives
\begin{equation}
    T = R \frac{1}{\beta - \alpha}[\exp(-\alpha\tau)-\exp(-\beta\tau)] \quad \text{if  } \alpha \neq \beta
\end{equation}
To get the thin layer approximation expand both of the exponentials and use $a=\alpha \tau$ and $b=\beta \tau$
\begin{align}
    T = R\tau (b-a)^{-1}[&1 - a +\frac{a^2}{2!} - \frac{a^3}{3!}+\frac{a^4}{4!} +... \\
    - & 1 + b -\frac{b^2}{2!} + \frac{b^3}{3!} -\frac{b^4}{4!} +...]
\end{align}

Reorganising these terms gives the expression
\begin{multline}
    T = R\tau\Big[ c_1 - \frac{1}{2}\Big(c_2 -\frac{1}{3}\Big(c_3 - \frac{1}{4}\Big(c_4 - \frac{1}{5}\Big(c_5-\\\frac{1}{6}\Big(c_6-\frac{1}{7}\Big(c_7 - \frac{c_8}{8}\Big)\Big)\Big)\Big)\Big)\Big)\Big] 
\end{multline}
    
where
\begin{align}
    c_1 &= 1\\
    c_2 &= a + b\\
    c_3 &= a^2 + ab + b^2\\ 
    c_4 &= (a^2 + b^2)\times c_2 \\
    c_5 &= a^4 + a^3b + b^2a^2 +ab^3 + b^4 \\
    c_6 &= (a^3 + b^3)\times c_3 \\
    c_7 &= a^6 + a^5b + a^4b^2 + a^3b^3 + b^2a^4 +ab^5 + b^6 \\
    c_8 &= (a^4 + b^4)\times c_4 \\
\end{align}
If $\alpha=\beta$, then the differential equation simplifies to
\begin{equation}
    T = R\tau\exp(-\alpha\tau) = R\tau\exp(-a)
\end{equation}
Expanding this to 8th order gives 
\begin{multline}
    T = R\tau\Big[1-a\Big(1-\frac{a}{2}\Big(1-\frac{a}{3}\Big(1-\frac{a}{4}\Big(1-\frac{a}{5}\Big(1-\\ \frac{a}{6}\Big(1-\frac{a}{7}\Big(1-\frac{a}{8}\Big)\Big)\Big)\Big)\Big)\Big)\Big)\Big]
\end{multline}

\section{Modelling IPH brightnesses }\label{sec: iph model brightness}
Photons from the interplanetary hydrogen background enter the atmosphere from all directions. Consequently, we integrate over all incidence angles when computing the brightness of IPH scattered by Saturn's atmosphere, rather than the delta function in incidence angle used for the solar case. As outlined in Section \ref{sec: IPH model description}, the IPH brightness is fit with respect to $\cos\theta_{max}$, the angle to the direction of maximum brightness. At 9-10 au, the direction of maximum IPH brightness is closely aligned with the sun direction. 

\begin{figure*}
    \centering
    \includegraphics[width =\textwidth]{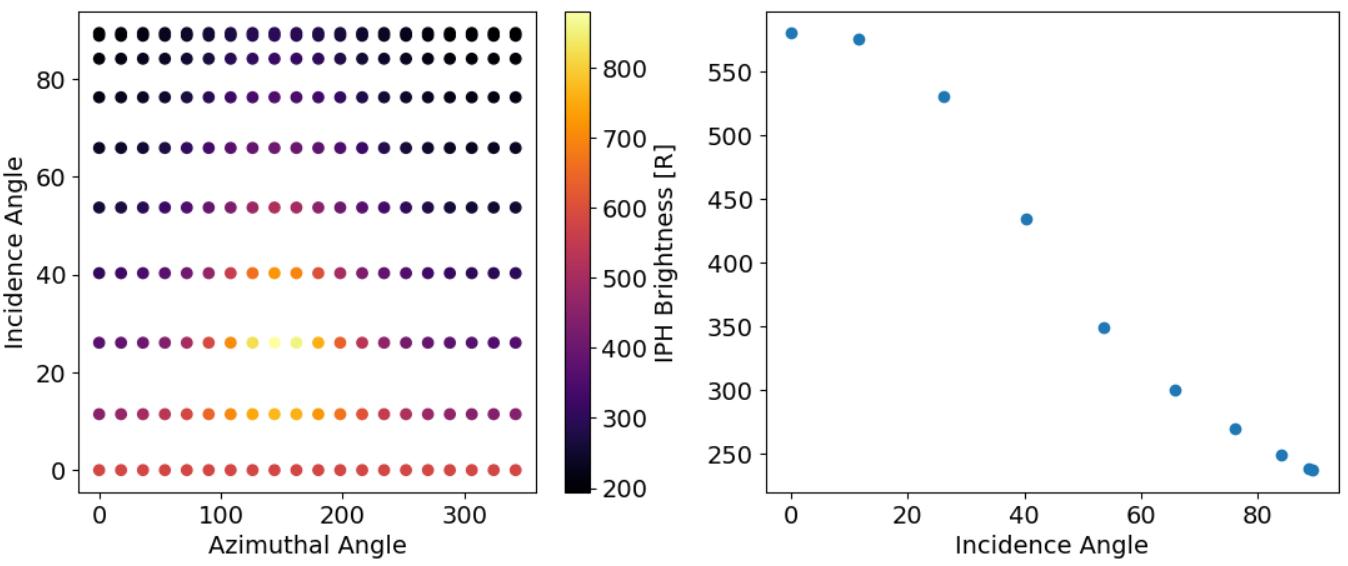}
    \caption{(a) Modelled IPH background brightness as a function of incidence and azimuthal angle for one observation pixel, using the fit in Figure \ref{fig: IPH map 2016}. (b) Azimuthally averaged model IPH brightnesses, which are incorporated into the RT model in Eq. \ref{eq: RT  IPH brightness}. }
    \label{fig: IPH grid brightness}
\end{figure*}

For each pixel, a grid of points is generated around the surface normal (see Fig. \ref{fig: IPH grid brightness}, spanning both incidence angle and azimuthal direction. The angle of each point to the direction of maximum brightness is computed, and used to calculate the IPH brightness in that direction. This brightness of the IPH incident on the Saturn's atmosphere is scaled with the 28-day averaged solar flux at the time of observation. We use a Gaussian lineshape at a temperature of 16,000K (Figure \ref{fig: lya lineshape at 1au}) in the radiative transfer model. 
The scattered brightness from the scattered IPH Lyman$\alpha$ photons is given by  
\begin{multline}\label{eq: RT  IPH brightness}
4\pi I_{IPH}(\mu_{em}) = \Delta\nu_{D, Sat} \times10^{-6} \int_0^{2\pi}d\phi_{in}\times\\\int_0^1d\mu_{in}\int_{-\infty}^{\infty} dx_{in}\int_{-\infty}^{\infty}dx_{em} S(\tau;\bm{x}_{em};\bm{x}_{in}) F_{IPH}(\bm{x}_{in})
\end{multline}
where we have integrated over all incidence angles.
The incident IPH flux is direction dependent such that
\begin{multline}\label{eq: IPH lineshape equation}
F_{IPH}(\bm{x}_{in}) = B_{IPH}(\mu_{in}, \phi_{in}) \times f_{IPH} (x_{in}, \mu_{in}, \phi_{in}),
\end{multline}
 where $f_{IPH} (x_{in}, \mu_{in}, \phi_{in})$ is the lineshape, which is normalised to 1. 
While this does not explicitly depend on the incidence angle, the direction of the normal vector relative to the sunward direction does impact the scattered IPH brightness. 

We tested the impact of including a variable IPH Lyman-$\alpha$ lineshift and temperature into the radiative transfer model. We found that this resulted in scattered brightnesses varying by several rayleighs ($\sim5$\% of the IPH brightness), which in the context of Saturn is negligible and small compared to the observational uncertainties. Therefore, we simplify $f_{IPH} (x_{in}, \mu_{in}, \phi_{in})= f_{IPH} (x_{in})$ in Eq, \ref{eq: IPH lineshape equation}.

\section{Regression of the Lyman-$\alpha$ brightnesses}
As described in Section \ref{sec: MVR outline} and presented in Section \ref{sec: MVR results}, we have analysed the extensive data set of Lyman-$\alpha$ emission brightnesses throughout the Cassini mission. In this section, we present supplementary material and figures from the regression analyses for the different mission periods.
\subsection{Northern hemisphere summer}
\begin{figure*}
    \centering
    \includegraphics[width=\textwidth]{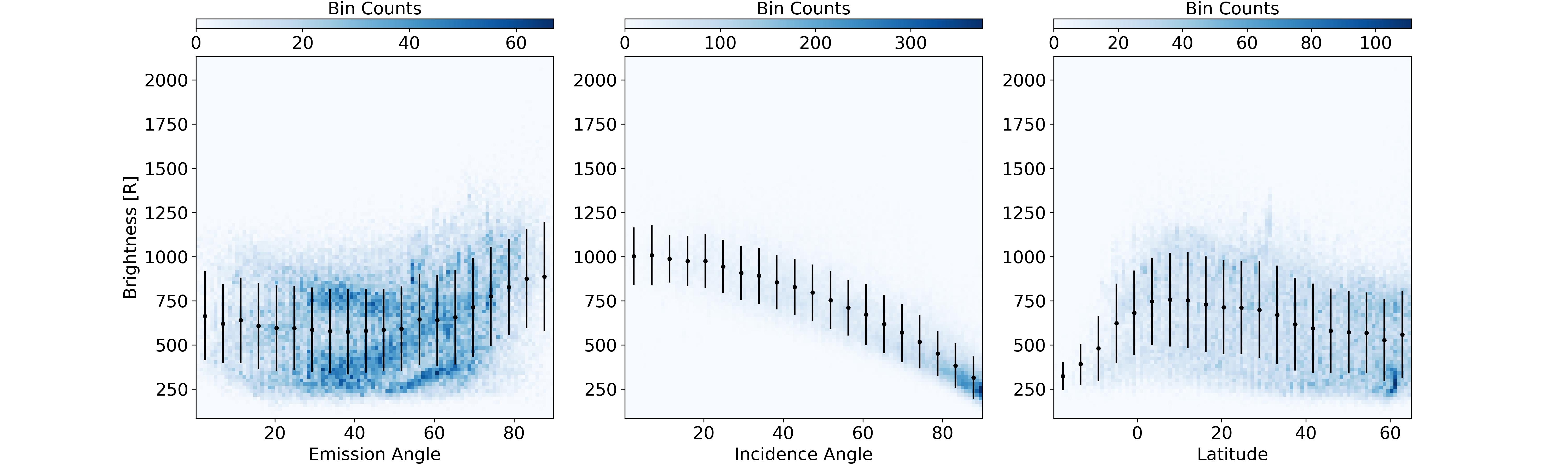}
    \caption{Brightness of the training dataset in the late mission period (2014-2016; see Section \ref{sec: MVR results}) against emission angle, solar incidence angle and latitude. The median and standard deviations of the observed brightnesses are given in bins of each independent variable. }
    \label{fig: MVR dataset late mission}
\end{figure*}

\begin{figure*}
    \centering
    \includegraphics[width=\textwidth]{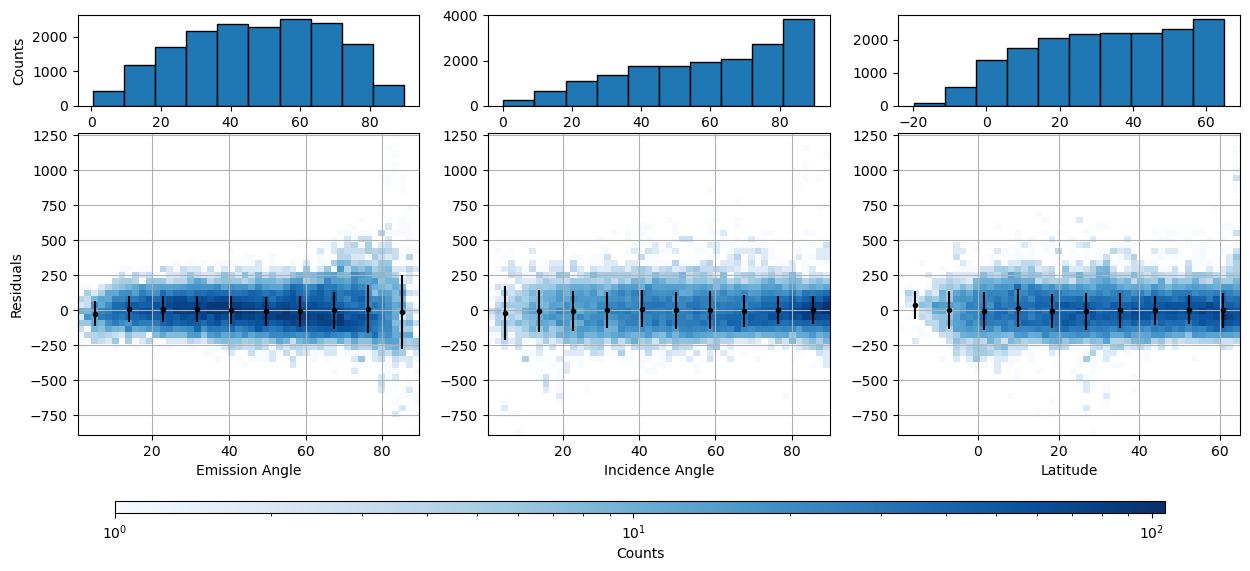}
    \caption{Residuals of the MVR analysis for the testing dataset against each independent variable used in the model for the late mission period (2014-2016). The black points give the mean and standard deviations of the residual brightness.}
    \label{fig: MVR residual vs var late}
\end{figure*}

For the northern hemisphere summer (here taken as 2014-2016), the observed brightnesses show correlations with the emission angle, incidence angle and latitude (see Figure \ref{fig: MVR dataset late mission}). 
The comparison of the observed and predicted brightnesses for 2014-2016 are shown in Figure \ref{fig: MVR prediction vs actual}. We noted that the residuals in Figure \ref{fig: MVR prediction vs actual}d show quadratic behaviour, being positive at small and large predicted brightnesses. However, the quadratic behaviour seen in Fig. \ref{fig: MVR residual vs var late}b, is not seen with respect to each independent variable (Figure \ref{fig: MVR residual vs var late}a-c). Therefore, the choice of a combination of quadratic and cubic fits with respect to the independent variables (see Eq. \ref{eq: MVR fit equation}) is sufficient to describe the brightnesses. 

\begin{figure*}
    \includegraphics[width =\textwidth]{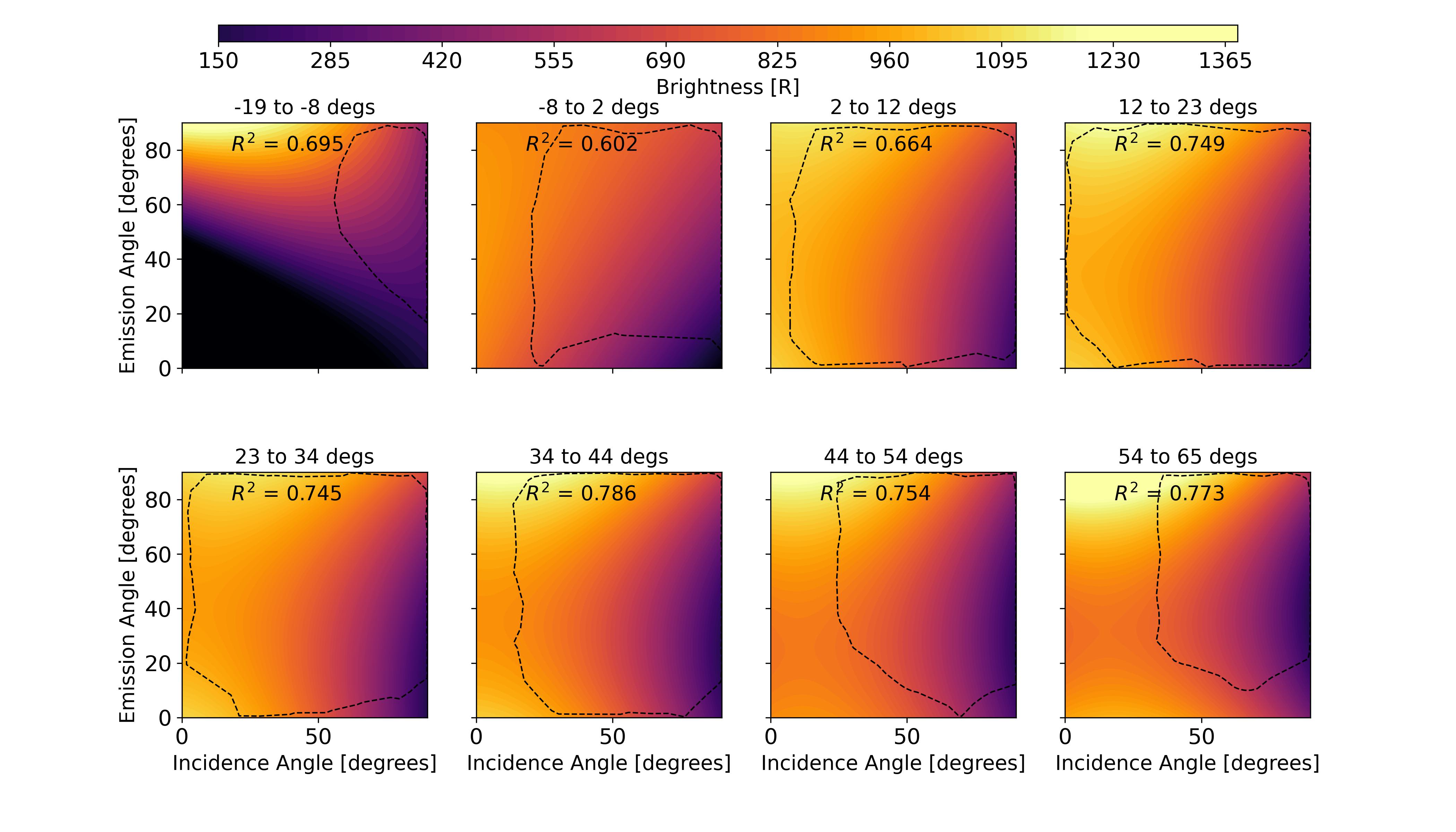}
    \caption{Predicted Lyman-$\alpha$ brightness vs incidence and emission angles in bins of latitude during the northern hemisphere summer (2014-2016). The MVR analysis is applied using a quadratic expression in incidence and emission angles and is trained independently for each latitude bin. The observation coverages in the training datasets are shown in Figure \ref{fig: NH summer counts vs inc em lat}.}
    \label{fig: NH summer B vs inc em lat}
\end{figure*}
\begin{figure*}
    \includegraphics[width =\textwidth]{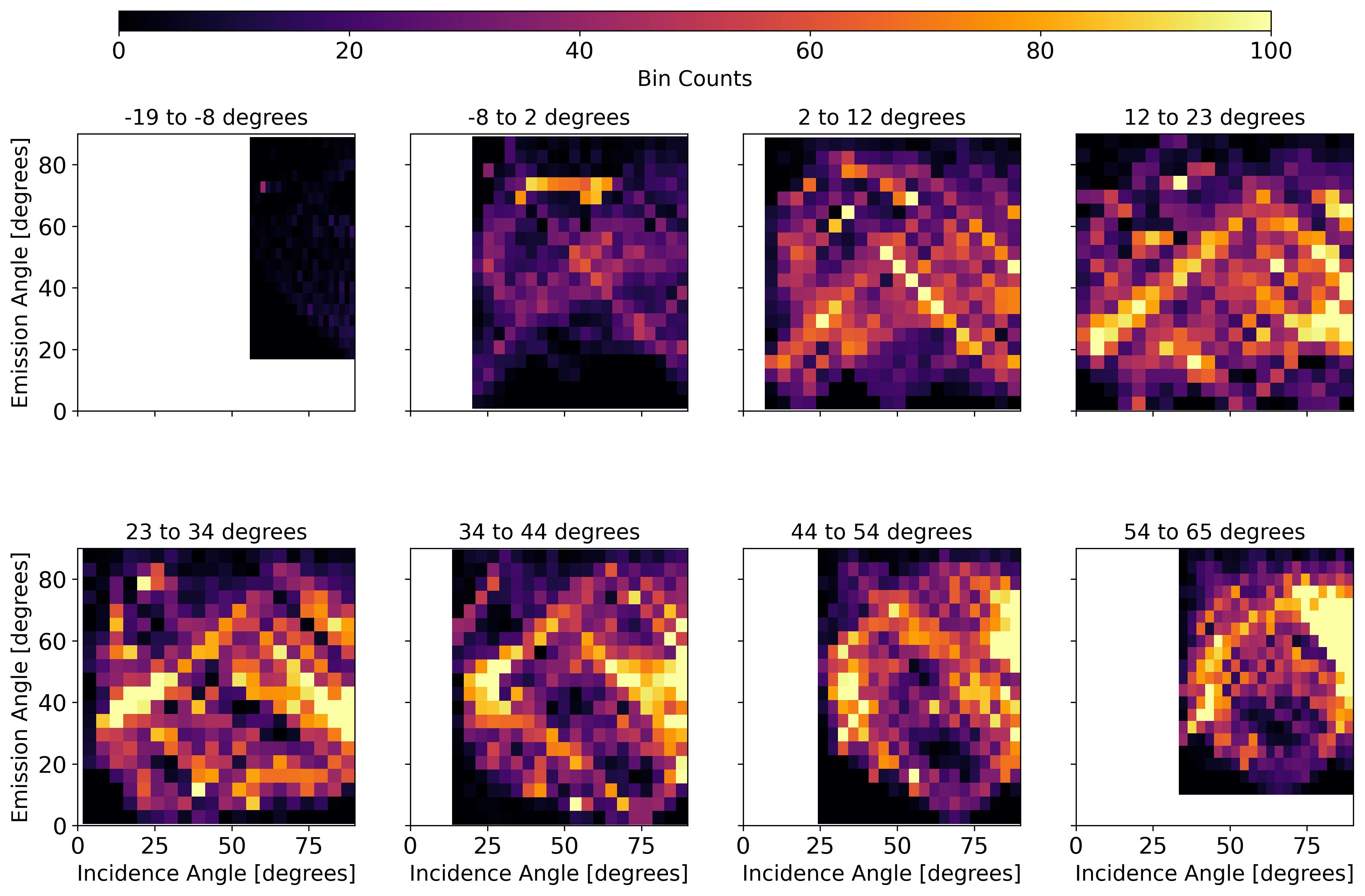}
    \caption{Counts of Lyman-$\alpha$ brightness observations vs incidence and emission angles in bins of latitude during the northern hemisphere summer (2014-2016). The predicted brightnesses from each dataset are shown in Figure \ref{fig: NH summer B vs inc em lat}. }
    \label{fig: NH summer counts vs inc em lat}
\end{figure*}

Figure \ref{fig: NH summer B vs inc em lat} shows the fits of the multi-variate regression in bins of latitude during the northern hemisphere summer. The variation of the observed brightness with incidence and emission angle is consistent across the full latitude range, and similar to the prediction of the radiative transfer model (see Fig. \ref{fig: MVR radiation field}b). The regions of difference with the radiative transfer model can be attributed to poor coverage in the latitude bin (see Figure \ref{fig: NH summer counts vs inc em lat}), so the regression model is not well constrained.


\subsection{Southern hemisphere summer}

\begin{figure*}
    \centering
    \includegraphics[width=\textwidth]{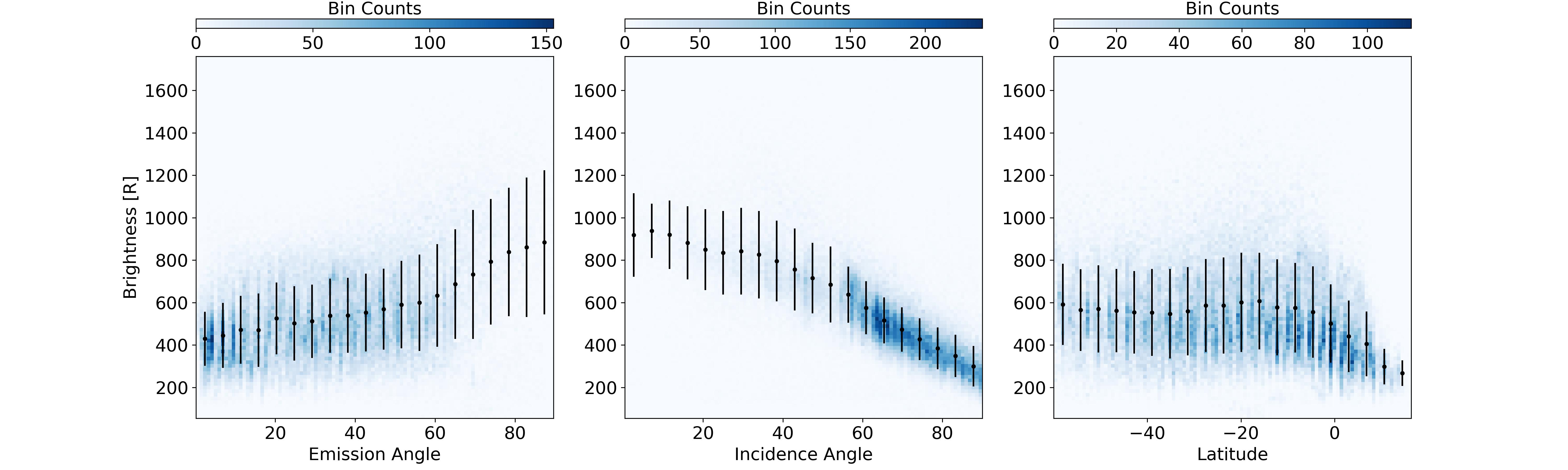}
    \caption{Brightness of the training dataset in the southern hemisphere summer (2004-2007) against emission angle, solar incidence angle and latitude. The median and standard deviations (black) of the observed brightnesses are given in bins of each independent variable.}
    \label{fig: MVR dataset early mission}
\end{figure*}

\begin{table*}[htbp]
\centering
\caption{Coefficients for the fit used in the multi-variate regression (see Eq. \ref{eq: MVR fit equation}), for the 2004-2006 observations.}
\label{tab: MVR coefficients early}
\begin{tabular}{cccc}
\hline
Variable & Coefficient & Mean Value & Confidence Interval \\
\hline
Constant & $p_0$ & $5.95 \times 10^{2}$ & ($5.93 \times 10^{2}$, $5.96 \times 10^{2}$) \\
$\theta_{em}$ & $p_1$ & $2.61 \times 10^{0}$ & ($2.55 \times 10^{0}$, $2.66 \times 10^{0}$) \\
$\theta_{in}$ & $p_2$ & $-8.85 \times 10^{0}$ & ($-8.89 \times 10^{0}$, $-8.81 \times 10^{0}$) \\
$\phi_{lat}$ & $p_3$ & $1.26 \times 10^{0}$ & ($1.15 \times 10^{0}$, $1.37 \times 10^{0}$) \\
$\theta_{em}^2$ & $p_4$ & $9.28 \times 10^{-2}$ & ($9.07 \times 10^{-2}$, $9.50 \times 10^{-2}$) \\
$\theta_{em} \cdot \theta_{in}$ & $p_5$ & $7.08 \times 10^{-3}$ & ($4.96 \times 10^{-3}$, $9.19 \times 10^{-3}$) \\
$\theta_{in}^2$ & $p_6$ & $-5.28 \times 10^{-2}$ & ($-5.45 \times 10^{-2}$, $-5.11 \times 10^{-2}$) \\
$\phi_{lat}^2$ & $p_7$ & $-1.35 \times 10^{-2}$ & ($-1.61 \times 10^{-2}$, $-1.09 \times 10^{-2}$) \\
$\phi_{lat}^3$ & $p_8$ & $-1.35 \times 10^{-3}$ & ($-1.44 \times 10^{-3}$, $-1.25 \times 10^{-3}$) \\
\hline
\end{tabular}
\end{table*}

\begin{figure*}
    \centering
    \includegraphics[width=0.8\textwidth]{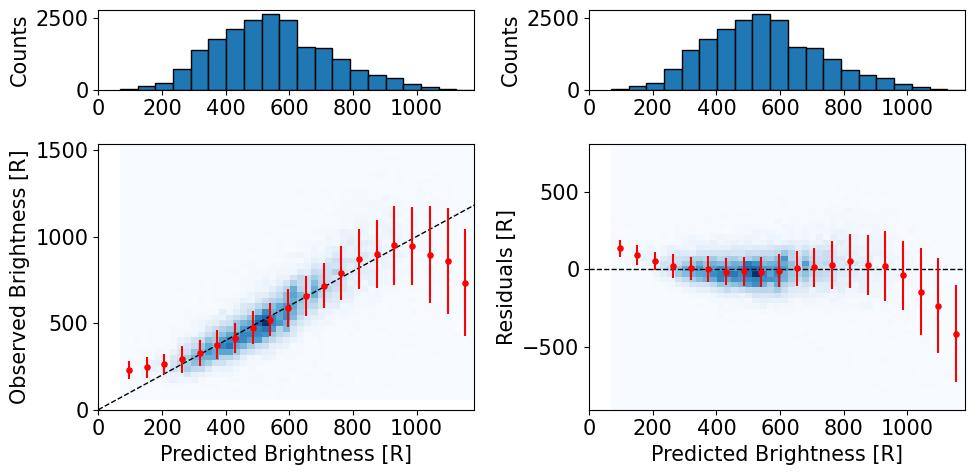}
    \caption{(a,b) Testing data counts vs predicted brightness from the MVR for 2004-2007. (c) Observed brightnesses and (d) residuals vs predicted brightnesses for the testing dataset (blue) with the averages and standard deviations  shown in red. }
    \label{fig: SH summer MVR obs vs predicted}
\end{figure*}

We apply the multi-variate regression model (see Section \ref{sec: MVR outline}) to the observations in the southern hemisphere summer (2004-2007). The fit coefficients are listed in Table \ref{tab: MVR coefficients early}. Figure \ref{fig: MVR dataset early mission} shows the observed Lyman-$\alpha$ brightness during this period, showing similar behaviour with incidence angle and emission angle to that observed in the northern hemisphere summer (see Figure \ref{fig: MVR dataset late mission}). The dependence of the Lyman-$\alpha$ brightness on latitude has reversed, with a peak at $-20^\circ$ and a sharp decrease in brightness in the northern hemisphere. Figure \ref{fig: SH summer MVR obs vs predicted} compares the observed brightnesses with predicted brightnesses and residuals. The model accurately reproduces the observed brightnesses between 200\,R and 900\,R, but overestimates some large brightnesses. There are few counts where the residuals are large, whereas the statistics are much stronger where the residuals are small ($<200$\,R).

\begin{figure*}
    \centering
    \includegraphics[width=\textwidth]{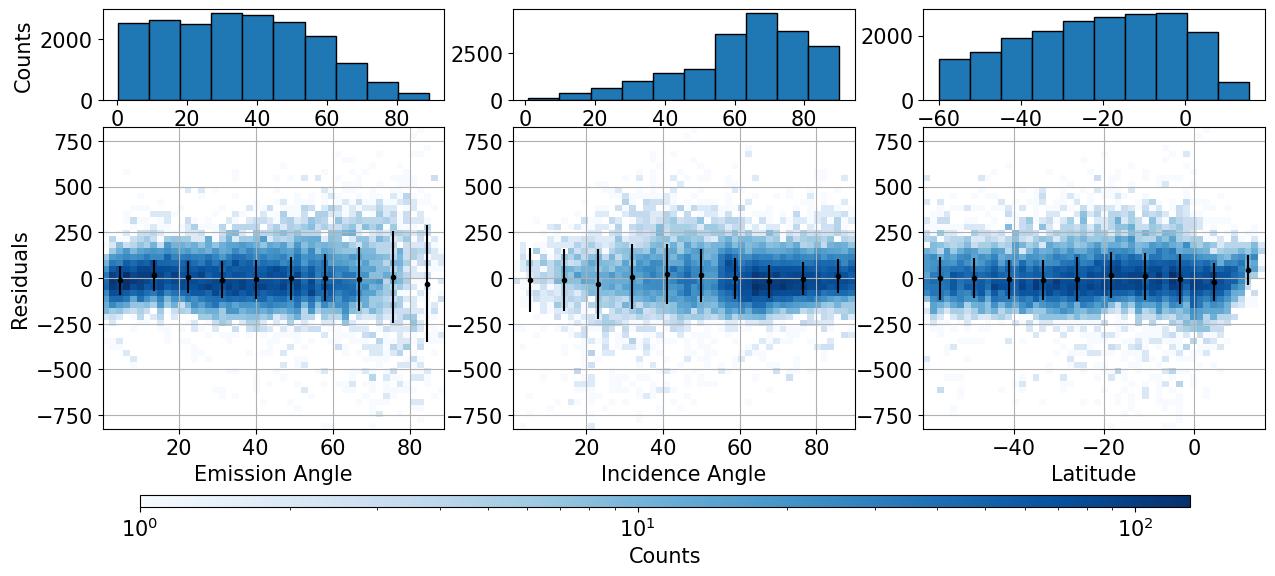}
    \caption{Residuals of the MVR analysis for the testing dataset for 2004-2007 against each independent variable used in the model. The black points give the mean and standard deviations of the residual brightness.}
    \label{fig: MVR residuals early}
\end{figure*}

Similar to the northern summer, the quadratic nature of the residuals in Figure \ref{fig: SH summer MVR obs vs predicted}b are not replicated in the dependences on each independent variable (see Figure \ref{fig: MVR residuals early}). The residuals show little dependence on each of the independent variables, suggesting higher order terms are not required to improve the fits.

\begin{figure*}
\centering

\begin{tikzpicture}
    \node at (0,0) {\includegraphics[width=0.45\textwidth]{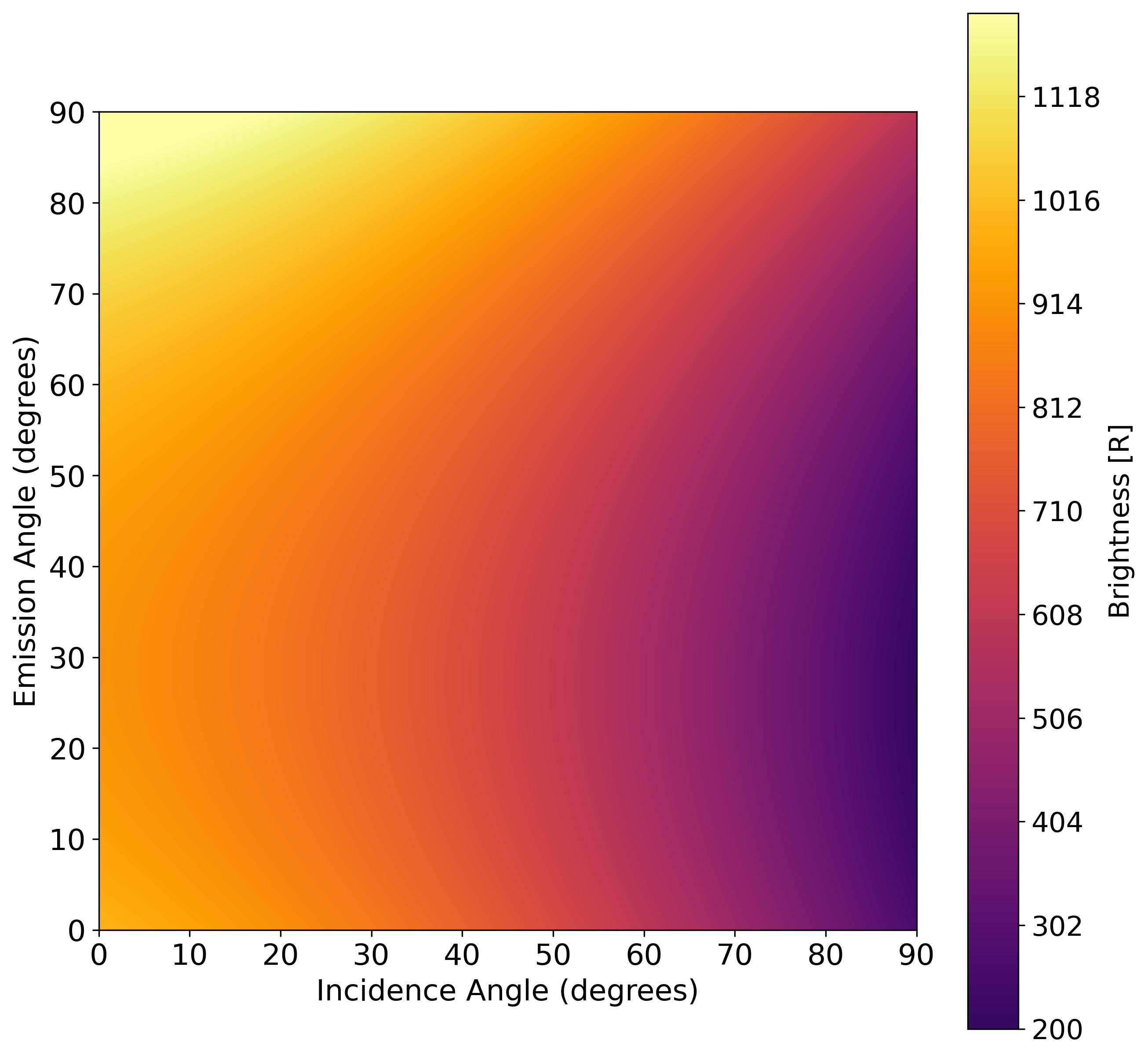}};
    \node at (9,0) {\includegraphics[width=0.45\textwidth]{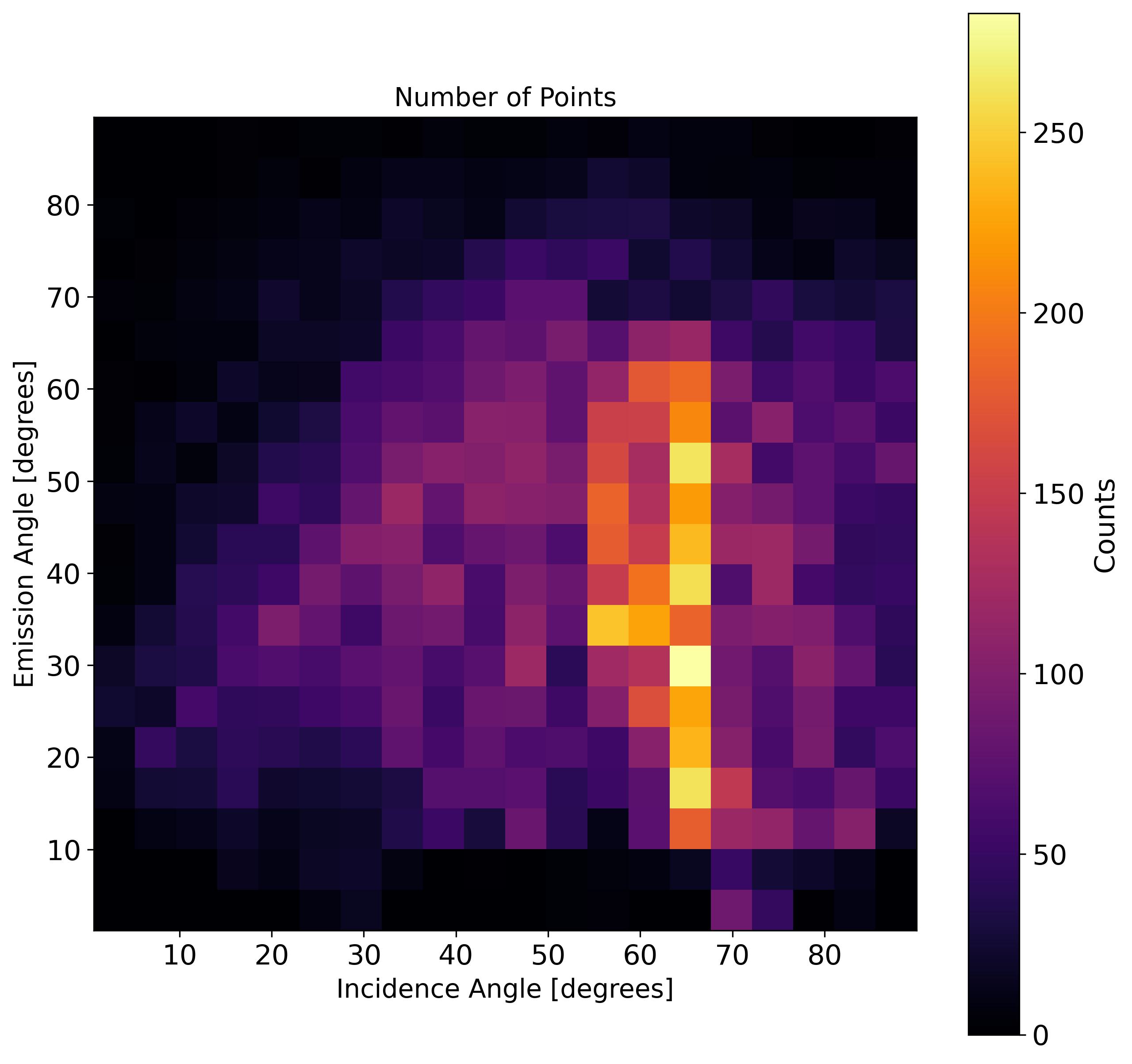}};
\end{tikzpicture}

    \caption{(left) Predicted brightness as a function of solar incidence and emission angles from the MVR analysis at a latitude of -20$^\circ$N for the 2004-2007 dataset. (right) Observation counts as a function of solar incidence and emission angles used to train the MVR analysis for the 2004-2007 dataset. }
    \label{fig: SH summer radiation field}
\end{figure*}

The dependence of the predicted brightness on the incidence and emission angles at $\phi_{lat}=-20^\circ$ (see Figure \ref{fig: SH summer radiation field}) is very similar to that seen in the northern hemisphere summer, and to the predictions of the radiative transfer model (see Figure \ref{fig: MVR radiation field}).

The same structure in the predicted brightness is retrieved for each latitude bin with a good fit ($R^2>0.6$) for all bins in the southern hemisphere, where more observations were available (see Figure \ref{fig: SH summer counts vs inc em lat}). Some bins, such as $\phi_{lat}=[-60, -49]$, predict large brightnesses at low emission angles, contrary to the radiative transfer model. In each of these cases, the data coverage does not extend to low emission angles, so this should not be interpreted physically.

\begin{figure*}
    \includegraphics[width =\textwidth]{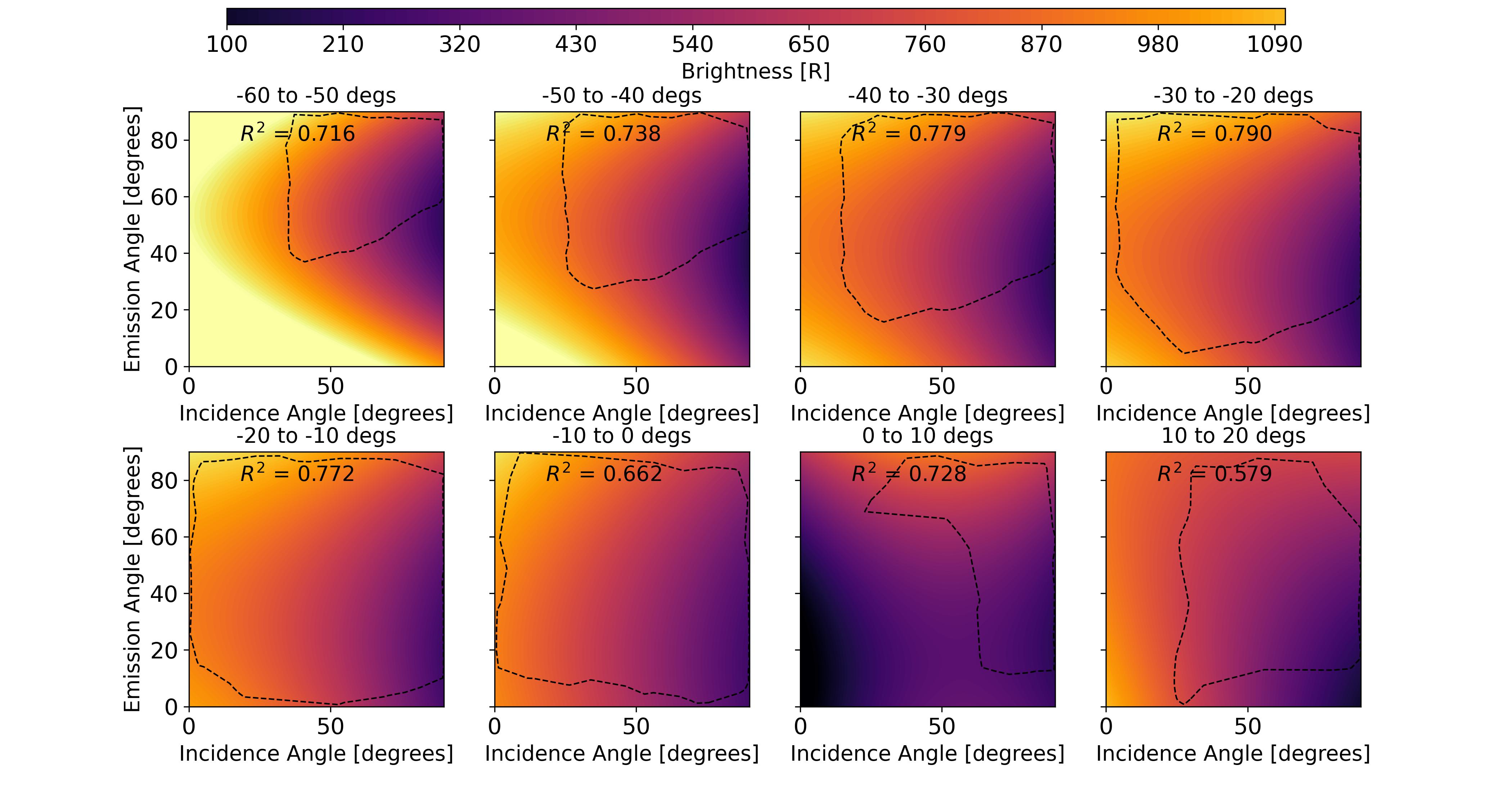}
    \caption{Predicted Lyman-$\alpha$ brightness vs incidence and emission angles in bins of latitude during the southern hemisphere summer (2004-2007). The MVR analysis is applied using a quadratic expression in incidence and emission angles and is trained independently for each latitude bin. The observation coverage in the training dataset are shown in Figure \ref{fig: SH summer counts vs inc em lat}. }
    \label{fig: SH summer B vs inc em lat}
\end{figure*}
\begin{figure*}
    \includegraphics[width =\textwidth]{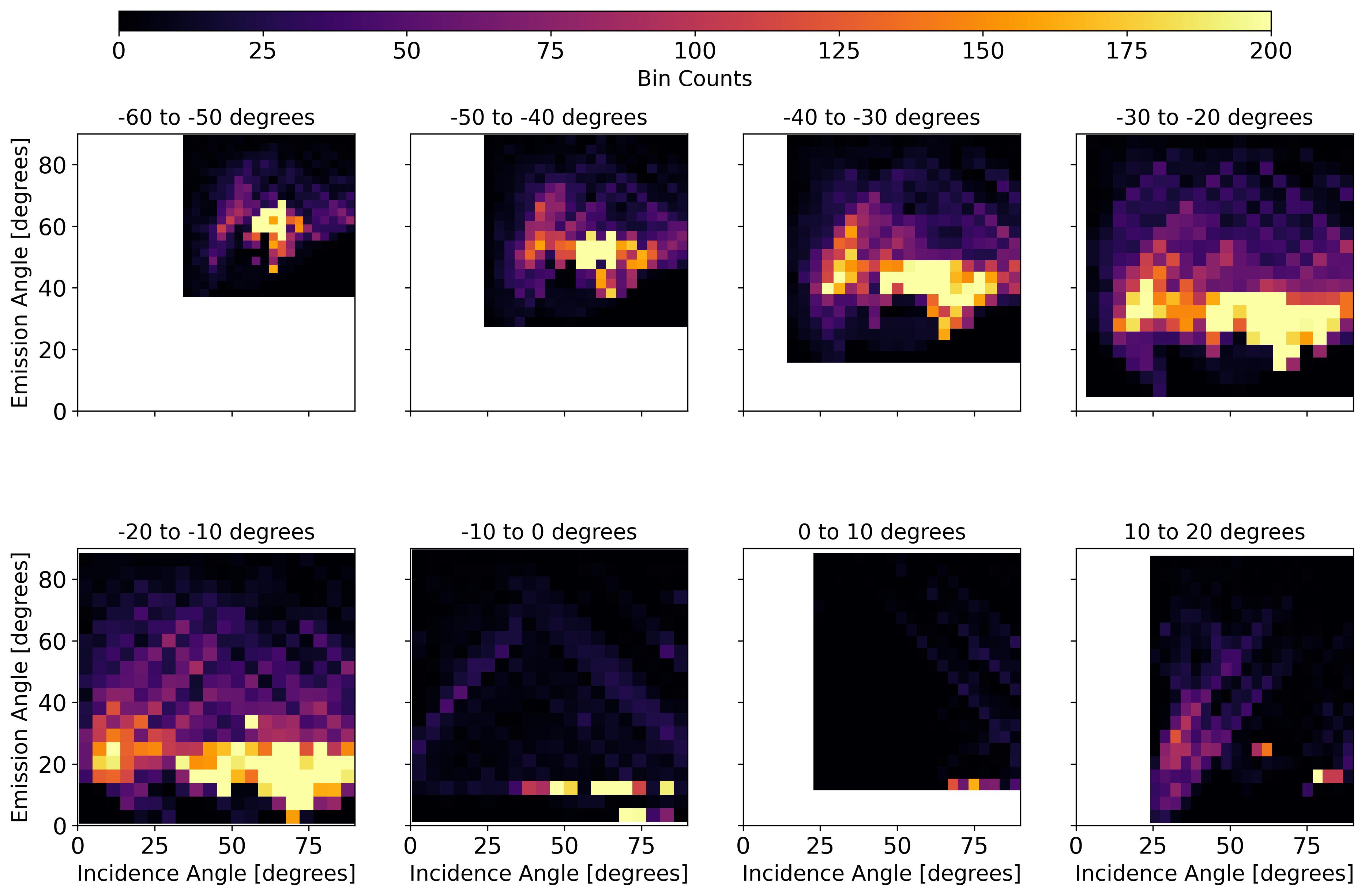}
    \caption{Counts of Lyman-$\alpha$ brightness observations vs incidence and emission angles in bins of latitude during the southern hemisphere summer (2004-2007). The predicted brightnesses from each dataset are shown in Figure \ref{fig: SH summer B vs inc em lat}. }
    \label{fig: SH summer counts vs inc em lat}
\end{figure*}

\section{Absorption by the interplanetary Lyman-$\alpha$}
Between 1 au, where the Lyman alpha profile has been measured \citep{Lemaire2005Variation23}, and Saturn, the interplanetary hydrogen background absorbs some of the Lyman $\alpha$, particularly near the line-centre, reducing the line-centre flux by up to 13\%.
As the flux at line centre is important for resonance scattering, we model the absorption of the IPH background using the IPH model of \cite{Quemerais2013CrosscalibrationHeliosphere, Izmodenov2013DistributionLyman-}, which is discussed in Section \ref{sec: IPH model description}.

We calculate the absorption for 36 radial profiles distributed along Saturn's orbit, relative to the flow of the local interstellar medium. The optical depth of the IPH background is integrated radially to a heliocentric distance of 9.5\,au, including the Doppler shift and width of the absorption profile resulting from the bulk IPH velocity. 

Using the optical depth, we correct the Lyman-$\alpha$ lineshape and average the absorption profiles around the linecenter. It is assumed that resonance scattering is symmetric about the linecenter in the radiative transfer model, so the absorption must also be made symmetric. 

Figure \ref{fig: lya lineshape at 1au} shows the corrected Lyman-$\alpha$ lineshape for the mean absorption profile (dashed black line) with the absorption variability (black shaded region), compared to the flux shape without IPH absorption.

\section{IPH observations}
The observations of the Lyman-$\alpha$ IPH background we use in Section \ref{sec: UVIS IPH comparison} are listed in Table \ref{tab: IPH observations}.
\begin{table*}[h]
\centering
\caption{Observations of the IPH Lyman-$\alpha$ background by {Cassini/UVIS} in 2006, with heliocentric distance ($D_H$) and background brightness. The right ascension and declination of the observation and the sunward direction ($\odot$) are given. }
\label{tab: IPH observations}
\begin{tabular}{ccccccccccccccccc}
\hline
Day of year & Lead & $D_H$ & Background [R] & Ly-$\alpha$ [R] & Scaled B [R] &RA & DEC & RA$_\odot$ & DEC$_\odot$ \\
\hline

014 & CIRS & 9.08 & 4.04$\pm$2.20 & 251$\pm$72 & $229\pm65$ & 101.7 & -2.75 & 309.91 & -19.04 \\
015 & UVIS & 9.08 & 5.50$\pm$1.10 & 557$\pm$61 & $508\pm55$ & 279.1 & 3.01 & 309.92 & -19.02 \\
    & ISS  & 9.08 & 5.74$\pm$0.69 & 596$\pm$39 & $544\pm35$ & 277.4 & 3.18 & 309.93 & -19.02 \\
    & CIRS & 9.08 & 5.80$\pm$0.94 & 348$\pm$47 & $317\pm42$ & 101.7 & -2.75 & 309.91 & -19.03 \\
    & CIRS & 9.08 & 5.75$\pm$1.66 & 202$\pm$14 & $184\pm12$ & 101.2 & -2.81 & 309.92 & -19.03 \\
057 & VIMS & 9.10 & 5.53$\pm$1.22 & 183$\pm$12 & $172\pm11$ & 134.2 & 0.75 & 311.46 & -18.67 \\
058 & UVIS & 9.10 & 5.45$\pm$1.67 & 619$\pm$52 & $582\pm48$ & 309.6 & -0.24 & 311.53 & -18.65 \\
    & CIRS & 9.10 & 5.40$\pm$0.64 & 673$\pm$45 & $633\pm42$ & 309.0 & -0.17 & 311.54 & -18.65 \\
    & ISS  & 9.10 & 5.91$\pm$1.04 & 286$\pm$74 & $269\pm69$ & 133.2 & 0.64 & 311.49 & -18.66 \\
077 & CIRS & 9.10 & 5.47$\pm$1.44 & 304$\pm$27 & $288\pm25$ & 73.5 & -5.12 & 312.28 & -18.52 \\
    & ISS  & 9.10 & 6.24$\pm$2.71 & 341$\pm$36 & $323\pm34$ & 72.2 & -5.2  & 312.29 & -18.52 \\
078 & CIRS & 9.09 & 5.46$\pm$1.70 & 433$\pm$32 & $411\pm30$ & 248.8 & 5.42 & 312.30 & -18.51 \\
    & VIMS & 9.09 & 4.97$\pm$1.51 & 444$\pm$15 & $421\pm14$ & 249.5 & 5.38 & 312.30 & -18.51 \\
121 & UVIS & 9.11 & 5.66$\pm$1.33 & 596$\pm$55 & $546\pm50$ & 282.1 & 2.72  & 313.84 & -18.12 \\
140 & UVIS & 9.11 & 5.27$\pm$1.54 & 274$\pm$37 & $250\pm33$ & 44.0 & -6.31 & 314.62 & -17.97 \\
183 & UVIS & 9.12 & 5.78$\pm$1.15 & 445$\pm$28 & $416\pm26$ & 255.3 & 4.92 & 316.14 & -17.58 \\
250 & CIRS & 9.14 & 5.44$\pm$0.92 & 333$\pm$21 & $310\pm19$ & 19.2 & -15.1 & 318.67 & -16.97 \\
    & UVIS & 9.14 & 4.85$\pm$2.06 & 330$\pm$27 & $307\pm25$ & 20.6 & -15.8 & 318.67 & -16.96 \\
    & CIRS & 9.14 & 5.52$\pm$1.15 & 247$\pm$84 & $229\pm78$ & 25.9 & -18.0 & 318.68 & -16.96 \\
    & ISS  & 9.14 & 6.08$\pm$1.12 & 249$\pm$43 & $231\pm40$ & 204.7 & 17.6 & 318.69 & -16.96 \\
    & ISS  & 9.14 & 5.51$\pm$1.37 & 237$\pm$52 & $220\pm48$ & 205.3 & 17.8 & 318.67 & -16.96 \\
266 & UVIS & 9.14 & 5.10$\pm$0.94 & 288$\pm$38 & $267\pm35$ & 27.2 & -20.6 & 319.26 & -16.81 \\
282 & CIRS & 9.14 & 4.73$\pm$1.79 & 300$\pm$8 & $279\pm7$ & 28.5 & -28.5 & 319.84 & -16.66 \\
    & CIRS & 9.14 & 5.40$\pm$1.01 & 337$\pm$15 & $313\pm13$ & 210.2 & 33.8 & 319.86 & -16.66 \\
283 & CIRS & 9.14 & 5.01$\pm$1.12 & 292$\pm$17 & $271\pm15$ & 208.0 & 32.3 & 319.89 & -16.64 \\
297 & UVIS & 9.15 & 5.43$\pm$0.88 & $317\pm11$ & $298\pm10$ & 31.0 & -35.3 & 320.40 & -16.52 \\
299 & VIMS & 9.15 & 5.73$\pm$1.55 & $315\pm31$ & $296\pm29$ & 219.4 & 36.3 & 320.44 & -16.50 \\
346 & UVIS & 9.16 & 5.00$\pm$1.18 & $293\pm33$ & $271\pm30$ & 39.6 & -38.7 & 322.15 & -16.04 \\
    & CIRS & 9.16 & 5.10$\pm$2.15 & $289\pm28$ & $267\pm25$ & 215.1 & 42.9 & 322.18 & -16.02 \\
361 & CIRS & 9.16 & 5.37$\pm$1.01 & $308\pm12$ & $288\pm11$ & 34.8 & -44.6 & 322.72 & -15.88 \\
362 & CIRS & 9.16 & 5.21$\pm$0.51 & $300\pm25$ & $281\pm23$ & 218.3 & 48.5 & 322.74 & -15.87 \\
    & CIRS & 9.16 & 5.21$\pm$0.84 & $286\pm22$ & $268\pm20$ & 219.2 & 49.6 & 322.75 & -15.87 \\

\end{tabular}
\end{table*}

\bibliography{references}{}
\bibliographystyle{aasjournal}
\end{document}

%% file: preamble.tex
\usepackage{mhchem}
\usepackage{tikz, mathtools}
\usetikzlibrary{shapes.geometric, arrows,arrows.meta,positioning,quotes,automata,decorations.text}
\usepackage{graphicx, xcolor, wrapfig}
\usepackage{esint,bm}
\usepackage{outlines}

\newcommand{\comments}[1]{{\color{black}#1}}

